\documentclass[reprint,aps,prb,amsmath,amssymb,superscriptaddress]{revtex4-2}
\pdfoutput=1
\usepackage{graphicx}
\usepackage{dcolumn}
\usepackage{bm}
\usepackage{subfigure}
\usepackage{tikz}
\usepackage{lipsum}
\usepackage{braket}
\usepackage{float}
\usepackage{hyperref}
\usepackage{stackengine}
\graphicspath{ {./images/} }

\begin{document}

\title{The Crossover from Ordinary to Higher Order van Hove Singularity \\
 in a Honeycomb System:
A Parquet Renormalization Group Analysis}

\author{Yueh-Chen Lee}
\email{lee02892@umn.edu}
 \affiliation{School of Physics and Astronomy and William I. Fine Theoretical Physics Institute, University of Minnesota, Minneapolis, Minnesota 55455, USA}
\author{Dmitry V. Chichinadze}
\email{chichinadze@magnet.fsu.edu}
\affiliation{National High Magnetic Field Laboratory, Tallahassee, Florida 32310}
\author{Andrey V. Chubukov}
\email{achubuko@umn.edu}
\affiliation{School of Physics and Astronomy and William I. Fine Theoretical Physics Institute, University of Minnesota, Minneapolis, Minnesota 55455, USA}
\date{\today}
\begin{abstract}
We investigate the crossover from an ordinary van Hove singularity  (OVHS) to a higher order van Hove singularity (HOVHS) in a model applicable to Bernal bilayer graphene and rhombohedral trilayer graphene in a displacement field.   At small doping, these systems possess three spin-degenerate  Fermi pockets near each Dirac point $K$ and $K'$;  at larger doping, the three pockets  merge into a single one.  The transition is of Lifshitz type and includes van Hove singularities.   Depending on system parameters, there are either 3 separate OVHS or a single HOVHS.
 We model this behavior by a one-parameter dispersion relation, which interpolates between  OVHS and HOVHS.
 In each case, the diverging density of states triggers various electronic orders (superconductivity, pair density wave, valley polarization, ferromagnetism, spin and charge density wave).
   We apply parquet renormalization group (pRG) technique and analyze how the ordering tendencies evolve between
   OVHS and HOVHS.  We report rich system behavior caused by disappearance/reemergence and pair production/annihilation of the fixed points of the pRG flow.
\end{abstract}

\maketitle

\section{Introduction}

One of the key  topics in the field of quantum materials is the interplay of different ordering tendencies
for interacting electrons. Usually, the most interesting results come from regions of the phase diagram where multiple phases are nearby and compete for the place on the phase diagram.   The well-known and well-studied examples are cuprates and Fe-pnictides/chalcogenides, in which the competitors are antiferromagnetism, superconductivity, pair-density-wave, charge density wave order (both real and current-type) and nematicity
    (see e.g., \cite{Keimer2015,fernandes2016low,Varma2020}).
    Some more recent examples of systems with multiple ordered states are twisted bilayer graphene (TBG) \cite{Cao2018SC,Cao2018insulator,Cao2020NematicSC,Zondiner2020,ali_2,Wu2021,Nadj_Perge_cascade}, Bernal bilayer graphene (BBG) \cite{Zhou2022,Holleis2023}, rhombohedral trilayer graphene (RTG) \cite{Zhou2021SC,Zhou2021}, and other hexagonal/honeycomb systems \cite{Park2021TTG}.
  From theory perspective, the interplay between different ordering instabilities is a complex problem. In strongly correlated
   electron systems one has to rely on numerical methods  and on a comparison of the energies of particular ordered states within, e.g., Hartree-Fock approximation (see, e.g., ~\cite{White2021,Xie2021}.
   For weakly/moderately correlated metallic systems, the competition can be analyzed in a controllable way
    if the polarizations in several different channels (particle-particle and particle-hole ones) are logarithmically singular.  The examples are the interplay between superconductivity and magnetism in Fe-pnictides, where superconducting and antiferromagnetic polarizations are  both logarithmic~\cite{CEE,Khodas2016,Classen2017}
    (superconducting  because of Cooper logarithm and antiferromagnetic because of different signs of dispersion
    of fermions near a hole and an electron pocket), and  the interplay between superconductivity and diagonal $(Q,Q)$ charge order in the spin-fermion model for the cuprates, where both polarizations are
    again logarithmic~\cite{Metlitski2010,Metlitski_2010,Sachdev_2012,Wang2014}.

When perturbation theory contains logarithms, one can  attempt to sum up infinite perturbative series in $(g_i L)$, where $g_i$ are the couplings and $L$ is the logarithmic factor, and neglect terms that contain additional powers of $g_i$  without logarithms. For the case when only a superconducting channel is logarithmic, this amounts to summing up Cooper logarithms and neglecting non-logarithmic corrections.   When there is more than one logarithmic channel, the computational procedure gets  more involved as there are several "directions" in which perturbative series in $(g_i L)$ have to be extended.  It turns out, however, that these series can be expressed as a set of coupled differential equations, known as parquet Renormalization Group (pRG) equations~\cite{Dzyaloshinskii_1987,Schulz_1987,Chubukov2009Review} (for a general review on parquet RG for metals, see \cite{Shankar1994RMP,Rice2009}).
 By solving the coupled set of differential pRG equations and analyzing the pRG flow,  one can identify the channel (particle-particle or particle-hole) in which the instability develops first.
A conceptually similar approach, called functional RG (fRG)  is to divide a Fermi surface into patches, instead of channels, and solve the set of coupled differential equations for the running couplings within patches (for a review, see~\cite{Metzner2012}, for fRG for RTG see \cite{Qin2023}).
\begin{figure*}
\centering
\includegraphics[width=15cm,height=7cm]{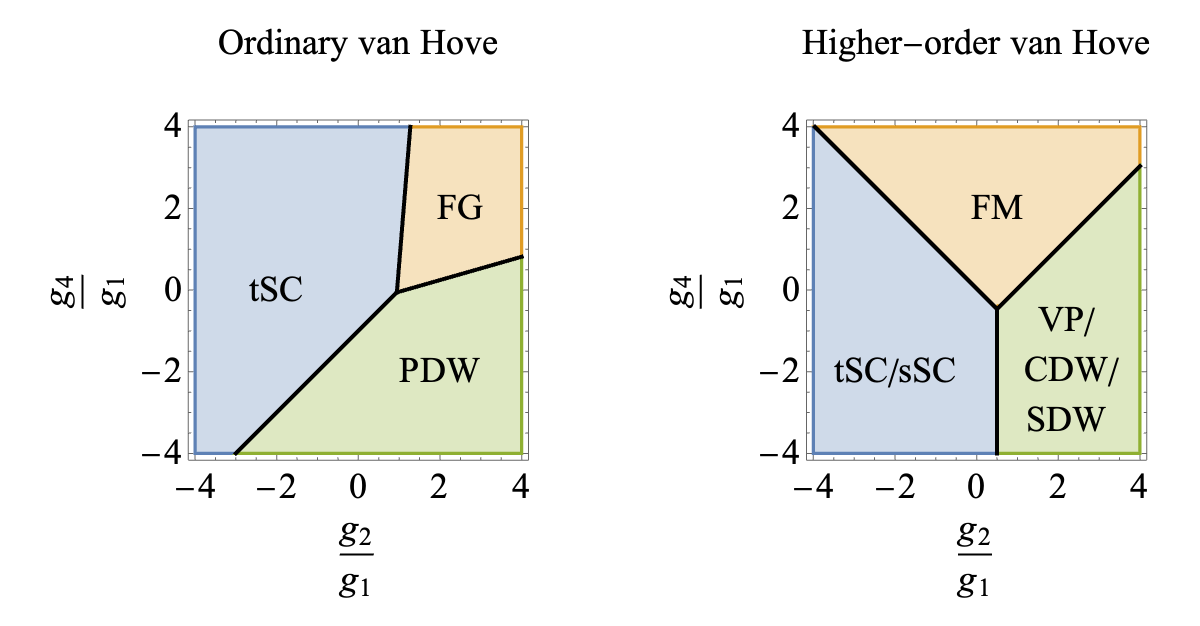}
    \caption{The phase diagram in the OVHS limit (a)  and  HOVHS limit (b).  The variables are the ratios of the bare couplings defined in Eq. (\ref{a_1}).
    In the OVHS limit, the ordered states are spin-triplet superconductivity (tSC) and pair density wave (PDW), and there is a parameter range of asymptotically free Fermi gas with no order (FG).  In the HOVHS limit, the ordered states are ferromagnetism (FM), superconductivity, degenerate between spin-triplet (tSC) and spin-singlet (sSC), and valley polarization (VP), degenerate with spin density wave (SDW) and charge density wave (CDW).}
    \label{fig:upperpanel}
\end{figure*}

\begin{figure}
    \centering
    \includegraphics[width=6cm,height=6cm]{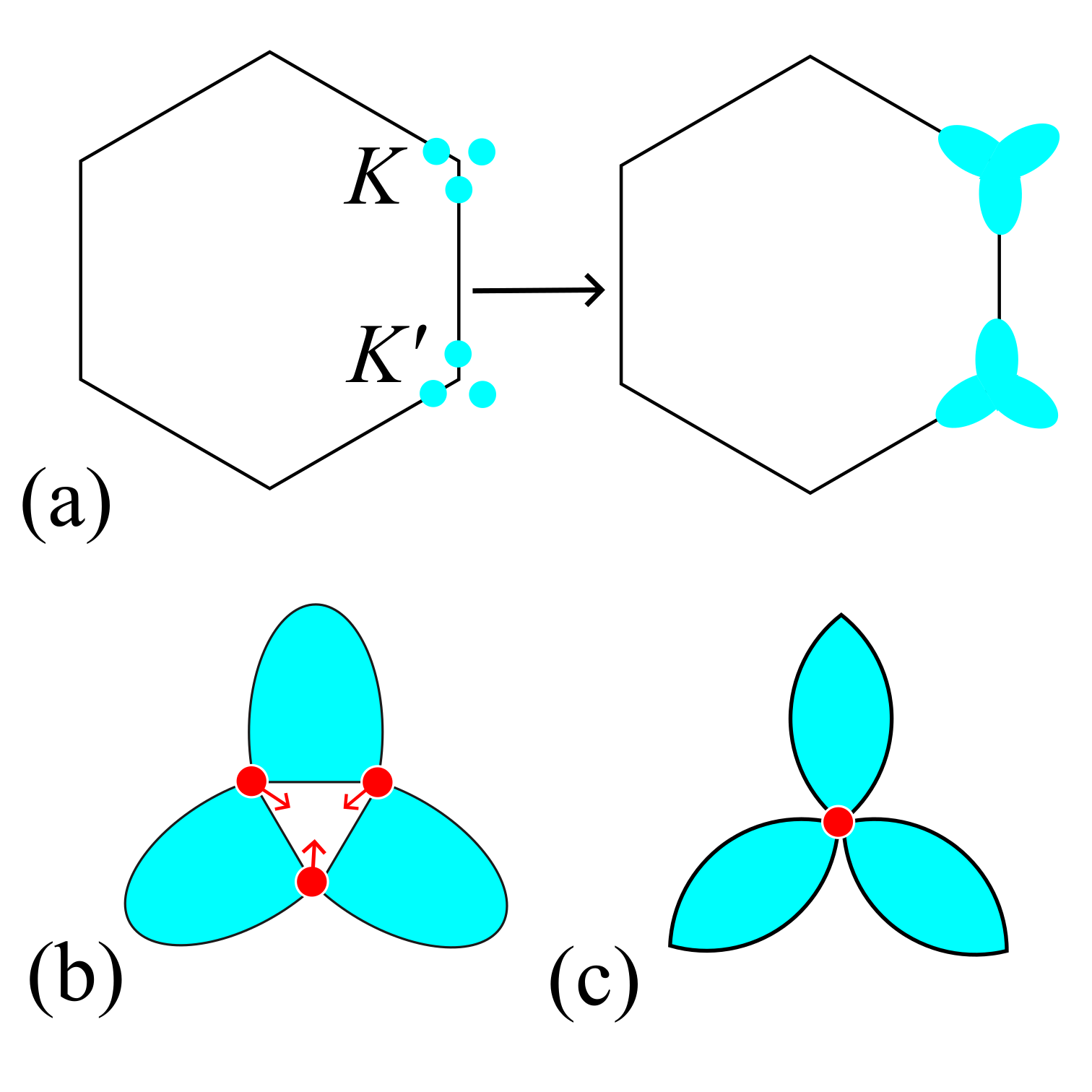}
    \caption{(a) Band dispersion near $K$ and $K'$ points. At low doping (left), there are three smaller electron pockets (cyan circles) around $K$ and three around $K'$. At high doping (right), the three pockets merge into a single Fermi surface.
    (b, c) A zoomed-in sketch of the dispersion at the VH point in the OVHS case (b) and in the case of HOVHS (c).
      The red dots represent the VH points, and the arrows denote the direction in which they move throughout the crossover from OVHS to HOVHS.   The dispersion at $K'$ can be obtained by rotating (b) and (c) by 60 degrees.}
    \label{fig:van hove points}
\end{figure}
  \begin{figure*}
\includegraphics[width=18.635cm,height=15.525cm]{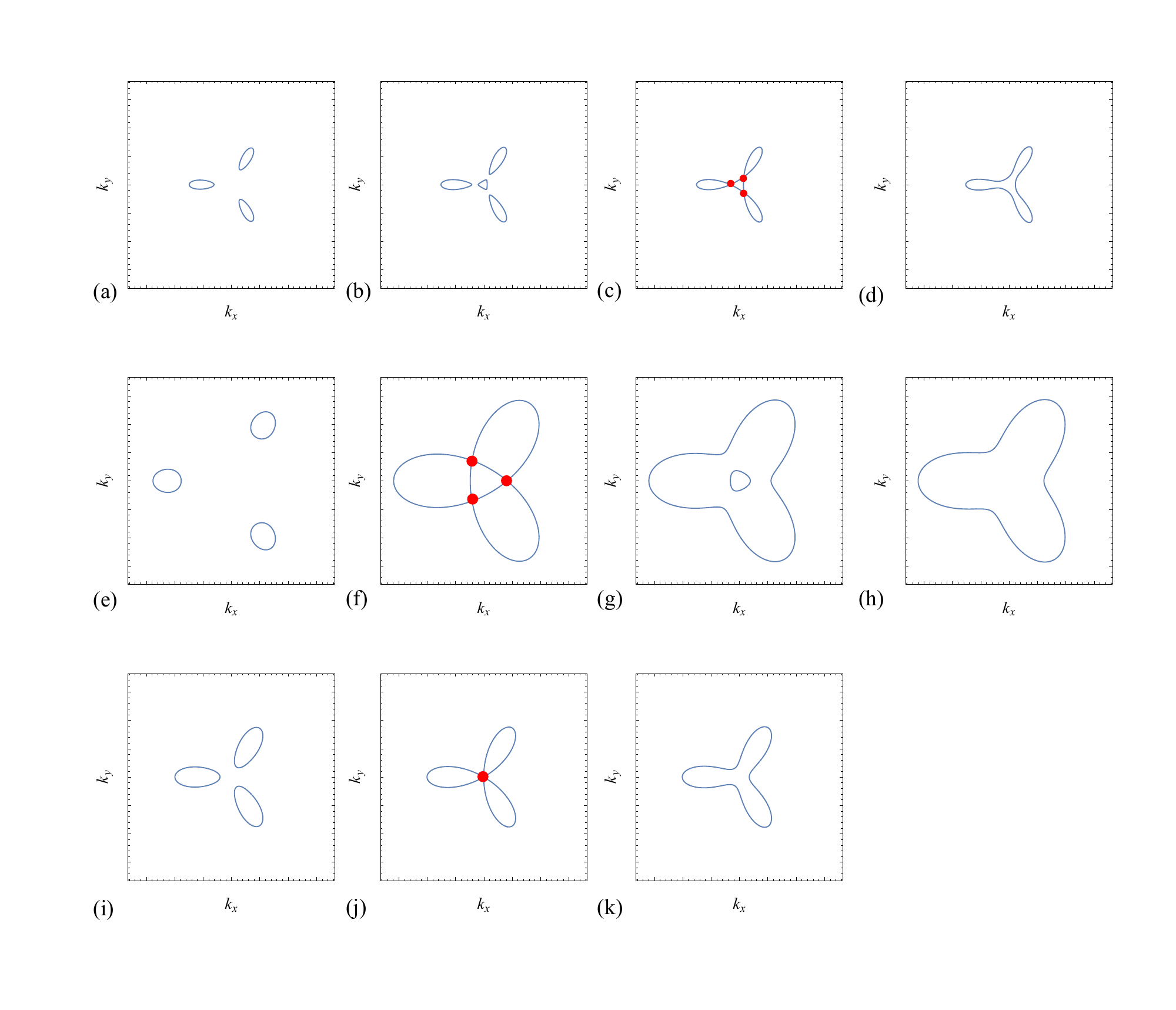}
     \caption{The sketch of the three scenarios for the Fermi surface evolution between three small pockets  near $K (K')$ and one larger pocket at  $K (K')$, (three panels, depending on the strength of the displacement field).
     In each row, the doping increases from left to right, and  van Hove  points are labeled as the red dots.
      In a weak displacement field (the top panel), as doping increases, a new pocket emerges at $K (K')$, and at van Hove doping, the three original pockets touch the center one, creating three OVHS. At larger doping, there is a
       singular Fermi surface. In a strong  displacement field (middle panel) the three pockets increase upon doping and touch each other at van Hove filling, creating again three OVHS.  At larger doping, there are
       two annular  Fermi surfaces.
        Bottom panel: at a critical displacement field, the three pockets touch each other at a single point, creating a HOVS.}
     \label{fig:FS_evolution}
 \end{figure*}
The RG analysis (both pRG and fRG) has been applied
 to fermions near a Lifshitz transition, in which its topology changes
 \cite{Furukawa1998PRL,Nandkishore2012,FaWang2013,Isobe2018PRX,SherkunovPRB,Ashvin,Nandkishore2014}. At the transition, the fermionic density of states diverges logarithmically, and polarizations in various channels become logarithmically singular.
   This divergence is known as van Hove singularity~\cite{VanHovePhysRev}.
  The superconducting polarization generally diverges faster than other polarizations, as $L^2$, however,  if there is a nesting, some particle-hole polarizations also diverge as $L^2$.  To treat the case of a  non-perfect nesting, Furukawa, Rice, and Salmhofer  \cite{Furukawa1998PRL} suggested treating $O(L)$ terms  as $L^2$ terms with phenomenological prefactors, $\alpha_j <1$.  This way one can reduce logarithmical series to the set of differential equations with variable $L^2$ rather than $L$, solve them, and identify the leading instabilities. If the same instability wins in a sizable range of $\alpha_j$, one can reasonably expect that this is the correct result, particularly because the leading instability develops at a finite $L \sim 1/\sqrt{g}$, when the difference between $L^2$ and $L$ with a large prefactor is not  crucial.

In recent years, several groups\cite{Shtyk2017PRB,Yuan2019,Classen2020PRB,Isobe2019,Link2019,castro2023emergence} extended the pRG analysis to higher-order Van Hove singularity (HOVHS), in which the density of states diverges by a power law.
   Shtyk et al argued~\cite{Shtyk2017PRB} that by varying tight-binding parameters of several hexagonal/honeycomb systems, one can move from ordinary van Hove singularity (OVHS)  to HOVHS.  For a HOVHS, superconducting  channel is no longer special as the power-law divergent polarization
   in the particle-particle channel is comparable to polarizations in particle-hole channels. It has been proposed~\cite{Shtyk2017PRB,Classen2020PRB,castro2023emergence} that one can still write a set of pRG equations, keeping the same processes, which gave rise to a logarithmic flow of the couplings for an OVHS, and taking the derivative with respect to the power of a typical scale instead of its logarithm.  This approach is
 less rigorous
          because the contributions from the processes, previously neglected as non-logarithmic, are of the same order as the ones from formerly logarithmic processes, As a result, the one-to-one correspondence with the perturbation theory, which exists for pRG near an OVHS,  is lost for a HOVHS. Still,  a comparison between the solution of HOVHS pRG equations and the direct perturbation theory shows~\cite{Classen2020PRB}  that the difference is only quantitative and can be small numerically. This justifies, at least partly,  the use of pRG for systems with HOVHS.

 In this paper, we analyze the interplay between ordering tendencies near OVHS and HOVHS for a system of fermions on a honeycomb lattice.  Specifically, we consider the model of low-energy fermions in the vicinity of two valleys, $K$ and $K'$. Such an electronic configuration emerges in BBG and RTG in the presence of a finite displacement field~\cite{Zhou2022,Zhou2021SC,Zhou2021,Holleis2023}. In both systems, electronic configuration at small hole/electron doping consists of three spin-degenerate small pockets near $K$ and three near $K'$, while at larger doping there is one spin-degenerate Fermi surface near $K$ and one near $K'$ (see Fig.~\ref{fig:van hove points}(a)).
For non-interacting fermions, a topological transformation from three small pockets to one larger Fermi surface occurs via a Lifshitz-type transition and is accompanied by a van Hove singularity. Depending on the strength of the displacement field, it can be an OVHS, where the three small pockets touch at three different $k-$points, or a single HOVHS,
 in which case the DOS has a power-law singularity. As explained in Ref.~\cite{Shtyk2017PRB}, the case of HOVHS separates two qualitatively different OVHS: the one at which three small pockets merge at three van Hove points at some critical doping, leading to two annular  Fermi surfaces at larger doping,  and the one at which another Fermi surface, centered at $K$($K'$), develops before the small pockets merge, and OVHS emerge where this new Fermi surface touches the three original pockets, see Fig.~\ref{fig:FS_evolution}. The HOVHS emerges at the boundary separating two different types of OVHS, when three OVHS merge at one point  (see Fig.~\ref{fig:van hove points}(b,c),Fig.~\ref{fig:FS_evolution}, and Ref.~\cite{Shtyk2017PRB}).  A van-Hove singularity,  which can be either ordinary or higher order,  also emerges in the electronic spectrum in  twisted bilayer WSe$_2$ either under a displacement field \cite{PanPRR2020}, or in the presence of Dzyaloshinskii-Moriya interaction \cite{ZangPRB2021}.

 For interacting fermions, the singularity in the density of states near  an OVHS or an  HOVHS
  gives rise to the emergence of electronic orders.
 We show that the ordering tendencies for a system with OVHS significantly differ from those in a system with HOVHS
   (see Fig. \ref{fig:upperpanel}). Specifically, for HOVHS, the ordered states, which develop in different parts of the phase diagram,  are superconductivity, degenerate between spin-triplet and spin-singlet, and valley polarization, degenerate with spin and charge density waves~\cite{Shtyk2017PRB,Classen2020PRB,castro2023emergence}. For OVHS,
   the ordered states are spin-triplet superconductivity and  pair-density wave
    We analyze how the ordering tendencies evolve as the system is tuned from an OVHS towards a HOVHS.

To carry out this analysis, we consider a somewhat simplified model of the electronic dispersion, in which there is a single VHS at $K$ and at $K'$, and tune the VHS between OVHS and HOVHS.  Within such a model,  the system can develop superconductivity, spin and valley polarizations, charge and spin-density wave orders with momenta $K-K'\equiv 2K$, and pair-density wave with the same momentum.  In the full model with three pockets,
  these are the orders that preserve $C_3$  rotational symmetry between the pockets.
 The order that we cannot detect within our approximation, is  nematic-like  symmetry breaking between the three small pockets. Such symmetry breaking has been reported in the experiments on BBG~\cite{Szabo2022,Holleis2023,Zhou2022}.
 The data, however, indicate that nematicity  develops only when there is a stronger spin or valley order of one of the types that we can detect within our model.

We model the dispersion by $\epsilon_{k}^{(\pm)}= \gamma k^2 \mathrm{cos}2\theta\pm \sqrt{(1-\gamma^2)}k^3\mathrm{cos}3\theta$, where the $+$ sign is $K$ and $-$ for $K'$,
 $k$ is the deviation from either $K$ or $K'$, and $\theta$ is the angle with respect to, say $k_x$ direction. We
 use $\gamma$ as the tuning parameter.  The limits $\gamma =1$ and $\gamma =0$ describe an OVHS and a HOVHS, respectively.
In our analysis, we focus on the intermediate regime $\gamma  \leq 1$, where the electronic dispersion has sizable contributions from both momentum-even ($k^2 \cos 2\theta$) and momentum-odd ($k^3 \cos 3\theta$) terms.
We calculate the susceptibilities in different channels, derive and solve the pRG equations for the flow of the couplings, and determine how the leading and subleading ordering tendencies evolve as the system is tuned from OVHS to HOVHS. We find that the changes in the phase diagram are topological in nature and originate from multiple appearances and disappearances
 and pair production and annihilation of the fixed and saddle points of the RG flow.

\section{The model and order parameters}\label{The model and the ordered parameters}

We consider a system on a honeycomb lattice, consisting of two independent patches located at $K$ and $K'$ points in the hexagonal Brillouin zone. We assume that at a given displacement field, the system is at critical doping, when there is a van Hove singularity at the Fermi level.
 The
 free-fermion Hamiltonian is
 \begin{equation}
\begin{aligned}
H_2&=\sum_{k}\left( \epsilon_{k}^{(+)} c^\dagger_{k+K} c_{k+K} +\epsilon_{k}^{(-)} d^\dagger_{k+K'} d_{k+K'} \right)
\end{aligned}
\end{equation}
where $c$ and $d$ operators correspond to fermions near the $K$ and $K'$ points, respectively, and $k$ is the deviation from either $K$ or $K'$, For an OVHS, the fermionic dispersion is  around a van Hove point, $ \epsilon_{k}^{(\pm)}$, starts as $k^2$, and is symmetric between $K$ ($\epsilon_{k}^{(+)}$) and $K'$ ($\epsilon_{k}^{(-)}$).
  For a HOVHS, the dispersion starts as $k^3$ and  is antisymmetric between $K$ and $K'$.
   We model the  dispersion in the intermediate case by introducing a single tunable parameter $\gamma$:
  \begin{equation}
    \epsilon_{k}^{(\pm)}= c_1\gamma k^2 \mathrm{cos}2\theta\pm  c_2\sqrt{(1-\gamma^2)}k^3\mathrm{cos}3\theta,
    \label{dispersion}
\end{equation}
where $\theta$ is the angle with respect to, say $k_x$ direction,
 and $c_1, c_2$ are constants, which we  set to one below by re-scaling the momentum $k$.
  The dimensionless parameter $\gamma$ varies in the range $0 \leq \gamma \leq 1$, where $\gamma=1$ corresponds to OVHS and $\gamma=0$ to HOVHS.

The full Hamiltonian includes four-fermion  interactions, allowed by  symmetry.
A simple experimentation shows that there are three different interaction terms:  interaction between fermionic densities within each valley, interaction between densities in different valleys, and inter-valley exchange (Refs.
 \cite{Shtyk2017PRB,Classen2020PRB,castro2023emergence,Zhou2021,Dong2023a,*Dong2023,ChichinadzePRL2022}
 \begin{equation}
\begin{aligned}
H_4&= \sum_{k,k',q,q'} g_1 c_{k+K}^\dagger d_{q+K'} d_{k'+K'}^\dagger c_{q'+K}\\& +g_2 c_{k+K}^\dagger c_{q+K} d_{k'+K'}^\dagger d_{q'+K'} \\& +g_4 \left(c_{k+K}^\dagger c_{q+K} c_{k'+K}^\dagger c_{q'+K}+d_{k+K'}^\dagger d_{q+K'} d_{k'+K'}^\dagger d_{q'+K'} \right)
\label{a_1}
\end{aligned}
\end{equation}
The momentum conservation in each term is implied  ($k+k' = q +q'$).  We use the same notations for the couplings $g_i$ with $i=1,2,4$,  as in~\cite{Shtyk2017PRB,Nandkishore2012}.
 The pair hopping interaction in the form $g_3 (c^\dagger_{k+K} c^\dagger_{k' +K} d_{q+K'} d_{q' + K'} + H.c)$ is forbidden because $K-K' =2K$ is not a reciprocal lattice vector.   For generality, we do not assume a particular sign of the couplings, nor set the coupling $g_1$ for the exchange interaction with momentum transfer $Q$ to be smaller than the couplings $g_2$ and $g_4$ for intra-pocket and inter-pocket density-density interactions with small momentum transfer.
  The results for the BBG and RTG, in which $g_i >0$ and $g_1 \ll g_2, g_4$, can be extracted from our generic phase diagrams below.

To study the ordering tendencies within the pRG, we introduce all  possible order parameters involving fermions near $K$ and $K'$. In the particle-hole channel, they are given by the expectation values of fermionic bilinears
\begin{equation}
    \Delta=\langle f^\dagger_{i,k} T_{ij} S_{kl} f_{j,l} \rangle,
\end{equation}
where $f_{i,j}$ are the fermion operators near $K$ or $K'$. Here
$i,j =1,2$ are spin indices, and $T_{ij}$ is either a Kronecker delta $\delta_{ij}$ or a spin Pauli matrix $\vec{\sigma}_{ij}$ acting in spin space, while  $k, l =1,2$ are valley indices, and $S_{kl}$ is also either a Kronecker delta $\delta_{kl}$ or an isospin Pauli matrix $\vec{\sigma}_{kl}$ acting in the valley space.
These order parameters can also be re-expressed in terms of the generators of SU(4), as direct products of Pauli matrices acting in spin and valley spaces \cite{ChichinadzePRL2022,Chichinadze2022}.  The total number of possible order parameters  is 15: 7 with zero momentum transfer and 8 with momentum transfer $\pm \textbf{Q}$.  The order parameters with zero momentum transfer are
 valley polarization $\Delta_{VP}= \langle d^{\dagger}_\alpha d_\alpha \rangle - \langle c^{\dagger}_\alpha c_\alpha \rangle$ and $3\times 2 =6$ intra-valley ferromagnetic orders $\vec{\Delta}_{KM}= \langle c^{\dagger}_\alpha \vec{\sigma}_{\alpha,\beta} c_{\beta} \rangle$ and $\vec{\Delta}_{K'M}= \langle d^{\dagger}_\alpha \vec{\sigma}_{\alpha,\beta} d_{\beta} \rangle$. The latter two are coupled by $g_1$, hence the global ferromagnetic
  and global antiferromagnetic orders $\vec{\Delta}_{KM} \pm \vec{\Delta}_{K'M}$ are competitors.
   The order parameters with momentum $\pm Q$ are 2 complex charge density wave (CDW) ${\Delta}^{CDW}=\langle c^{\dagger}_\alpha d_{\alpha} \rangle$ and $3 \times 2 =6$ complex spin density wave (SDW) $\vec{\Delta}^{SDW}=\langle c^{\dagger}_\alpha \vec{\sigma}_{\alpha \beta} d_\beta \rangle$.

\begin{table}[]
    \centering
    \begin{tabular}{||c|c||}
        \hline

    t/sSC & $ \langle c_\alpha d_\beta \rangle \pm \langle c_\beta d_\alpha  \rangle$ \\[0.5ex]
    \hline
    $s$PDW & $i \sigma^y_{\alpha \beta} \langle c_\alpha c_\beta \rangle;  i \sigma^y_{\alpha \beta} \langle d_\alpha d_\beta \rangle$\\[0.5ex]
    \hline
    r/iCDW & $ \langle c_\alpha^\dagger\delta_{\alpha\beta}d_\beta \rangle \pm \langle d_\alpha^\dagger\delta_{\alpha\beta}c_\beta \rangle$\\[0.5ex]
        \hline
    r/iSDW & $\langle c_\alpha^\dagger\vec{\sigma}_{\alpha\beta}d_\beta \rangle \pm \langle d_\alpha^\dagger\vec{\sigma}_{\alpha\beta}c_\beta \rangle$\\[0.5ex]
        \hline

    VP & $\langle c_\alpha^\dagger c_\alpha \rangle - \langle d_\alpha^\dagger d_\alpha \rangle$ \\[0.5ex]
        \hline

    FM/AFM & $\langle c_\alpha^\dagger \vec{\sigma}_{\alpha\beta}c_\beta \rangle \pm \langle d_\alpha^\dagger \vec{\sigma}_{\alpha\beta}d_\beta \rangle$\\ [0.5ex]
        \hline

    \end{tabular}
    \caption{Order parameters in particle-particle and particle-hole channels. The total number of order parameters is 19: 15 in the particle-hole channel and 4 in the particle-particle channel. }
     \label{Table:interaction channels}

\end{table}

In the particle-particle channel, we define
$\Delta^{SC}_{\alpha \beta}=\langle c^\dagger_\alpha d^\dagger_\beta \rangle$ for Cooper pairs with zero total momentum and $\Delta^{KPDW}_{\alpha \beta}=\langle c^\dagger_\alpha c^\dagger_\beta \rangle$ and $\Delta^{K'PDW}_{\alpha \beta}=\langle d^\dagger_\alpha d^\dagger_\beta \rangle$ for pair density wave (PDW) with net momentum $\pm\mathbf{Q}$, where $\mathbf{Q}=K-K'\equiv 2K$. The spin triplet/singlet superconducting order parameters can be obtained by adding/subtracting $\Delta^{SC}_{\alpha \beta}$ and
$\Delta^{SC}_{\beta \alpha}$.  The order parameters $\Delta^{SC}_{\alpha \beta}$ are coupled, hence
 spin-singlet/valley-triplet and spin-triplet/valley-singlet states are competitors.
 The two PDW order parameters $\Delta^{KPDW}$ and $\Delta^{K'PDW}$ are not coupled in the absence of pair hopping and develop simultaneously.  Because the interactions are momentum-independent, only spin-singlet $s$-wave PDW is possible.
The total number of particle-particle order parameters is 4 ($\Delta^{SC}_{\alpha \beta}, \Delta^{SC}_{\beta\alpha}, \Delta^{KPDW}_{\alpha \beta}, \Delta^{K'PDW}_{\alpha \beta}$).

We present 4+15=19 order parameters in Table.~\ref{Table:interaction channels}.

\section{Phase diagram in the two limits}
\label{sec:phase_diag}

To set the stage for our pRG analysis, we first present the results in the two limiting cases of OVHS ($\gamma=1$) and  HOVHS ($\gamma=0$).  The phase diagram in the HOVHS limit has been previously obtained in \cite{Shtyk2017PRB,Classen2020PRB,castro2023emergence}.  The phase diagram in the OVHS limit has been obtained in ~\cite{Ashvin}. Our phase diagram disagrees with their (see the next section for the reasoning).

 The phase diagram in the two limits is presented in Fig. \ref{fig:upperpanel}.
 Here and in subsequent figures we use the ratios of the bare couplings $g_2/g_1$ and $g_4/g_1$ as the two variables for the phase diagram.

  In the OVHS case,
 there are three regions in the phase diagram. In the top right corner, all the couplings flow to zero, and the system asymptotically behaves like a free Fermi gas (FG). In the rest of the phase diagram the leading instability
   is either $s$-wave pair-density-wave (PDW) or spin-triplet superconductivity (tSC).

In the HOVHS case, order develops for all ratios of $g_2/g_1$ and $g_4/g_1$. There are three types of order, depending on the bare couplings:  ferromagnetism (FM),  superconductivity, degenerate between spin-triplet and spin-singlet (sSC), or valley polarization (VP) degenerate with spin-density-wave (SDW) and charge-density-wave (CDW).

We see that the ordered states are very different in the two cases.  Our goal is to understand how the ordering tendency evolves as our tuning parameter $\gamma$ changes between $\gamma =0$ (OVHS) and $\gamma =1$ (HOVHS).  To address this, below we apply pRG procedure at varying $\gamma$.

\section{General parquet RG scheme}

The general pRG scheme is the two-stage procedure~\cite{Vafek2014,Khodas2016,Classen2017,Xing2017}. In the first stage, one obtains and solves the set of coupled pRG equations for the coupling $g_i$. The renormalizations hold in both particle-particle and particle-hole channels, and can be graphically represented as running in two orthogonal directions (this is why the method is called  "parquet" RG).
In the second stage, one introduces infinitesimally small bare order parameters $\Delta^j_0$ and  obtains and solves pRG equations for  running $\Delta^j$ using  the running couplings $g_i$ as inputs.  This allows one to obtain
the  corresponding susceptibilities.   The strongest ordering tendency is for the order parameter with the largest  susceptibility.

The pRG equations for $g_i$ are expressed in terms of  polarization bubbles -- the fermionic loops in  particle-particle and particle-hole channels.  We start our discussion of the pRG flow by analyzing the polarization bubbles and their dependence on the parameter $\gamma$.

\subsection{Polarization bubbles}
 There are four  relevant polarization bubbles $\Pi_{ph}(0)$, $\Pi_{ph}(\textbf{Q})$, $\Pi_{pp}(0)$ and $\Pi_{pp}(\textbf{Q})$.
  For our purposes, it is convenient to define all $\Pi_i (\textbf{k})$ as positive.  We have
\begin{widetext}
\begin{equation}
\begin{gathered}
\Pi_{ph} (0) = -\frac{T}{(2\pi)^2} \sum_{\omega} \int  \frac{d^2 \textbf{k}}{(i \omega - \epsilon_{\textbf{k}}^+)^2} = -\frac{T}{(2\pi)^2} \sum_{\omega} \int \frac{d^2 \textbf{k}}{(i \omega - \epsilon_{\textbf{k}}^-)^2}, \\
\Pi_{ph} (\textbf{Q}) = -\frac{T}{(2\pi)^2} \sum_{\omega} \int \frac{d^2 \textbf{k}}{(i \omega - \epsilon_{\textbf{k}}^+)(i \omega - \epsilon_{\textbf{k}}^-)}, \\
\Pi_{pp} (0) = \frac{T}{(2\pi)^2} \sum_{\omega} \int \frac{d^2 \textbf{k}}{(i \omega - \epsilon_{\textbf{k}}^+)(-i \omega - \epsilon_{-\textbf{k}}^-)}, \\
\Pi_{pp} (\textbf{Q}) = \frac{T}{(2\pi)^2} \sum_{\omega} \int \frac{d^2 \textbf{k}}{(i \omega - \epsilon_{\textbf{k}}^+)(-i \omega - \epsilon_{-\textbf{k}}^+)} = \frac{T}{(2\pi)^2} \sum_{\omega} \int \frac{d^2 \textbf{k}}{(i \omega - \epsilon_{\textbf{k}}^-)(-i \omega - \epsilon_{-\textbf{k}}^-)}.
\end{gathered}
\label{bubbles_definition}
\end{equation}
\end{widetext}
\begin{figure}
\begin{subfigure}
    \centering
    \includegraphics[width=8cm,height=4cm]{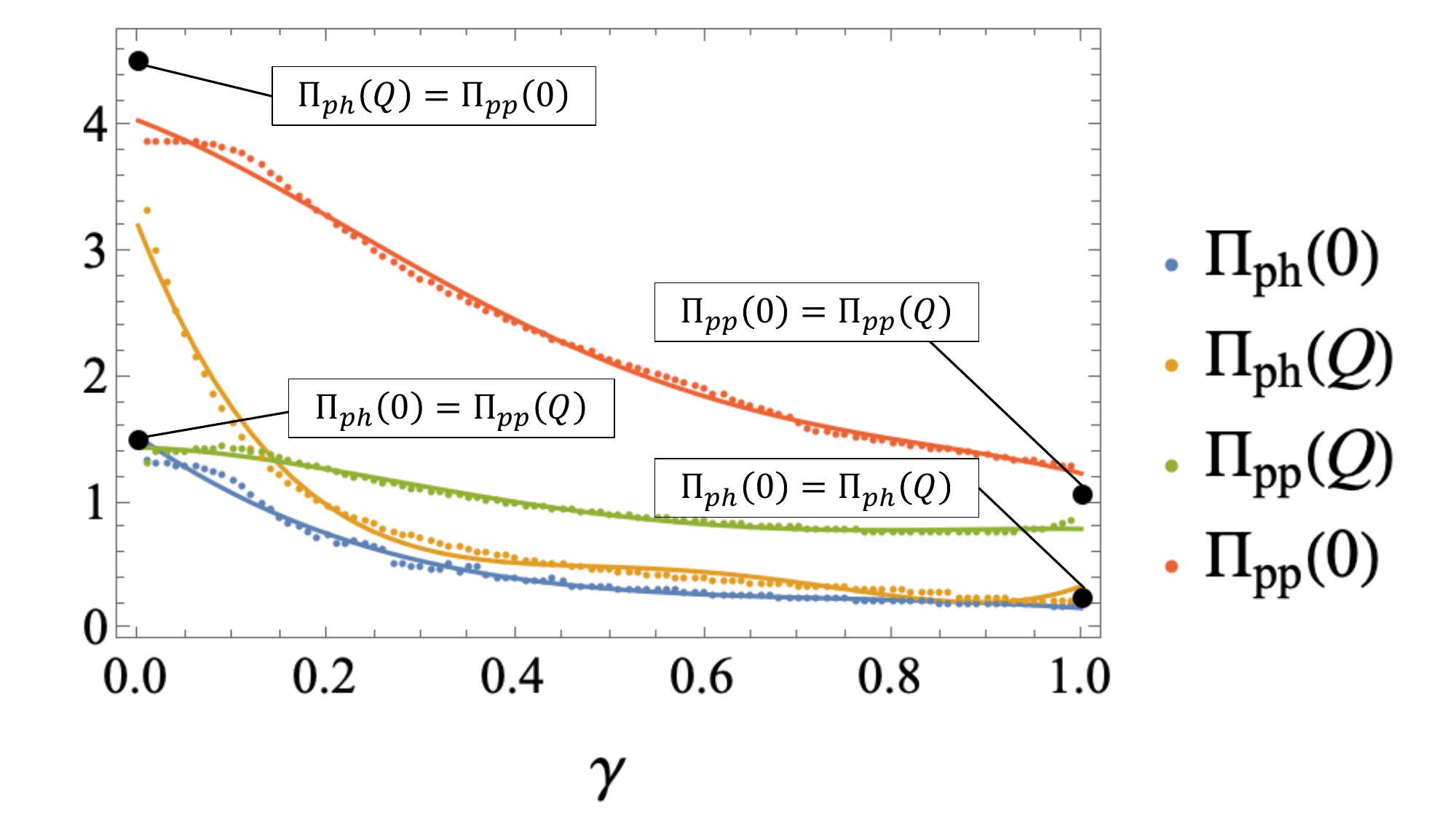}
\end{subfigure}
\begin{subfigure}(b)
    \centering
    \includegraphics[width=8cm,height=4cm]{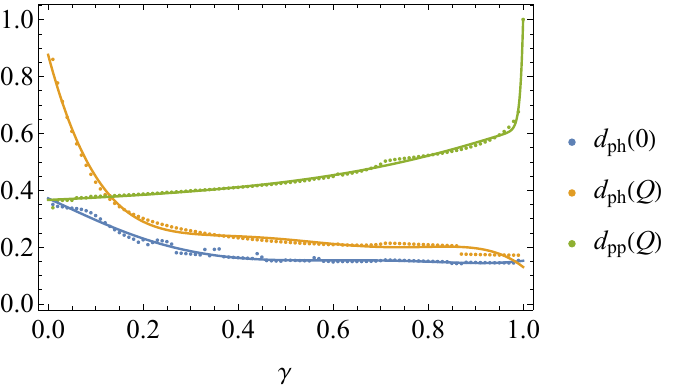}
\end{subfigure}
    \caption{(a) Polarization bubbles evaluated at $T=10^{-4}$ as functions of $\gamma$ -- the control parameter for the crossover from OVHS to HOVHS
    ($\gamma =1$ for OVHS and $\gamma =0$ for HOVHS). (b) The polarizations normalized to $\Pi_{pp}(0)$, $d_i(\textbf{k})=\Pi_i(\textbf{k})/\Pi_{pp}(0)$, which we used in the pRG equations.
     Scattered dots are the numerical results and solid curves are the fits to approximate analytical expressions (see App. \ref{app:C}). Black dots -- analytic results in the two limits, Eqs. (\ref{a_3}) and (\ref{a_2}).}
    \label{fig:bubbles}
\end{figure}
Because the dispersion starts with $k^2$, or even $k^3$,  the momentum integral is formally singular and has to be regularized.
 We use temperature regularization, i.e., evaluate the momentum integral at $q =0$ or $Q$ and a finite $T$, and set either $\log 1/T$ or $1/T^{1/3}$ (for HOVHS) as our pRG scale (see below).   Evaluating the momentum integral at a finite $T$, we obtain in the OHVS limit ($\gamma = 1$)
\begin{equation}
\begin{aligned}
    \Pi_{ph}(\mathbf{Q})&=\Pi_{ph}(0)=\frac{1}{4\pi^2}\log(2.26/T)\\
    \Pi_{pp}(\mathbf{Q})&=\Pi_{pp}(0)=\frac{1}{8\pi^2} \log^2(a/T), ~~ a = O(1)
\end{aligned}
\label{a_3}
\end{equation}
and in the HOVHS limit ($\gamma =0$)
\begin{equation}
    \begin{aligned}
        \Pi_{ph}(\mathbf{Q}) = \Pi_{pp}(\mathbf{0}) = 3X,
        \Pi_{ph}(0)=\Pi_{pp}(\mathbf{Q}) = X,\\
       X  \approx \frac{0.07}{T^{\frac{1}{3}}}
       \end{aligned}
       \label{a_2}
\end{equation}
\\
(see Appendices ~\ref{appendix: Polarization Operators For HOVHS Dispersions} and ~\ref{appendix: Calculations for the dispersion with both types of Van Hove singularities} for details).
The authors of Ref. ~\cite{Ashvin} evaluated the polarization bubbles $\Pi (q)$  at $q=0$ and $q=Q$ right at $T=0$ and obtained $\Pi_{ph}(0) =0$ because of the double pole.   We believe that the polarization bubbles relevant to pRG have to be computed at a finite external energy scale.   As a check, we computed the polarization bubbles $\Pi_{ph/pp} (q)$ at $T=0$ but $q$ different from $0$ or $Q$ and used $q$ as the pRG parameter.  We obtained the same results as with temperature regularization.

For $0<\gamma<1$, we evaluate polarization bubbles numerically at
  $T=10^{-4}$.
 We present the results for the four $\Pi$'s in Fig. \ref{fig:bubbles} as functions of $\gamma$. For convenience of calculations, we fitted
the numerical results by analytical functions (solid lines in the fits) (see Appendix \ref{app:C}).  The use of scaling functions is advantageous as the ratios of polarization bubbles are given by ratios of analytic functions and can be easily evaluated without extensive numerical integration.

\subsection{pRG equations}
We see Fig.~\ref{fig:bubbles} that the inter-valley particle-particle bubble $\Pi_{pp}(0)$
is the largest  for all values of $\gamma$. It is then convenient to define $L=\Pi_{pp}(0)$ as our RG scale, \emph{i.e.} define $\dot{g}_i=\frac{dg_i}{dL}=\frac{dg_i}{d\Pi_{pp}(0)}$.  For other $\Pi$'s, we introduce the
  ratios $d_i(\mathbf{k})=\frac{d{\Pi}_i(\mathbf{k})}{d{\Pi}_{pp}(0)}$.

The one-loop pRG equations for the running $g_i$ are obtained in a standard way, by evaluating one-loop diagrams for the renormalization of the 4-fermion vertices, differentiating with respect to $L$, and  replacing the bare internal vertices by the running ones, at the scale  $L$.
The equations are
\begin{equation}
\begin{aligned}
    \dot{g}_1&=2g_1(g_2-g_1)d_{ph}(\mathbf{Q})+2g_1g_4 d_{ph}(0)-2g_1g_2\\
    \dot{g}_2&=g_2^2d_{ph}(\mathbf{Q})+2(g_1-g_2)g_4d_{ph}(0)-(g_1^2+g_2^2)\\
    \dot{g}_4&=-g_4^2d_{pp}(\mathbf{Q})+(g_1^2+2g_1g_2-2g_2^2+g_4^2)d_{ph}(0).
    \label{RG_equation_g}
\end{aligned}
\end{equation}
The boundary condition is at $L=0$, where the couplings are bare.

We see that the coupling $g_1$ is self-generated, $i.e.$ $\dot{g}_1=0$ if $g_1=0$. For this reason, $g_1$ does not change sign under pRG. This property allows us to study the pRG flow in the space of  $x_2=g_2/g_1$ and $x_4=g_4/g_1$. The equations for running  $x_2$ and $x_4$  are
\begin{equation}
    \begin{aligned}
        \dot{x}_2&=g_1\{x_2^2d_{ph}(\mathbf{Q})+2(1-x_2)x_4d_{ph}(0)\\&-(1+x_2^2)\\&-x_2[2(x_2-1)d_{ph}(\mathbf{Q})+2x_4 d_{ph}(0)-2x_2]\}\\
        \dot{x}_4&=g_1\{-x_4^2d_{pp}(\mathbf{Q})\\&+(1+2x_2-2x_2^2+x_4^2)d_{ph}(0)\\&-x_4[2(x_2-1)d_{ph}(\mathbf{Q})+2x_4 d_{ph}(0)-2x_2]\}.
        \label{RG_equation_x}
    \end{aligned}
\end{equation}

  As defined, $d_i(\mathbf{k})$ in Eqs. (\ref{RG_equation_g}) and (\ref{RG_equation_x}) are functions of $L$.
  The exception is the case of HOVHS ($\gamma =0$), where all $\Pi_i(\mathbf{k})$ scale as $1/T^{1/3}$, and $d_i(\mathbf{k})$ are just numbers  ($d\Pi_i(\mathbf{k})/d\Pi_{pp} (0) = \Pi_i(\mathbf{k})/\Pi_{pp} (0)$).  Below we follow Refs.\cite{Furukawa1998PRL,Rice2009,Nandkishore2012}  and
   replace $d\Pi_i(\mathbf{k})/d\Pi_{pp} (0)$ by $\Pi_i(\mathbf{k})/\Pi_{pp} (0)$ for all values of $\gamma$. For the
    OHVS this amounts to replacing the ratios that scale as $1/\sqrt{L}$ by some small numbers from Fig. \ref{fig:bubbles}.  We refer a reader to Refs. \cite{Furukawa1998PRL,Rice2009,Nandkishore2012} for justification.

    \subsection{Fixed trajectories of pRG flow}

    In general, the pRG flow in the $x_2-x_4$ plane is determined by the location of fixed trajectories, when all $g_i$ either diverge or vanish, but the ratios of the couplings tens to finite values.  In terms of  $x_2 =g_2/g_1$ and $x_4 =g_4/g_1$, which we will be using, these fixed trajectories are fixed points, satisfying $\dot{x}_2=0$ and $\dot{x}_4=0$. There are three generic possibilities for fixed points: a stable fixed point, an unstable fixed point, and a saddle point (half-stable fixed point).
The stability of a fixed point is determined by
the eigenvalues of the matrix
\begin{equation}
    \begin{pmatrix}
        d\dot{x}_2/dx_2 & d\dot{x}_2/dx_4 \\
        d\dot{x}_4/dx_2 & d\dot{x}_4/dx_4
    \end{pmatrix}.
    \label{eq:stability}
\end{equation}
A stable fixed point has two negative eigenvalues, an unstable fixed point has two positive eigenvalues, and a half-stable fixed point has one positive and one negative eigenvalue.

We label the fixed points in the ($x_2,x_4$) plane as $(c_2,c_4)$. Substituting $g_2 = c_2 g_1$ and $g_4 = c_4 g_1$ into  Eq. \eqref{RG_equation_g},  we obtain three equations in the form
\begin{equation}
    \dot{g}_1=f_i(c_2,c_4)g_1^2, ~~i=1,2,4
    \label{eq:f}
\end{equation}
The set of equations $f_1 (c_2,c_4) = f_2 (c_2,c_4) =f_4 (c_2,c_4)$ determines the values $c_2$ and $c_4$.
 Then, if  $f_i (c_2, c_4) >0$, all the couplings diverge upon pRG at some scale $L_0$.  Near $L=L_0$, $g_i \sim 1/(L_0-L)$.
 If $f_i (c_2, c_4) <0$, all the couplings tend to zero under pRG.

Besides fixed points, there can be also fixed lines on the $(x_2,x_4)$ plane, specified by $x_4=\xi x_2$. Such lines define the asymptotic direction or pRG flow  at $x_2, x_4 \to \pm \infty$, i.e., when the magnitude of $g_2$ and $g_4$ becomes parametrically larger than $g_1$. The slope $\xi$ can be obtained by
substituting $\dot{x}_4=\xi\dot{x}_2$ into Eq. (\ref{RG_equation_x}) and taking the limit  $|x_2|, |x_4| \rightarrow\infty$. We find
\begin{widetext}
\[
\xi=\frac{d_{ph}(\mathbf{Q})-1\pm\sqrt{24d_{ph}(0)^2+(d_{ph}(\mathbf{Q})-1)^2-8d_{ph}(0)d_{pp}(\mathbf{Q})}}{6d_{ph}(0)-2d_{pp}(\mathbf{Q})}.
\]
\end{widetext}

\subsection{The flow of the order parameters}\label{test}
The candidate order parameters $\Delta^i$ are listed in Sec.~\ref{The model and the ordered parameters}.
The generic equation for the flow of $\Delta^i$ is~\cite{Vafek2014,Khodas2016}
\begin{equation}
    \dot{\Delta}^i=\lambda_i \Delta^i,
\end{equation}
where the couplings $\lambda_i = d_i(\textbf{k}) g_i$ are linear combinations of the products of $g_1$, $g_2$ and $g_4$ and the  ratios of the polarization bubbles.  These couplings can be obtained by solving ladder equations for the order parameters~\cite{Dong2023}.  At the boundary, $\Delta^i_0$ is an infinitesimally small bare order parameter.
  An instability towards a finite $\Delta^i$ occurs when one or more couplings $g_i$ diverge as $g_i \sim 1/(L_0 -L)$.  The order parameter $\Delta_i$ then generally diverges as
$$\Delta^i\sim\frac{\Delta^i_0}{(L_0-L)^{\beta_i}}$$. The corresponding susceptibility $\chi_i$ is obtained
 by solving $\frac{d\chi_i}{dL}=\Delta^2$ (Refs. \cite{Vafek2014,Khodas2016}) and scales as
   $\chi_i\sim\frac{1}{(L_0-L)^{2\beta_i-1}}$. The leading instability is determined by the largest critical exponent $\beta_i$.

In the particle-particle channel, the order parameters with momentum $q=0$ satisfy~\cite{Dong2023}
\begin{equation}
\begin{aligned}
    \Dot{\Delta}^{SC}=g_2\Delta^{SC}+g_1(\Delta^{SC})^*  \\
    \end{aligned}
\end{equation}
By adding/subtracting
the equations for $\Dot{\Delta}^{SC}$ and its complex conjugate, we find the
couplings in the sSC / tSC channels are
$$\lambda_{t/sSC}=-(g_2\mp g_1).$$
For PDW, we have
\begin{equation}
\Dot{\Delta}^{PDW}=g_4 d_{pp} (Q) \Delta^{PDW}
\end{equation}
The equation is the same for PDW with $2K$ and $2K'$.

\begin{figure*}
    \centering
    \includegraphics[width=15cm,height=20.5cm]{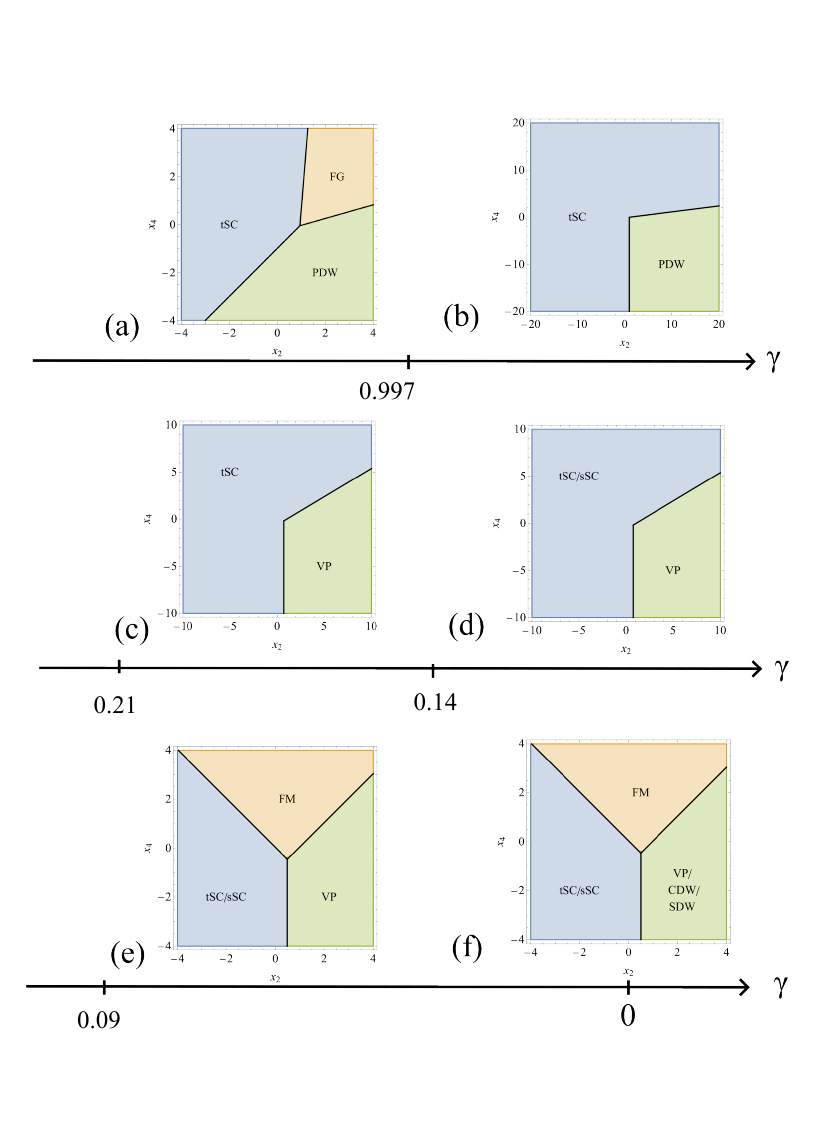}
    \caption{The evolution of the phase diagram between OVHS and HOVHS. We describe the evolution in the text.}
    \label{fig:motion}
\end{figure*}

 \begin{table}[]
    \centering
    \begin{tabular}{||c|c||}
        \hline
    t/sSC & $-(g_2\mp g_1)$ \\[0.5ex]
    \hline
    sPDW & $-d_{pp}(\mathbf{Q})g_4$\\[0.5ex]
        \hline
    CDW & $d_{ph}(\mathbf{Q})(g_2-2g_1)$\\[0.5ex]
        \hline
    SDW & $d_{ph}(\mathbf{Q})g_2$\\[0.5ex]
        \hline

    VP & $d_{ph}(0)(-g_4-g_1+2g_2)$ \\[0.5ex]
        \hline

    FM/AFM & $d_{ph}(0)(g_4\pm g_1)$\\ [0.5ex]
        \hline

    \end{tabular}
    \caption{The couplings for various order parameters in terms of running $g_i$. An order parameter may develop only when the corresponding coupling is positive.}
 \label{Table: critical exponents}

\end{table}

In the particle-hole channel the test order parameter vertices satisfy~\cite{Chichinadze2020,Dong2023}
\begin{equation}
\begin{aligned}
    &\Dot{\Delta}^{SDW}= d_{ph}(\mathbf{Q})g_2 \Delta^{SDW} \\
    &\Dot{\Delta}^{CDW}=d_{ph}(\mathbf{Q})(g_2-2g_1) \Delta^{CDW}\\
    &\Dot{\Delta}^{VP}= d_{ph}(0)(2g_2-g_1-g_4)  \Delta^{VP}\\
    &\Dot{\Delta}^{KM}= d_{ph}(0)(g_4\Delta^{KM}+g_1\Delta^{K'M})\\
    &\Dot{\Delta}^{K'M}= d_{ph}(0)(g_4\Delta^{K'M}+g_1\Delta^{KM}).\\
\end{aligned}
\end{equation}
We then obtain
$$\lambda^{CDW}=d_{ph}(\mathbf{Q})(g_2-2g_1)$$
$$\lambda^{SDW}=d_{ph}(\mathbf{Q})g_2$$
Note that the SDW and CDW order parameters can be either real or imaginary. The two are degenerate because the pRG equations do not involve their complex conjugates.

\begin{figure*}
    \centering
    \includegraphics[width=17cm,height=21cm]{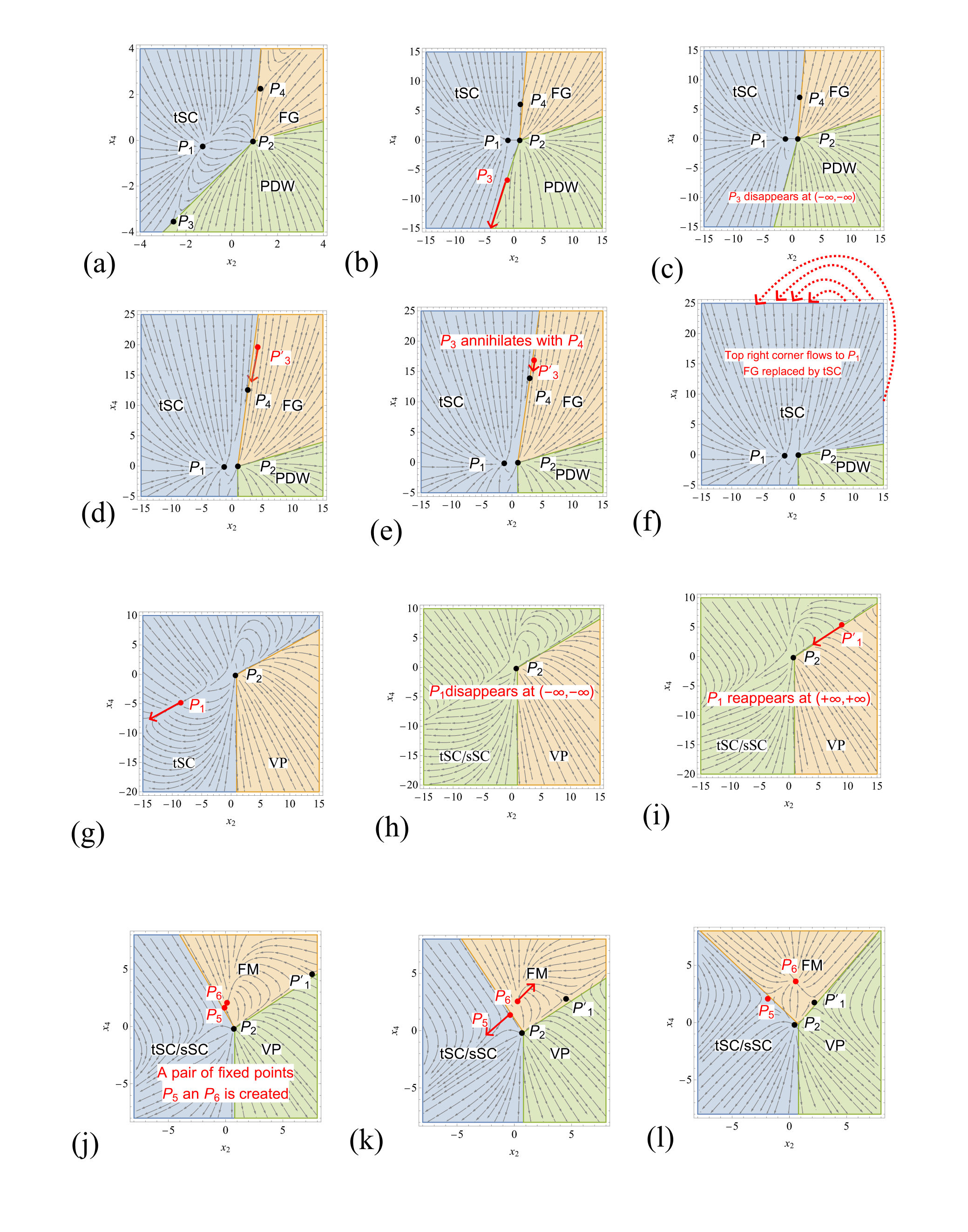}
    \caption{The evolution of the pRG flow when $\gamma$ becomes smaller than $1$.
     (a) $\gamma=1$ (b) $\gamma = 1^-$, (c) $\gamma=0.9984$  (d) $\gamma=0.998$ (e) $\gamma=0.997$ (f) $\gamma=0.996$ (g) $\gamma=0.15$ (h) $\gamma=0.14$ (i) $\gamma=0.13$ (j) $\gamma=0.09$ (k) $\gamma=0.04$ (l) $\gamma=0^+$}
    \label{fig:P_3 disappears}
\end{figure*}

Further,
 $$\lambda_{VP}=d_{ph}(0)(2g_2 -g_4-g_1),$$and
  $$\lambda_{FM/AFM}=d_{ph}(0)(g_4\pm g_1),$$for the couplings of global ferromagnetism and global
   antiferromagnetism.
We list the couplings in Table.~\ref{Table: critical exponents}. \par

\section{Results}

We solved the pRG equations for the running couplings $g_i$ for arbitrary $\gamma$ between $0$ and $1$, used $g_i$ to obtain susceptibilities for various order parameters, found the largest exponent $\lambda_i$ and identified the leading ordering tendencies for various $x_2$ and $x_4$.

We first present the results and discuss how we obtained them below and in Appendix \ref{app:B}.
 The behavior at $\gamma=1$ and $\gamma =0$ is shown in Fig. \ref{fig:upperpanel}.
We show the evolution of the phase diagram with $\gamma$ in Fig. \ref{fig:motion}.  Each modification of the phase diagram develops at a particular $\gamma$, listed in the Figure.  We see that the evolution starts at $\gamma$ very close to 1,  i.e., already a small deviation from the OVHS changes the phase diagram and ordering tendencies. Another  change of the ordering tendencies happens at $\gamma$ near $0.1\sim0.2$.
Comparing with Fig.~\ref{fig:bubbles}(b), we see that the first change is caused by rapid drop of  $d_{pp}(\textbf{Q})$, while the second is caused by crossing of  $d_{pp}(\textbf{Q})$ and $d_{ph}(\textbf{Q})$.

  In practical terms, as $\gamma$ decreases from $1$, first the boundary between tSC and PDW states rotates until it becomes vertical, and simultaneously the FG region is replaced by tSC order. Next, the PDW order is replaced by the VP order. Then
 susceptibilities for tSC and sSC become degenerate. At even smaller $\gamma$, the FM phase is created at the top of the phase diagram. Finally, at $\gamma = 0^+$,  the VP phase in the bottom right corner becomes degenerate with  CDW and SDW.

We now show the flow of the couplings, from which the evolution of the phase diagram has been extracted.
 We will see that the
crossover between OVHS and HOVHS involves the disappearance and emergence of fixed points of the RG flow via (i) fixed points approaching infinity and (ii) pair creation or annihilation of fixed points.

We display the evolution of the flow in various panels in   Fig. \ref{fig:P_3 disappears}.
For $\gamma =1$ (panel a), there is a stable fixed point $P_1$, an unstable fixed point $P_2$, and two half-stable fixed points $P_3$ and $P_4$.  Once $\gamma$ decreases, the half-stable fixed point $P_3$ quickly moves towards the lower left corner (panel b) and at $\gamma = 0.9984$ reaches $(-\infty,-\infty)$ (panel c). At infinitesimally smaller $\gamma$, $P_3$ re-emerges at $(+\infty,+\infty)$, now as a stable fixed point  (panel d) [the positive eigenvalue changes sign and becomes negative]. At $\gamma = 0.997$, the fixed points $P_3$ and $P_4$ annihilate (panel e). We show how this happens in more detail in Fig.  \ref{fig:schematic} in Appendix \ref{app:B}.   After annihilation, the pRG flow everywhere in the upper part of the phase diagram is towards the stable fixed point $P_1$ (panel f). Next, at smaller $\gamma$, the stable fixed point $P_1$ starts moving  towards $(-\infty,-\infty)$ (panel g)
  It reaches $(-\infty, -\infty)$ at $\gamma = 0.14$ (panel h),
   re-appears at $(+\infty, +\infty)$ as a half-stable fixed point [one of the eigenvalues becomes positive, another remains negative], and starts moving towards the unstable point $P_2$ (panel i).
   After that, the pRG flow in the left part of the phase diagram is towards $(-\infty,-\infty)$. In this flow, $g_2$ becomes parametrically larger than $g_1$, which leads to degeneracy between sSC and tSC (see Table. ~\ref{Table: critical exponents}).
 At even  smaller $\gamma$, a pair of fixed points $P_5$ and $P_6$ is created at $\gamma = 0.09$ (panel j),
 and starts moving apart (panel k).  After that, the structure of fixed points becomes the same as in the limit of HOVHS
  (panel l).

\begin{figure*}

    \begin{subfigure}(a)
    \centering
        \includegraphics[width=5cm,height=3.5cm]{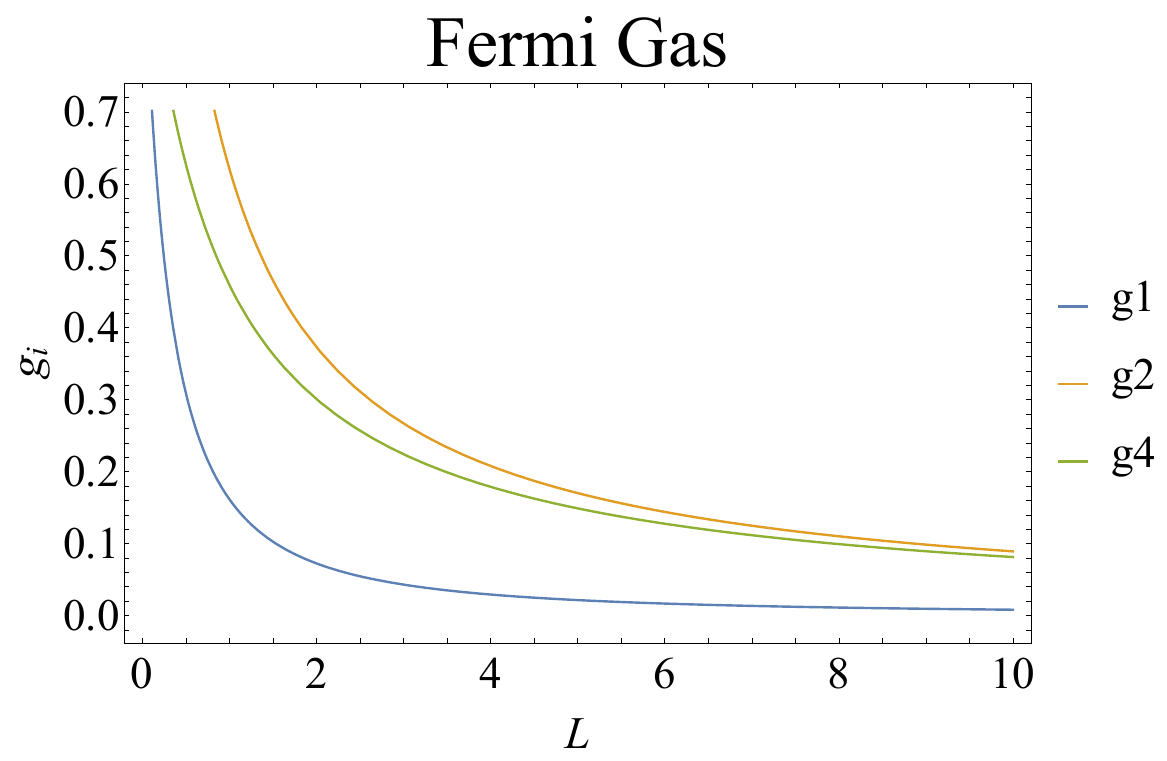}
    \end{subfigure}
    \begin{subfigure}(b)
    \centering
        \includegraphics[width=5cm,height=3.5cm]{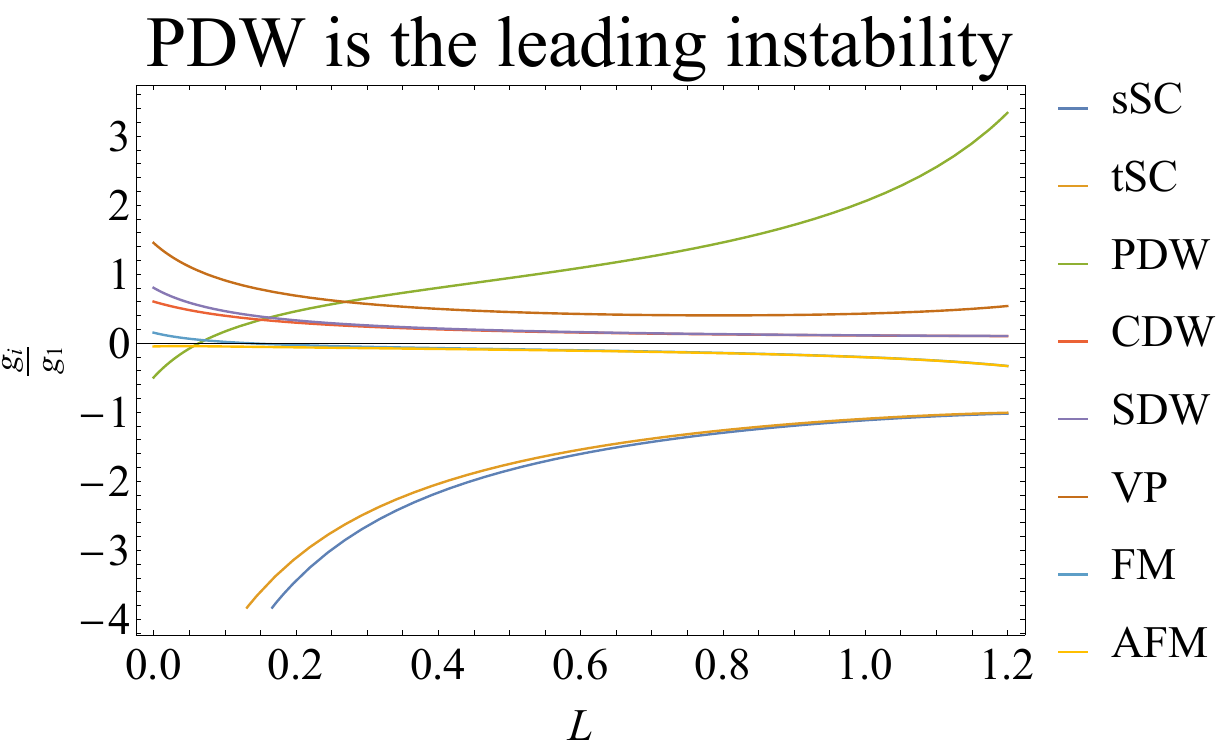}
    \end{subfigure}
    \begin{subfigure}(c)
    \centering
        \includegraphics[width=5cm,height=3.5cm]{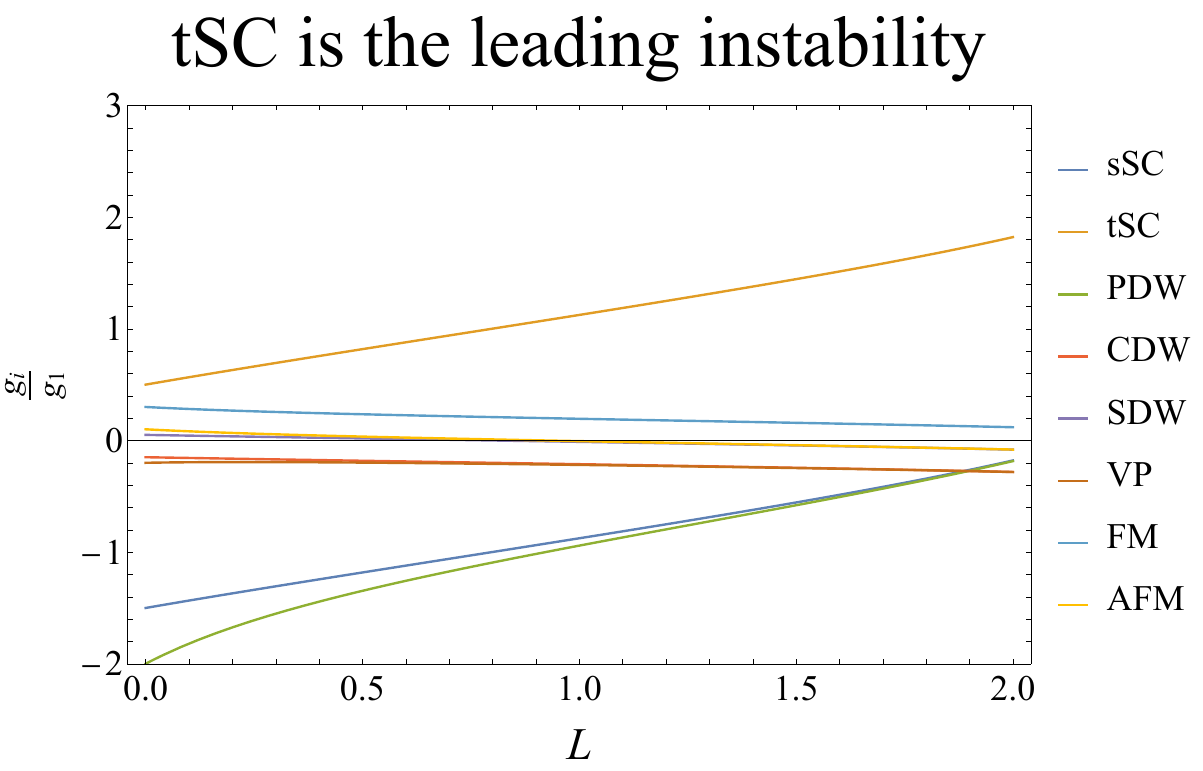}
    \end{subfigure}
    \caption{The pRG flow as a function of L in the OHVS limit for the bare values (a) $x_2^0 = 2$, $x_4^0 =1$,
    (b) $x_2^0=8$, $x_4^0=0.5$, (c)  $x_2^0=0.5$ and $x_4^0=2$.}
    \label{fig:OVHS}
\end{figure*}
\begin{figure*}
    \begin{subfigure}(a)
        \includegraphics[width=5cm,height=3.5cm]{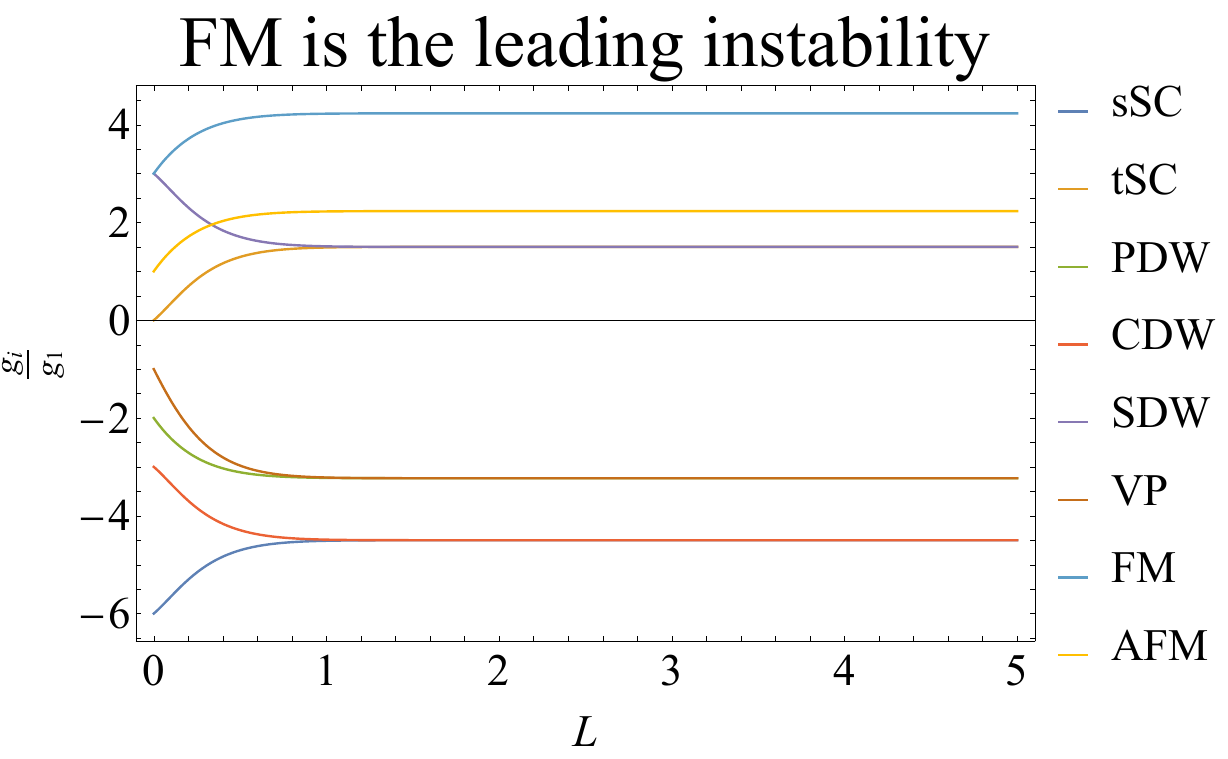}
    \end{subfigure}
    \begin{subfigure}(b)
        \includegraphics[width=5cm,height=3.5cm]{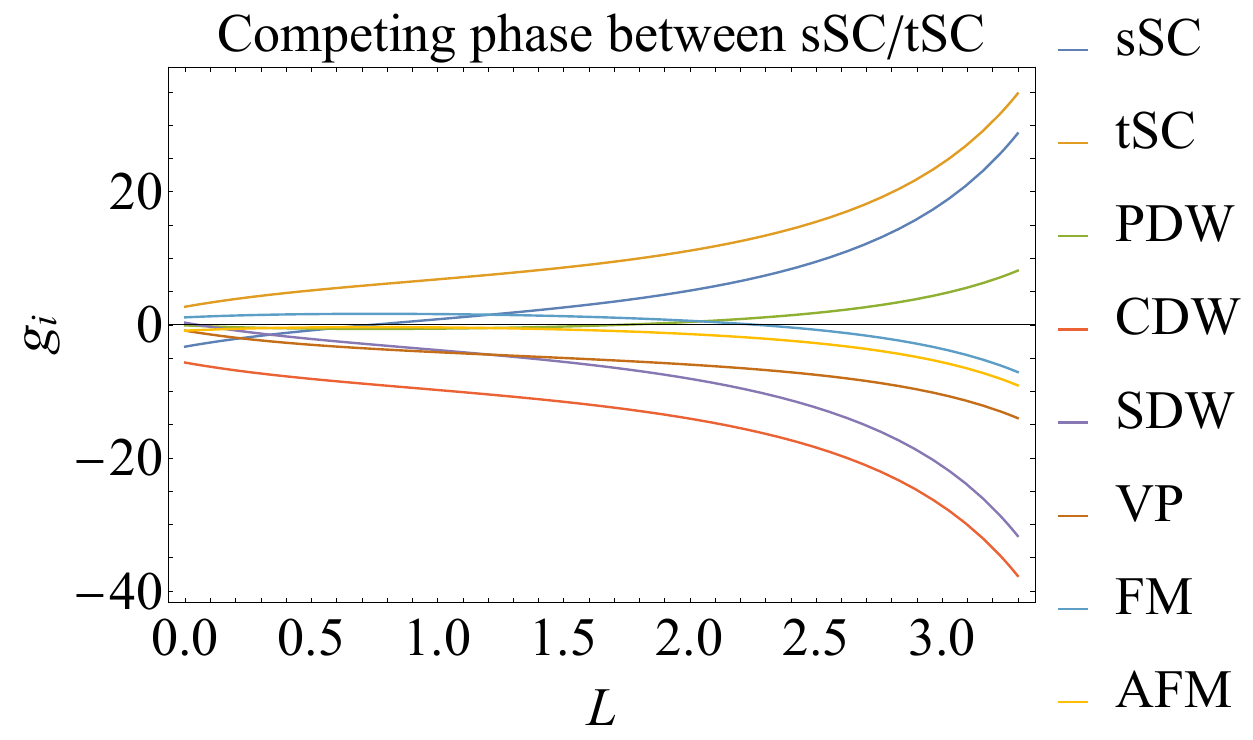}
    \end{subfigure}
    \begin{subfigure}(c)
        \includegraphics[width=5cm,height=3.5cm]{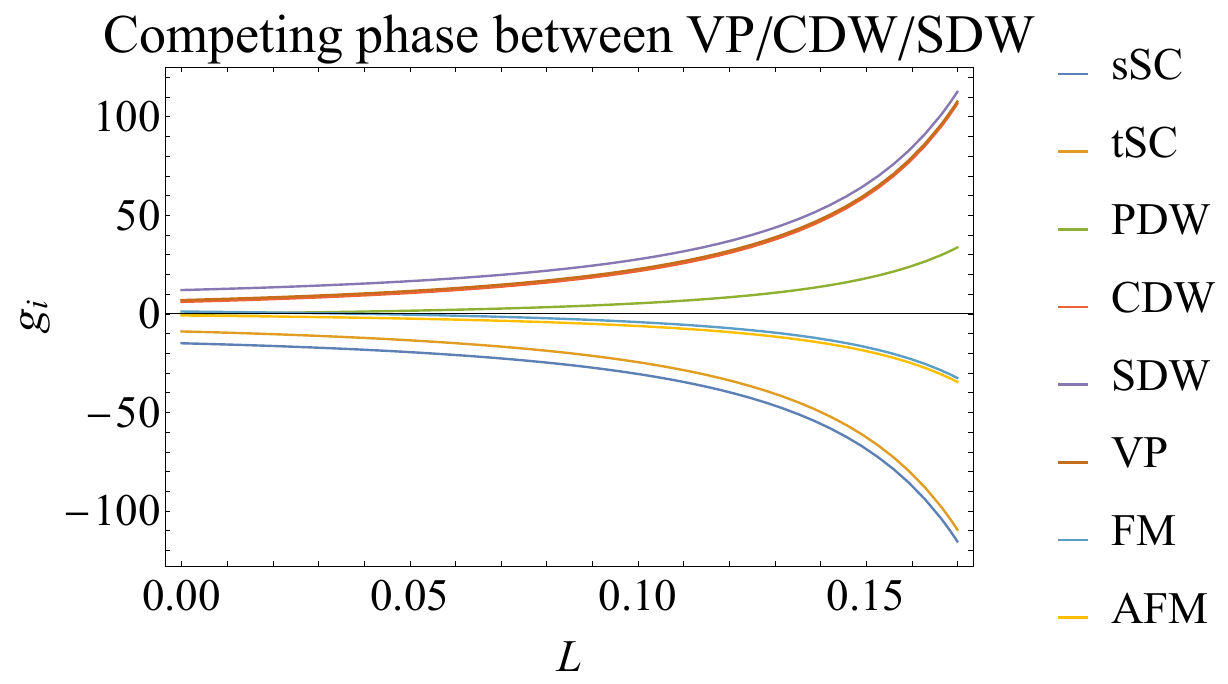}
    \end{subfigure}

    \caption{The pRG flow as a function of L in the HOHVS limit for the bare values (a) $x_2^0=1$, $x_4^0=2$ (b) $x_2^0=0.1$, $x_4^0=0.1$ (c) $x_2^0=4$, $x_4^0=0.1$.}
    \label{fig:HOVHS}
\end{figure*}
The evolution of the fixed points gives rise to the evolution of the ordering tendencies.  The leading ordering tendency, the one with the largest coupling $\lambda_i$, is displayed by the corresponding color in the phase diagram in
 Fig. \ref{fig:P_3 disappears}.  The computation of $\lambda_i$ based on the pRG results for $g_i$ is straightforward. We discuss the details in Appendix \ref{app:B} and here show, as an example, the flow of the couplings $\lambda_i$ in various channels in the OVHS  limit, $\gamma =1$ (Fig. \ref{fig:OVHS})  and in the HOVHS limit, $\gamma =0$ (Fig.
  \ref{fig:HOVHS}).

 \begin{figure}
      \centering
      \includegraphics[width=8cm,height=5cm]{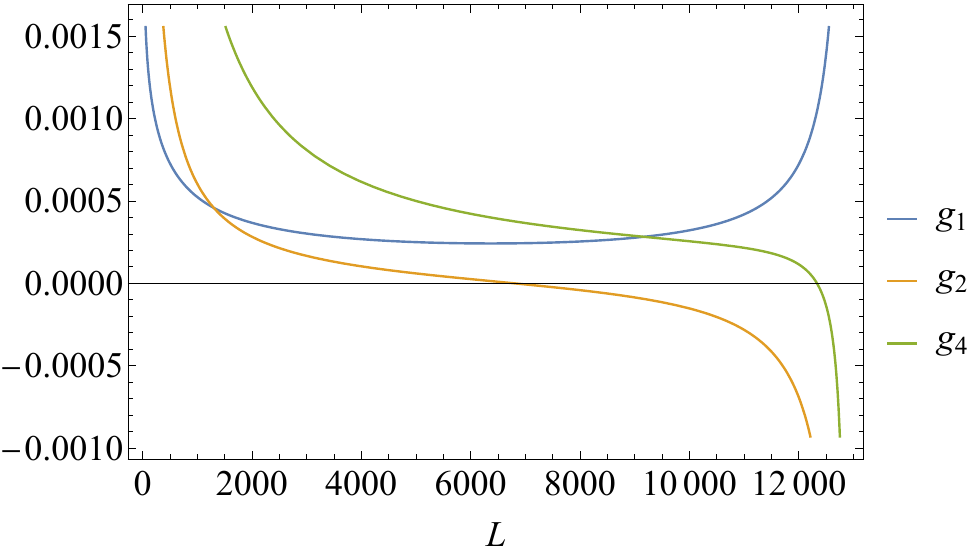}
      \caption{The pRG flow of the couplings with $x_2^0=x_4^0=100$ at $\gamma=0.9$.}
      \label{fig:flow}
  \end{figure}

\subsection{Application to BBG and RTG}

For practical applications to BBG and RTG, we use the fact that bare couplings $g_2$ and $g_4$ are positive and  nearly equal and are larger than bare $g_1$, as $g_2$ and $g_4$  are the interactions with near-zero momentum transfer, while $g_1$ is the interaction with momentum transfer $2K$.  The corresponding points on the phase diagram are near the diagonal in the upper-right quadrant (we show an example of pRG flow at large initial $x_2$ and $x_4$ in Fig. 
\ref{fig:flow}).  We see from the phase diagram in Fig. \ref{fig:upperpanel} that for OVHS, the system is not ordered, while in the case of HOVHS, the order is ferromagnetic,  and there is a wide intermediate regime, where the ordered state is a  triplet superconductor.  The reason why there is no order in the OVHS limit is due to the strongest effect from the renormalization in the particle-particle channel, which for positive (repulsive) interactions drives $g_i$ to zero.  In the HOVHS limit, ferromagnetism is the consequence of the fact that the diagonal direction in the upper-right quadrant is in the basin of attraction of the fixed point $P_6$  in Fig. \ref{fig:P_3 disappears}(l).  This fixed point is at $x_4 \gg x_2$ i.e., the dressed $g_4$ is much larger than dressed $x_2$. In this case, the ordered state is a FM (see Table II). At intermediate $\gamma$, the pRG flow in the upper-right corner of the phase diagram  is either towards the fixed point $P_1$ in Fig.~\ref{fig:P_3 disappears}(f), or towards $(-\infty, -\infty)$. In both cases, $x_2$ and $x_4$ become negative, which  results in either tSC (flow towards $P_1$)  or tSC degenerate with sSC (flow towards $(-\infty, -\infty)$). We also emphasize that to get an instability towards FM one has to keep $g_1$ in the pRG equations. It is small initially, but grows under pRG and eventually determines the pRG flow of the other two couplings $g_2$ and $g_4$. Indeed, using $d_{ph}(0)=d_{pp}(\textbf{Q})=1/3$ and $d_{ph}(Q)=1$ at HOVHS, we obtain from (\ref{RG_equation_g})
\begin{equation}
    \begin{aligned}
    \dot{g}_1&=\frac{2}{3}g_1g_4 -2 g^2_1\\
    \dot{g}_2&=\frac{2}{3}(g_1-g_2)g_4 - g^2_1\\
    \dot{g}_4&=\frac{2}{3}g_2(g_1-g_2) + \frac{1}{3}g^2_1
\end{aligned}
\end{equation}
If we neglected $g_1$, we would find that $g_2$ and $g_4$ flow to zero.  However, for a non-zero positive bare $g_i$,
 the couplings flow towards the fixed trajectory at $x_4 = g_4/g_1 = (3+ 2 \sqrt{3})/2$ and $x_2 = g_2/g_1 =1/2$, and the
  overall flow $\dot{g}_1 = g^2_1/(3+2\sqrt{3})$ is towards strong coupling, i.e., towards an ordered state.  This fixed trajectory (fixed point on $(x_2,x_4)$ plane) is $P_6$ in Fig.  \ref{fig:P_3 disappears}(l).  Along the same lines one can also verify the flow of $x_4$ and $x_2$ at intermediate $\gamma$ from initially large positive values towards the tSC fixed point $P_1$ (Fig.~\ref{fig:P_3 disappears}(f)) or towards $(-\infty, -\infty)$ (Fig.~\ref{fig:P_3 disappears}(h)). In both cases, the initially repulsive pairing interactions becomes attractive in the process of the pRG flow.

\section{Discussion and conclusions}

The purpose of this work was to investigate leading ordering tendencies in a system of interacting fermions on a honeycomb lattice, which, upon doping, undergoes a Lifshitz-type transition from small pockets to a single Fermi surface  via van Hove singularity. Such a behavior has been detected in Bernal bilayer graphene and rhombohedral trilayer graphene under the displacement field~\cite{Zhou2022,Zhou2021SC,Zhou2021,Holleis2023}. Our main goal was to analyze the evolution of the ordering tendencies  upon the change of fermionic dispersion under which there is a transformation from
 an OVHS, with logarithmic divergence of the density of states, to HOVHS, when the density of states diverges by a power law. To address this issue, we introduced a toy model with van Hove points at $K$ and $K'$ points in the Brillouin zone.  The dispersion around van Hove points  is governed by a single parameter $\gamma$, by changing which one can tune from OVHS ($\gamma =1$) to HOVHS ($\gamma =0$).

To study the phase diagram of the system and its evolution between OVHS and HOVHS, we employed the pRG approach. We found that phase diagrams at $\gamma =0$ and $\gamma =1$  differ significantly. In the OVHS limit the two possible ordered phases are triplet superconductivity, $s$-wave pair-density wave state, and there is a region in the phase diagram with no order.
In the HOVHS limit, the superconducting phase still exists, but spin-triple and spin-singlet states are degenerate.
 Other orders that develop for different bare values of the interactions  are ferromagnetism and valley polarization, degenerate with spin and charge density wave.

We applied the pRG procedure to $\gamma$ between zero and one and found that the evolution between the two limits is a multi-stage process with severe changes in the structure of the pRG flow due to the annihilation and recreation of individual fixed points, once they reach the  boundary of the phase diagram at infinity, and annihilation and creation of pairs of fixed points within the phase diagram.   In terms of the ordered states, as $\gamma$ decreases from one, first the Fermi gas phase is replaced by  spin-triplet superconductivity, then  PDW order loses to valley polarization, then spin-singlet superconducting state becomes attractive and degenerate with spin-triplet state, then a part of superconducting region becomes a ferromagnet, and, finally,
  spin/charge density wave channels become attractive and at $\gamma = 0^+$ become degenerate with valley polarization.

All ordered states develop at a finite energy.  In this regard, our results show how different orders compete and replace each other as the fermionic dispersion is modified  at a finite energy offset from the van Hove singularity.
 This offset provides a scale that separates the region where the effects of higher-order van Hove singularity become important from the region where they are negligible. We expect our results to be applicable, besides BBG and RTG,  to a number of twisted and untwisted systems,  which undergo Lifshitz-type transition upon doping, e.g., to twisted  WSe$_2$ under displacement field~\cite{Klebl2023,Wang2020}.

\section{Acknowledgments}

The authors acknowledge with thanks useful discussions with L. Classen, R. Fernandes,  Y. Huang,   K. R. Islam, R. D. Mayrhofer,
 L. Santos, D. Shaffer and B. Shklovskii.
The work by Y-C.L.  and A.V.C. was supported by the U.S. Department
of Energy, Office of Science, Basic Energy Sciences,
under Award No. DE-SC0014402. D.V.C. acknowledges financial support from the National High Magnetic Field Laboratory through a Dirac Fellowship, which is funded by the National Science Foundation (Grant No. DMR-1644779) and the State of Florida.

\appendix

\section{Calculation of polarization operators in the HOVHS limit}\label{appendix: Polarization Operators For HOVHS Dispersions}

In  the HOVHS limit, $\gamma =0$, the dispersion for two patches is given by
$$
\varepsilon_{\textbf{k}}^{\pm} = \pm k^3 \cos 3\theta,
$$
where indices $\pm$ labels valleys (patches). We will use temperature as a regularizer.

Like we said in the main text, we use the sign convention in which  all polarization bubbles are positive.
Following (\ref{bubbles_definition})  and using
$$
\varepsilon_{\textbf{k}}^{+} = - \varepsilon_{\textbf{k}}^{-}, \;\;\; \varepsilon_{-\textbf{k}}^{+} = - \varepsilon_{\textbf{k}}^{+}
$$
we have
\begin{equation}
\begin{gathered}
\Pi_{ph}(0) =- T \sum_{\omega} \int \frac{d^2 k}{4\pi^2} \frac{1} {(i \omega - \varepsilon_{\textbf{k}}^{+})^2} = I_1, \\
\Pi_{ph} (Q) =  -T \sum_{\omega} \int \frac{d^2 k}{4\pi^2} \frac{1}{(i \omega - \varepsilon_{\textbf{k}}^{+})(i \omega + \varepsilon_{\textbf{k}}^{+})} = I_2 , \\ 
\Pi_{pp}(0) = T \sum_{\omega} \int \frac{d^2 k}{4\pi^2} \frac{1}{(i \omega - \varepsilon_{\textbf{k}}^{+})(-i \omega - \varepsilon_{\textbf{k}}^{+})} = I_2, \\ 
\Pi_{pp} (Q) =  T \sum_{\omega} \int \frac{d^2 k}{4\pi^2} \frac{1}{(i \omega - \varepsilon_{\textbf{k}}^{+})(-i \omega + \varepsilon_{\textbf{k}}^{+})} = I_1, 
\end{gathered}
\end{equation}
where $I_1$ and $I_2$ are the two integrals that need to be calculated:
$$
I_1 = \frac{1}{16\pi^2 T} \int \frac{d^2 k}{\cosh^2 \frac{\varepsilon_{\textbf{k}}^{+}}{2T}}, \\
I_2 = \frac{1}{8\pi^2} \int \frac{d^2 k \tanh \frac{\varepsilon_{\textbf{k}}^{+}}{2T}}{\varepsilon_{\textbf{k}}^{+}}.
$$

\subsection{Calculation of $I_1$}

Consider $I_1$ first. There the integral can  be taken by performing an appropriate change of variables:
\begin{equation}
    \begin{gathered}
        I_1 = \frac{1}{16\pi^2 T} \int \frac{d^2 k}{\cosh^2 \frac{\varepsilon_{\textbf{k}}^{+}}{2T}} = \\
        = \frac{3 (2T)^{2/3}}{16 \pi^2 T} \int_{-\pi/6}^{\pi/6} \frac{d \theta}{(\cos 3\theta)^{2/3}} \int \frac{d (k^2 (\cos 3\theta)^{2/3} /(2T)^{2/3})}{\cosh^2 \frac{k^3 \cos 3\theta}{2T}} \\
        = \frac{3 \cdot 2^{2/3}}{4T^{1/3} (2 \pi)^2} \int_{-\pi/6}^{\pi/6} \frac{d \theta}{(\cos 3\theta)^{2/3}} \int_0^{\infty} \frac{d x}{\cosh^2 x^{3/2}}
    \end{gathered}
\end{equation}
Note, that we can set the upper limit of the $x$ integration to infinity as the integral is convergent; the integral over $\theta$ is convergent as well. Using
\begin{equation}
\begin{gathered}
\int_0^{\infty} \frac{dx}{\cosh^2 x^{3/2}} \simeq 0.958, \\
\int_{-\pi/6}^{\pi/6} \frac{d \theta}{(\cos 3\theta)^{2/3}} \simeq 2.43
\end{gathered}
\end{equation}
we obtain the final result
\begin{equation}
    I_1 \simeq 0.07 T^{-1/3}.
\end{equation}

We can also do an approximate calculation using Taylor-expanded dispersion near the points where $\cos{3\theta} =0$; such calculations will prove useful in the intermediate regime. Expanding near $\theta = \pi /6$ and multiplying by 6 to accommodate for all lines of zero energy, we obtain the approximate expression
\begin{equation}
\begin{gathered}
I_1 \approx \frac{1}{4T (2 \pi)^2} \int \frac{d^2 k }{\cosh^2 \frac{\varepsilon_{\textbf{k}}^{+}}{2T}} \simeq \frac{6}{4T (2 \pi)^2} \int \frac{k d k d\theta }{\cosh^2 \frac{3k^3 \theta}{2T}}  \\ = \frac{3^{1/3}}{2(2T)^{1/3} (2 \pi)^2} \int_0^1 \frac{d \theta}{\theta^{2/3}} \int_0^{\infty} \frac{d \left(\frac{3^{2/3} k^2 \theta^{2/3}}{(2T)^{2/3}} \right)}{ \cosh^2 \frac{3k^3 \theta}{2T}}   \\
= \frac{3^{1/3}}{2(2T)^{1/3} (2 \pi)^2} \int_0^1 \frac{d \theta}{\theta^{2/3}} \int_0^{\infty} \frac{dx}{\cosh^2 x^{3/2}}.
\end{gathered}
\end{equation}
Hence, we get for $I_1$
\begin{equation}
I_1  \simeq \frac{3^{1/3}}{2(2T)^{1/3} (2 \pi)^2} \cdot 3 \cdot 0.958 = 0.042 T^{-1/3}.
\end{equation}
This approximate result is about $0.6$ of the exact expression.

One can also calculate the same integral without taking into account the angular dependence of the dispersion. In this case $\varepsilon = k^3$. The analog of $I_1$, which we define as $I^*_1$,
is
\begin{equation}
\begin{gathered}
I^*_1 = \frac{1}{4T (2 \pi)^2} \int \frac{d^2 k }{\cosh^2 \frac{\varepsilon_{\textbf{k}}^{+}}{2T}} = \frac{1}{4T (2 \pi)^2} \int \frac{k d k d\theta }{\cosh^2 \frac{k^3 }{2T}}  \\ = \frac{1}{4 (2T)^{1/3} (2 \pi)} \int_0^{\infty} \frac{d \left(\frac{k^2 }{(2T)^{2/3}} \right)}{ \cosh^2 \frac{k^3 }{2T}}    \\
= \frac{1}{4 (2T)^{1/3} (2 \pi)} \int_0^{\infty} \frac{dx}{\cosh^2 x^{3/2}} \simeq 0.03 T^{-1/3} .
\end{gathered}
\end{equation}

\subsection{Calculation of $I_2$}

Now consider $I_2$. We perform the same change of variables to obtain
\begin{equation}
    \begin{gathered}
        I_2 = \frac{1}{2 (2 \pi)^2} \int \frac{d^2 k \tanh \frac{\varepsilon_{\textbf{k}}^{+}}{2T}}{\varepsilon_{\textbf{k}}^{+}} = \frac{3 (2T)^{2/3}}{2 (2 \pi)^2 (2T)} \cdot \\ \cdot \int_{-\pi/6}^{\pi/6} \frac{d \theta}{(\cos 3\theta)^{2/3}} \int \frac{\left(\frac{\cos 3\theta}{2T} \right)^{2/3} dk^2}{k^3 \cos 3\theta / 2T} \tanh \frac{k^3 \cos 3\theta}{2T} = \\
        = \frac{3 (2T)^{2/3}}{2 (2 \pi)^2 (2T)} \int_{-\pi/6}^{\pi/6} \frac{d \theta}{(\cos 3\theta)^{2/3}} \int_0^{\infty} \frac{dx}{x^{3/2}} \tanh x^{3/2}.
    \end{gathered}
\end{equation}
The integral over $x$ yields
\begin{equation}
\int_0^{\infty} \frac{dx \tanh x^{3/2}}{x^{3/2}} = 3 \int_0^{\infty} \frac{dx}{\cosh^2 x^{3/2}} \simeq 2.87,
\end{equation}
as can be verified by integration by parts. Therefore,
\begin{equation}
    I_2 = 3 I_1 \simeq 0.21 T^{-1/3}.
\end{equation}

A similar approximate calculation can be done here as well.
Expanding near $\theta \simeq \pi/6$, multiplying by $6$ contributions from 6 angles where $\cos{3\theta} =0$, we obtain
\begin{equation}
\begin{gathered}
I_2 \simeq \frac{6}{2 (2 \pi)^2} \int \frac{k dk d\theta \tanh \frac{3k^3 \theta}{2T}}{3k^3 \theta}  = \\ \frac{3^{1/3}}{2 (2 \pi)^2 (2T)^{1/3}} \int_0^1 \frac{d \theta}{\theta^{2/3}} \int \frac{d \left(\frac{ 3^{2/3} k^2 \theta^{2/3}}{ (2T)^{2/3}} \right) \tanh \frac{3k^3 \theta}{2T}}{3k^3 \theta /2T}  \\ = \frac{3^{1/3}}{2 (2 \pi)^2 (2T)^{1/3}} \int_0^1 \frac{d \theta}{\theta^{2/3}} \int_0^{\infty} \frac{dx \tanh x^{3/2}}{x^{3/2}} .
\end{gathered}
\end{equation}
Evaluating the last integral, we obtain
\begin{equation}
I_2 \simeq \frac{3^{1/3}}{2 (2 \pi)^2 (2T)^{1/3}} \cdot 3 \cdot 2.87 T^{-1/3} \simeq 0.126 T^{-1/3}.
\end{equation}
Evidently, $I_1$ and $I_2$ differ by a factor of 3.

One can also calculate $I_2$ for the cubic dispersion without the angular dependence, i.e., $\varepsilon = k^3$. The The analog of $I_2$, which we define as $I^*_2$,
is
\begin{equation}
\begin{gathered}
I^*_2 = \frac{1}{2 (2 \pi)^2} \int \frac{k dk d\theta \tanh \frac{k^3 }{2T}}{k^3 } = \frac{1}{4 (2 \pi)} \int \frac{d k^2 \tanh \frac{k^3 }{2T}}{k^3 }  \\ = \frac{1}{4 (2 \pi) (2T)^{1/3}} \int_0^{\infty} \frac{dx \tanh x^{3/2}}{x^{3/2}}  \simeq 0.09 T^{-1/3}.
\end{gathered}
\end{equation}
For the cubic dispersion without angular dependence, the results again differ by a factor of 3. So, the temperature dependence for cubic dispersion with and without HOVHS is identical.

\section{Calculation of polarization operators in the OVHS limit}

In  the OVHS limit, $\gamma =1$ and the dispersion for two patches is identical
$$
\varepsilon_{\textbf{k}}^{\pm} =  k^2 \cos 2\theta = \varepsilon_k,
$$
where $\pm$ label valleys (patches). As previously, we will use temperature as a regularizer. In our sign convention the four polarization are given by
\begin{equation}
\begin{gathered}
\Pi_{ph}(0) = \Pi_{ph} (Q) = -T \sum_{\omega} \int \frac{d^2 k}{4\pi^2} \frac{1}{(i \omega - \varepsilon_k)^2} = I_1, \\
\Pi_{pp}(0) = \Pi_{pp} (Q) =  T \sum_{\omega} \int \frac{d^2 k}{4\pi^2} \frac{1}{(i \omega - \varepsilon_k)(-i \omega - \varepsilon_k)} = I_2,
\end{gathered}
\end{equation}
where
$$
I_1 = \frac{1}{4T} \int \frac{d^2 k}{4\pi^2} \frac{1}{\cosh^2 \frac{\varepsilon_k}{2T}}, \\
I_2 = \frac{1}{2} \int \frac{d^2 k}{4\pi^2} \frac{\tanh \frac{\varepsilon_k}{2T}}{\varepsilon_k}.
$$

For both integrals we need to set the upper limit of integration over $k$. We set it at $k=1$.
Consider $I_1$ first.  Introducing $\varepsilon_k = k^2 \cos{2\theta}$ and $u = \cos{2\theta}$ as new variables and integrating over $\varepsilon_k$, we obtain
 \begin{equation}
 I_1 = \frac{1}{4\pi^2} \int_0^1 \frac{du \tanh{\frac{u}{2T}}}{u\sqrt{1-u^2}}
 \end{equation}
 Evaluating the integral, we obtain
 \begin{equation}
 I_1 =  \frac{1}{4\pi^2} \log{\frac{2.26}{T}}
 \end{equation}
Just as in the HOVHS case we can also evaluate the integral approximately by expanding around the angles $\theta$ for which $\cos{2\theta} =0$. Doing this, we obtain
\begin{equation}
    \begin{gathered}
        I_1 = \frac{1}{4T (2 \pi)^2} \int \frac{d^2 k}{\cosh^2 \frac{\varepsilon_k}{2T}} \simeq \frac{4}{4T (2 \pi)^2} \int \frac{k dk d\theta}{\cosh^2 \frac{2 k^2 \theta}{2T}} \simeq \\
        \simeq \int_{0}^{O(1)} \frac{d\theta}{\theta (2 \pi)^2} \int_0^{\theta/T} \frac{d \left( k^2 \theta /T \right) }{\cosh^2 \frac{k^2 \theta}{T}} = \\ = \frac{1}{4\pi^2} \int_0^{O(1)} d\theta \frac{\tanh{\theta/T}}{\theta} = \frac{1}{4 \pi^2} \ln{\frac{b}{T}}, ~~ b=O(1)
    \end{gathered}
\end{equation}
We see that to logarithmic accuracy, the approximate expression for $I_1$ coincides with the exact one.

For $I_2$, we obtain, by changing the variables in the integrand to $x=2\theta$ and $y = k^2 \cos{2\theta}/(2T)$,
\begin{equation}
 I_2 = \frac{1}{4\pi^2} \int_0^{\pi/2} \frac{dx}{x}  \int_0^{x/(2T)} dy \frac{\tanh{y}}{y}
 \end{equation}
Evaluating the integral with logarithmic accuracy, we obtain
\begin{equation}
 I_2 = \frac{1}{8\pi^2} \log^2{\left(\frac{a}{T}\right)}, ~~ a = O(1)
 \end{equation}
 To logarithmic accuracy, the same expression is obtained if we expand around the angles for which $\cos{2\theta} =0$.

\section{Calculation of polarization operators  close to the OVHS limit: $\gamma \leq  1$}
\label{appendix: Calculations for the dispersion with both types of Van Hove singularities}

In this section, we present approximate expressions for the polarization bubbles for $\gamma$, which on one hand are close to one, and on the other are such that $\sqrt{1 - \gamma^2} > T$.

\subsection{Calculation of $\Pi_{ph}(0)$}

The intra-valley particle-hole susceptibility is given by
\begin{equation}
\begin{aligned}
    \Pi_{ph}(0) &=  \frac{1}{4T (2\pi)^2} \int \frac{d^2 k}{\cosh^2 \frac{\varepsilon_{\textbf{k}}^{+}}{2T}}  \\ &=  \frac{1}{4T (2\pi)^2} \int \frac{d^2 k}{\cosh^2 \frac{\gamma k^2 \cos 2\theta +\sqrt{1-\gamma^2} k^3 \cos 3\theta}{2T}}.
\end{aligned}
\end{equation}
To calculate the integral we expand around $\theta=\pi/4, 3\pi/4, 5\pi/4, 7\pi/4$ since the main contribution comes from the vicinity of these angles. Around these angles,
$$
\varepsilon_{\textbf{k}}^{+} \simeq \pm 2 \gamma k^2 \theta \pm \frac{\sqrt{1-\gamma^2}}{\sqrt{2}} k^3,
$$
Since $\cosh$ is an even function one needs to consider only two angles, the other two will give the same result.
 Choosing $\theta$ near $\pi/4$ and $3\pi/4$, we obtain near these angles
\begin{eqnarray}
&&\varepsilon_{\textbf{k}}^{+} \simeq  2 \gamma k^2 \theta + \frac{\sqrt{1-\gamma^2}}{\sqrt{2}} k^3, \\ \nonumber
&&\text {and} \\ \nonumber
 &&\varepsilon_{\textbf{k}}^{+} \simeq  2 \gamma k^2 \theta - \frac{\sqrt{1-\gamma^2}}{\sqrt{2}} k^3. \nonumber
\end{eqnarray}
With this one can approximate the particle-hole susceptibility as
\begin{widetext}
    \begin{equation}
\begin{gathered}
\Pi_{ph}(0) \simeq  \frac{2}{4T(2\pi)^2} \int \frac{k dk d\theta}{\cosh^2 \frac{2 \gamma k^2 \theta + \frac{\sqrt{1-\gamma^2}}{\sqrt{2}}k^3 }{2T}} + \frac{2}{4T(2\pi)^2} \int \frac{k dk d\theta}{\cosh^2 \frac{2 \gamma k^2 \theta - \frac{\sqrt{1-\gamma^2}}{\sqrt{2}}k^3 }{2T}}.
\end{gathered}
\label{Pph0twoint}
\end{equation}
\end{widetext}

Consider the first of the two integrals:
\begin{equation}
\frac{1}{T} \int \frac{d^2 k }{\cosh^2 \frac{2 \gamma k^2 \theta + \frac{\sqrt{1-\gamma^2}}{\sqrt{2}}k^3 }{2T}}.
\end{equation}
It is convenient to introduce new variables:
$$
\bar k = k \frac{\left( 1 - \gamma^2 \right)^{1/6}}{\sqrt{2} T^{1/3}}, \; \bar{\alpha} =  \frac{2\gamma}{\left(1-\gamma^2 \right)^{1/3} T^{1/3}}.
$$
Then, the first integral can be rewritten in the following form
\begin{equation}
 \frac{ 2 }{(1 - \gamma^2)^{1/3} T^{1/3}}  \int \frac{d^2 \bar{k}}{\cosh^2 (\bar{\alpha} \bar{k}^2 \theta + \bar{k}^3)}.
\label{Pph0_1}
\end{equation}
The scaling function $f(\gamma)$ is given by the integral, where $\bar{\alpha} (\gamma) \gg 1$ for $\gamma$ slightly smaller that 1. We calculate the integral analytically for $\bar \alpha \gg1$ to logarithmic accuracy.
We will first integrate over the angle $\theta$ and then integrate over $\bar k$. Let us rewrite the integral (\ref{Pph0_1}):
\begin{equation}
\begin{gathered}
 \frac{ 2 }{(1 - \gamma^2)^{1/3} T^{1/3}} \int \frac{d^2 \bar{k}}{\cosh^2 (\bar{\alpha} \bar{k}^2 \theta + \bar{k}^3)}  \\ = \frac{ 2 }{(1 - \gamma^2)^{1/3} T^{1/3}} \int \frac{\bar{k} d \bar{k}}{\bar \alpha \bar{k}^2 } \int \frac{d ( \bar \alpha \bar{k}^2 \theta)}{\cosh^2 (\bar{\alpha} \bar{k}^2 \theta + \bar{k}^3)}.
\end{gathered}
\end{equation}
For $\theta$ the region of integration is $\theta \in [0, 1]$. The cubic terms can be neglected for $\bar \alpha \bar{k}^2 \theta > \bar{k}^3$. Then, for the integration over $x = \bar \alpha \bar{k}^2 \theta$, the limit is defined by the condition on the smallness of cubic terms:
$$
\bar \alpha \theta > \bar k \Rightarrow x < \bar \alpha \theta \cdot (\bar \alpha \theta)^2 = (\bar \alpha \theta)^3 .
$$
For maximal $\theta=1$ the upper limit for $\bar k$ is then just $\bar \alpha$.
We use those integration boundaries, introduce variable $x = \bar \alpha \bar{k}^2 \theta + \bar{k}^3$, and obtain
\begin{equation}
\begin{gathered}
 \frac{ 2 }{(1 - \gamma^2)^{1/3} T^{1/3}} \int_0^{\bar \alpha} \frac{\bar{k} d \bar{k}}{\bar \alpha \bar{k}^2 } \int_{\bar{k}^3}^{\bar \alpha \bar{k^2} + \bar{k}^3} \frac{d x}{\cosh^2 x}.
\end{gathered}
\end{equation}
The first integration is straightforward:
\begin{equation}
\begin{gathered}
 \frac{ 2 }{(1 - \gamma^2)^{1/3} T^{1/3}} \int_0^{\bar \alpha} \frac{\bar{k} d \bar{k}}{\bar \alpha \bar{k}^2 } \int_{\bar{k}^3}^{\bar \alpha \bar{k^2} + \bar{k}^3} \frac{d x}{\cosh^2 x}  \\ =  \frac{ 2 }{(1 - \gamma^2)^{1/3} T^{1/3}} \int_0^{\bar \alpha} \frac{\bar{k} d \bar{k}}{\bar \alpha \bar{k}^2 } \left[ \tanh(\bar{\alpha} \bar{k}^2 + \bar{k}^3) - \tanh(\bar{k}^3)  \right].
\end{gathered}
\label{P0int1}
\end{equation}
Now let us consider the second integral in (\ref{Pph0twoint}). After performing exactly the same transformations (except for the variable $x$, which now reads $x = \bar \alpha \bar{k}^2 \theta - \bar{k}^3$, the lower limit of integration over $x$ changes)
we get
\begin{equation}
\begin{gathered}
 \frac{ 2 }{(1 - \gamma^2)^{1/3} T^{1/3}} \int_0^{\bar \alpha} \frac{\bar{k} d \bar{k}}{\bar \alpha \bar{k}^2 } \left[ \tanh(\bar{\alpha} \bar{k}^2 + \bar{k}^3) + \tanh(\bar{k}^3)  \right].
\end{gathered}
\label{P0int2}
\end{equation}
As one can see now, terms with $\tanh \bar{k}^3$ in (\ref{P0int1}) and (\ref{P0int2}) cancel each other, so $\Pi_{ph} (0)$ is
\begin{equation}
\begin{gathered}
 \Pi_{ph}(0) =
 \frac{ 4 }{(1 - \gamma^2)^{1/3} T^{1/3} 8 \pi^2} \int_0^{\bar \alpha} \frac{\bar{k} d \bar{k}}{\bar \alpha \bar{k}^2 } \tanh(\bar{\alpha} \bar{k}^2 + \bar{k}^3).
\end{gathered}
\end{equation}
This integral can be evaluated with logarithmic accuracy. Neglecting $\bar{k}^3$  under the $\tanh$ one gets
\begin{equation}
\begin{gathered}
\Pi_{ph}(0) \simeq
 \frac{ 1 }{(1 - \gamma^2)^{1/3} T^{1/3} 2 \pi^2} \int_0^{\bar \alpha} \frac{\bar{k} d \bar{k}}{\bar \alpha \bar{k}^2 } \tanh(\bar{\alpha} \bar{k}^2) 
\\ =
 \frac{ 1 }{(1 - \gamma^2)^{1/3} T^{1/3} 4 \pi^2 \bar{\alpha}} \int_0^{\bar{\alpha}^3 } \frac{dx}{x} \tanh x \simeq \\
  \simeq \frac{ 3 }{ 8 \pi^2 \gamma}  \ln \frac{2 \gamma}{T^{1/3} (1 - \gamma^2)^{1/3}}
\end{gathered}
\end{equation}
Note that one cannot strictly take the limit $\gamma \rightarrow 1$ as we assumed $\sqrt{1-\gamma^2} > T$.

\subsection{Calculation of $\Pi_{ph}(Q)$}

The inter-valley particle-hole susceptibility is given by
\begin{equation}
\Pi_{ph}(Q) = -T \sum_{\omega} \int \frac{d^2 k}{(i \omega - \varepsilon_{\textbf{k}}^{+})(i \omega - \varepsilon_{\textbf{k}}^{-})},
\end{equation}
where    $\varepsilon_{\textbf{k}}^{+} = \gamma k^2 \cos 2\theta +\sqrt{1-\gamma^2} k^3 \cos 3\theta$ and
$\varepsilon_{\textbf{k}}^{-} = \gamma k^2 \cos 2\theta -\sqrt{1-\gamma^2} k^3 \cos 3\theta$.
We first sum over Matsubara frequencies and obtain
\begin{equation}
\begin{gathered}
\Pi_{ph}(Q) = -T \sum_{\omega} \int \frac{d^2 k}{(i \omega - \varepsilon_{\textbf{k}}^{+})(i \omega - \varepsilon_{\textbf{k}}^{-})}  \\ =- \frac{1}{2 (2\pi)^2} \int \frac{d^2 k \left( \tanh \frac{\varepsilon_{\textbf{k}}^{-}}{2T} - \tanh \frac{\varepsilon_{\textbf{k}}^{+}}{2T} \right)}{\varepsilon_{\textbf{k}}^{+} - \varepsilon_{\textbf{k}}^{-}}  \\ = -\frac{1}{4 (2\pi)^2} \int \frac{k d k d\theta \left( \tanh \frac{\varepsilon_{\textbf{k}}^{-}}{2T} - \tanh \frac{\varepsilon_{\textbf{k}}^{+}}{2T} \right)}{\sqrt{1-\gamma^2} k^3 \cos 3\theta }.
\end{gathered}
\label{PphQgen}
\end{equation}
We will now expand around $\theta \simeq 3\pi/4$ and proceed with an approximate dispersion. The 
reason for this approximation is two-fold. First, the double-log contribution comes from the $k^2$ dispersion near the zero energy lines (given by $\theta \simeq (2n+1) \pi /4$). Second, as we saw above, the cubic term gives the $\sim T^{-1/3}$ dependence regardless of whether there is angular dependence or not. Furthermore, $\tanh$ is an odd function, hence, we can again consider only the case near one angle to get the results to logarithmic accuracy (like we did for $\Pi_{ph} (0))$. Expanding near $\theta \simeq 3\pi/4$ and changing variables just like for $\Pi_{ph} (0)$ we get
    \begin{equation}
\begin{gathered}
    \Pi_{ph}(Q) \simeq - \int \frac{d^2 \bar{k} \left( \tanh (\bar\alpha \bar{k}^2 \theta - \bar{k}^3) - \tanh (\bar\alpha \bar{k}^2 \theta + \bar{k}^3) \right)}{ \bar{k}^3 (1 - \gamma^2)^{1/3} T^{1/3} (2 \pi)^2} \\ =
- \int \frac{d \bar k d\theta \left( \tanh (\bar\alpha \bar{k}^2 \theta - \bar{k}^3) - \tanh (\bar\alpha \bar{k}^2 \theta + \bar{k}^3) \right)}{\bar{k}^2 (1 - \gamma^2)^{1/3} T^{1/3} (2 \pi)^2},
\end{gathered}
\label{y}
\end{equation}
We are interested in the regime with large $\bar \alpha\gg1$ where quadratic dispersion dominates. This allows us to expand the difference between two hyperbolic tangents in Taylor series using $\bar k^3 / \bar \alpha \bar k^2 \theta$ as a small parameter. This means that $\bar k / \bar \alpha \theta <1$, i.e.,
$\bar k < \bar \alpha \theta$. Using this, we evaluate the integral as
\begin{equation}
\begin{gathered}
\int \frac{d \bar k d\theta \left( \tanh (\bar\alpha \bar{k}^2 \theta - \bar{k}^3) - \tanh (\bar\alpha \bar{k}^2 \theta + \bar{k}^3) \right)}{\bar{k}^2}  \\ \simeq
\int \frac{d \bar k d\theta \cdot 2 \bar \alpha \bar{k}^2 \theta \left( \tanh^2 (\bar \alpha \bar{k}^2 \theta)  -1 \right) \frac{\bar{k}^3}{\bar \alpha \bar{k}^2 \theta}}{\bar{k}^2} \\
= \int 2 \bar k d \bar k d\theta \cdot  \left( \tanh^2 (\bar \alpha \bar{k}^2 \theta)  -1 \right)  \\ = - \int_0^{\bar \alpha} \frac{2 \bar k d \bar k}{\bar \alpha \bar{k}^2}  \int_0^1  \frac{d(\bar \alpha \bar{k}^2 \theta)}{\cosh^2 (\bar \alpha \bar{k}^2 \theta)}  \\ =
 -2 \int_0^{\bar \alpha} \frac{\bar k d \bar k}{\bar \alpha \bar{k}^2}  \int_0^{\bar{\alpha} \bar{k}^2}  \frac{d x}{\cosh^2 x}.
\end{gathered}
\label{PphQ_1}
\end{equation}
Substituting into (\ref{y}), we obtain
\begin{equation}
\begin{gathered}
\Pi_{ph}(Q) =\frac{2}{(1 - \gamma^2)^{1/3} T^{1/3} (2 \pi)^2} \int_0^{\bar \alpha} \frac{\bar k d \bar k}{\bar \alpha \bar{k}^2} \tanh \bar\alpha \bar{k}^2  \\ =
\frac{1}{(1 - \gamma^2)^{1/3} T^{1/3} (2 \pi)^2 \bar{\alpha}}  \int_0^{\bar{\alpha}^3} \frac{dx}{x} \tanh x \simeq \\
\simeq \frac{ 3 }{ 8 \pi^2 \gamma}  \ln \frac{2 \gamma}{T^{1/3} (1 - \gamma^2)^{1/3}}
\end{gathered}
\end{equation}
 We see that $\Pi_{ph}(Q) \approx \Pi_{ph} (0)$ not only at $\gamma=1$ but also in the intermediate regime $\gamma \leq 1$.  This agrees with the numerical results in Fig. \ref{fig:bubbles}.

\subsection{Calculation of $\Pi_{pp}(Q)$}

Let us now calculate the intra-valley particle-particle susceptibility. It is given by
\begin{widetext}
    \begin{equation}
\begin{gathered}
\Pi_{pp}(Q) = \frac{1}{2 (2 \pi)^2} \int \frac{d^2 k \left( \tanh \frac{\varepsilon_{\textbf{k}}^{+}}{2T} + \tanh \frac{\varepsilon_{\textbf{k}}^{-}}{2T} \right)}{\varepsilon_{\textbf{k}}^{+} + \varepsilon_{\textbf{k}}^{-}} =
 \frac{1}{2 (2 \pi)^2} \int \frac{d^2 k \left( \tanh \frac{\gamma k^2 \cos 2\theta + \sqrt{1-\gamma^2}k^3 \cos 3\theta}{2T} + \tanh \frac{\gamma k^2 \cos 2\theta - \sqrt{1-\gamma^2}k^3 \cos 3\theta}{2T} \right)}{2 \gamma k^2 \cos 2\theta},
 \end{gathered}
 \label{Ppp0gen}
\end{equation}
\end{widetext}
where we used Eq. \eqref{bubbles_definition} and that $\varepsilon_{-\textbf{k}}^{+} = \varepsilon_{\textbf{k}}^{-}$.

Since we calculate the integral in the limit of small $\sqrt{1-\gamma^2}$, we can again consider only one case ($\tanh$ is an odd function, so every time $\cos 2\theta$ changes sign $\tanh$ will change the sign as well and the integral remains positive). We expand around $\theta \simeq 3 \pi/4$ (to have $\cos 2\theta$ being positive) and obtain
\begin{equation}
\begin{gathered}
\Pi_{pp}(Q) \simeq \frac{4}{2 (2 \pi)^2} \int \frac{d^2 k \; 2 \tanh \left(\frac{2 \gamma k^2 \theta}{2T} \right)}{4 \gamma k^2 \theta}   \\=
\frac{1}{(2 \pi)^2 T} \int \frac{k dk d \theta  \cdot \tanh \left(\frac{\gamma k^2 \theta}{T} \right)}{\gamma k^2 \theta / T}.
\end{gathered}
\label{Ppp0}
\end{equation}
Here we will integrate over momentum after we integrate over the angle. To do so, we first need to define the limits of integration. Integration over $\theta$ runs from $0$ to $1$, i.e., $\theta \in [0,1]$.
The cutoff on $k$ is defined through the condition $\gamma k^2 \theta \sim \sqrt{1 - \gamma^2} k^3$.  For $\theta \sim 1$ the maximal value of $k$ is of the order of $\frac{\gamma}{\sqrt{1 - \gamma^2}}$, hence $k \in [0, \frac{\gamma}{\sqrt{1 - \gamma^2}}]$. The lower cutoff in $\theta$ is governed by the condition  $\frac{\gamma}{\sqrt{1 - \gamma^2}} k^2 \theta > k^3$. With this, we can now proceed with the integration:
\begin{equation}
\begin{gathered}
\Pi_{pp}(Q) \simeq
\frac{1}{(2 \pi)^2 T} \int \frac{k dk d \theta  \cdot \tanh \left(\frac{\gamma k^2 \theta}{T} \right)}{\gamma k^2 \theta / T}  \\ =
\frac{1}{(2 \pi)^2} \int_0^{\frac{\gamma}{\sqrt{1-\gamma^2}}} \frac{ k d k}{ \gamma k^2} \int_{k \frac{\sqrt{1-\gamma^2}}{\gamma}}^1 \frac{ d \left( \gamma k^2 \theta / T \right)  \cdot \tanh \left(\frac{\gamma k^2 \theta}{T} \right)}{\gamma k^2 \theta / T}  \\ =
\frac{1}{(2 \pi)^2} \int_0^{\frac{\gamma}{\sqrt{1-\gamma^2}}} \frac{ k d k}{ \gamma k^2}  \int^{\gamma k^2 /T}_{\frac{\sqrt{1-\gamma^2} k^3}{T}} \frac{ dx  \cdot \tanh x}{x} .
\end{gathered}
\end{equation}
Let us now take the integral over $x$ to logarithmic accuracy. We obtain
    \begin{equation}
\begin{gathered}
\Pi_{pp}(Q) \simeq
\frac{1}{(2 \pi)^2} \int_0^{\frac{\gamma}{\sqrt{1-\gamma^2}}} \frac{ k d k}{ \gamma k^2}
\cdot \left[  \ln \left( \frac{\gamma k^2}{T} \right) \tanh \left( \frac{\gamma k^2}{T} \right) - \right. \\  \left. - \ln \left( \frac{\sqrt{1-\gamma^2} k^3}{T} \right) \tanh \left( \frac{\sqrt{1-\gamma^2} k^3}{T} \right) \right] .
\end{gathered}
\end{equation}
Consider the first contribution. Taking the resulting integral over $\bar k$ to logarithmic accuracy we get
\begin{equation}
\begin{gathered}
\frac{1}{(2 \pi)^2} \int_0^{\frac{\gamma}{\sqrt{1-\gamma^2}}} \frac{ k d k}{ \gamma k^2}
 \ln \left( \frac{\gamma k^2}{T} \right) \tanh \left( \frac{\gamma k^2}{T} \right) = \\
 = \frac{1}{2 (2 \pi)^2 \gamma} \int_0^{\frac{\gamma^3}{(1-\gamma^2)T}} \frac{ d y}{ y} \ln y  \tanh  y \simeq \\
 \simeq \frac{9}{4 (2 \pi)^2 \gamma} \ln^2 \frac{\gamma}{(1-\gamma^2)^{1/3} T^{1/3}}
\end{gathered}
\label{Pipp0_first_contribution}
\end{equation}

The second contribution is
\begin{equation}
\begin{gathered}
-\frac{1}{(2 \pi)^2} \int_0^{\frac{\gamma}{\sqrt{1-\gamma^2}}} \frac{ k d k}{ \gamma k^2}
\ln \left( \frac{\sqrt{1-\gamma^2} k^3}{T} \right) \tanh \left( \frac{\sqrt{1-\gamma^2} k^3}{T} \right) \\
= -\frac{1}{3 \gamma (2 \pi)^2} \int_0^{\frac{\gamma^3}{(1-\gamma^2)T}} \frac{ d y}{y} \ln y \tanh y \simeq \\
\simeq -\frac{9}{6 (2 \pi)^2 \gamma} \ln^2 \frac{\gamma}{(1-\gamma^2)^{1/3} T^{1/3}}
\end{gathered}
\end{equation}
The final result for $\Pi_{pp}(Q)$ is then given by
\begin{equation}
\begin{gathered}
\Pi_{pp}(Q) \approx \frac{3}{4 (2 \pi)^2 \gamma} \ln^2 \frac{\gamma}{(1-\gamma^2)^{1/3} T^{1/3}}.
\end{gathered}
\end{equation}

\subsection{Calculation of $\Pi_{pp}(0)$}

We perform the same calculation for $\Pi_{pp} (0)$. We first calculate the Matsubara sum:
\begin{equation}
\begin{aligned}
    \Pi_{pp} (0) &= \frac{1}{2 (2 \pi)^2} \int \frac{d^2 k  \tanh \frac{\varepsilon_{\textbf{k}}^{+}}{2T} }{\varepsilon_{\textbf{k}}^{+}} \\
    &= \frac{1}{2 (2 \pi)^2} \int \frac{d^2 k  \tanh \frac{\gamma k^2 \cos 2 \theta + \sqrt{1-\gamma^2}k^3 \cos 3\theta}{2T} }{\gamma k^2 \cos 2 \theta + \sqrt{1-\gamma^2}k^3 \cos 3\theta},
\end{aligned}
\end{equation}
where in Eq. \eqref{bubbles_definition} we used that $\varepsilon_{\textbf{k}}^{+} = \varepsilon_{-\textbf{k}}^{-}$.
Let us again work near $\theta = 3\pi /4$. Then, the integral can be rewritten as
\begin{equation}
\begin{gathered}
\Pi_{pp} (0) \simeq\frac{4}{2 (2 \pi)^2 T^{1/3} (1 - \gamma^2)^{1/3}} \int \frac{d^2 \bar{k}  \tanh (\bar{\alpha} \bar{k}^2 \theta + \bar{k}^3) }{\bar{\alpha} \bar{k}^2 \theta + \bar{k}^3}
\end{gathered}
\end{equation}
using the same change of variables
$$
\bar k = k \frac{\left( 1 - \gamma^2 \right)^{1/6}}{\sqrt{2} T^{1/3}}, \; \bar{\alpha} =  \frac{2\gamma}{\left(1-\gamma^2 \right)^{1/3} T^{1/3}}.
$$
We again will first integrate over the angle and only then proceed to the $\bar{k}$ integration. The integration limits are:
$\theta \in [-\bar{k}/\bar{\alpha},1]$ and $\bar{k} \in [0, \bar \alpha]$ (the main contribution here comes from very small angles).
Hence, the integral is
\begin{equation}
\begin{gathered}
\Pi_{pp}(0) \simeq
\frac{4(1 - \gamma^2)^{-1/3}}{2 (2 \pi)^2 T^{1/3} } \int \frac{\bar{k} d\bar{k} d \theta  \cdot  \tanh (\bar\alpha \bar{k}^2 \theta + \bar{k}^3)  }{\bar\alpha \bar{k}^2  \theta + \bar{k}^3 }  \\ =
\frac{4(1 - \gamma^2)^{-1/3}}{2 (2 \pi)^2 T^{1/3} } \int_0^{\bar \alpha} \frac{\bar{k} d\bar{k}}{\bar \alpha \bar{k}^2} \int_{-\bar{k}/\bar{\alpha}}^1 \frac{ (\bar \alpha \bar{k}^2) d\theta  \cdot  \tanh (\bar\alpha \bar{k}^2 \theta + \bar{k}^3)  }{\bar\alpha \bar{k}^2  \theta + \bar{k}^3 }  \\
= \frac{4(1 - \gamma^2)^{-1/3}}{2 (2 \pi)^2 T^{1/3} } \int_0^{\bar \alpha} \frac{\bar{k} d\bar{k}}{\bar \alpha \bar{k}^2} \int \frac{ d(\bar \alpha \bar{k}^2 \theta + \bar{k}^3)  \cdot  \tanh (\bar\alpha \bar{k}^2 \theta + \bar{k}^3)  }{\bar\alpha \bar{k}^2  \theta + \bar{k}^3 }  \\
= \frac{4(1 - \gamma^2)^{-1/3}}{2 (2 \pi)^2 T^{1/3} } \int_0^{\bar \alpha} \frac{\bar{k} d\bar{k}}{\bar \alpha \bar{k}^2} \int_{0}^{\bar \alpha \bar{k}^2 + \bar{k}^3} \frac{ dx  \cdot  \tanh x  }{x}.
\end{gathered}
\end{equation}
To logarithmic accuracy, the last integral reads
\begin{equation}
\begin{gathered}
\frac{4(1 - \gamma^2)^{-1/3}}{2 (2 \pi)^2 T^{1/3} } \int_0^{\bar \alpha} \frac{\bar{k} d\bar{k}}{\bar \alpha \bar{k}^2} \left[ \ln (\bar \alpha \bar{k}^2 + \bar{k}^3) \tanh (\bar \alpha \bar{k}^2 + \bar{k}^3) \right].
\end{gathered}
\end{equation}
This integral for large $\bar{\alpha}$ is almost exactly the same as in
\eqref{Pipp0_first_contribution}:
\begin{equation}
\begin{gathered}
\frac{4(1 - \gamma^2)^{-1/3}}{2 (2 \pi)^2 T^{1/3} } \int_0^{\bar \alpha} \frac{\bar{k} d\bar{k}}{\bar \alpha \bar{k}^2} \left[ \ln (\bar \alpha \bar{k}^2 ) \tanh (\bar \alpha \bar{k}^2 ) \right].
\end{gathered}
\end{equation}
We then obtain
\begin{equation}
\begin{gathered}
\Pi_{pp}(0) \approx \frac{9}{4 (2 \pi)^2 \gamma} \ln^2 \frac{ 2 \gamma}{(1-\gamma^2)^{1/3} T^{1/3}}.
\end{gathered}
\end{equation}
Comparing $\Pi_{pp} (Q)$ and $\Pi_{pp}(0)$ we see that the two now differ by a factor of 3.
 This is different from $\Pi_{pp} (Q) = \Pi_{pp}(0)$ at $\gamma =1$ (OVHS case). We see therefore that
 the ratio $\Pi_{pp}(0)/\Pi_{pp}(Q)$ rapidly increases from one to three as $\gamma$ decreases slightly down from one. This result is in agreement with the behavior which we found numerically in  Fig. \ref{fig:bubbles}.
Indeed, the ratio $\Pi_{pp}(0)/\Pi_{pp}(Q)$ is close to 3 for all $\gamma$ in this figure, down to
the  HOVHS value $\gamma =0$.

\section{Fitting of the polarization bubbles}
\label{app:C}
The numerical results for the polarization bubbles for arbitrary $\gamma$ are shown in Fig. \ref{fig:bubbles}.
 We fitted the numerical results  with the scaling functions
 \begin{widetext}
     \begin{equation}
    \begin{aligned}
    \Pi_{pp}(0)&=4.05-2.81\gamma-7.4\gamma^2+13.8508\gamma^3-6.446\gamma^4\\
\Pi_{ph}(0)&=(0.37-0.798\gamma-0.453\gamma^2+4.729\gamma^3-6.38\gamma^4+2.646\gamma^5)\Pi_{pp}(0)\\
\Pi_{ph}(Q)&=(0.876-6.32\gamma+24.774\gamma^2-47.15\gamma^3+42.817\gamma^4-14.868\gamma^5)\Pi_{pp}(0)\\
\Pi_{pp}(Q)&=\{0.618+0.382e^{166.7(\gamma-1)}+0.213\tan^{-1}[2.4486(\gamma-1)]\}\Pi_{pp}(0).\\
    \end{aligned}
    \label{eq:scaling_functions}
\end{equation}
 \end{widetext}

These are not the "exact" expressions, but they fit the numerical results rather well (see Fig. \ref{fig:bubbles})  For $\gamma \ll1$,  the results are quite consistent with Eq. (\ref{a_2}) in the sense that $\Pi_{pp} (0) \approx \Pi_{ph} (Q) \approx 3 \Pi_{pp} (Q) \approx 3 \Pi_{ph} (0)$.  For $\gamma \to 1$, $\Pi_{pp} (0)$ and $\Pi_{pp} (Q)$ are the largest, yet because singularities in the OVHS case are logarithmic, larger $\gamma$ are needed to match the results in Eq. (\ref{a_3}).

\section{Details of the evolution of the fixed points of the pRG flow and the ordering tendencies}
\label{app:B}

 In this Appendix, we discuss in some detail the evolution of the phase diagram with $\gamma$.

\subsection{The disappearance of the half-stable fixed point, its reemergence, and the disappearance of the Fermi gas region}
\label{subsection:disappearance}

\begin{figure*}
    \begin{subfigure}(a)
        \centering
        \includegraphics[width=6cm,height=5cm]{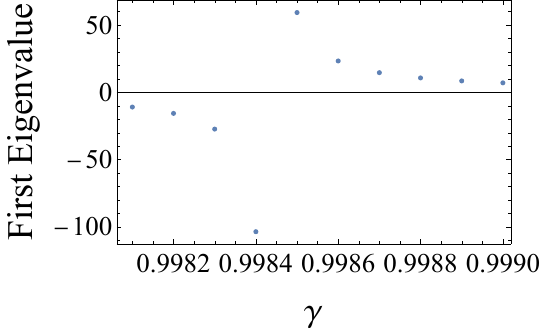}
    \end{subfigure}
    \begin{subfigure}(b)
        \centering
        \includegraphics[width=6cm,height=5cm]{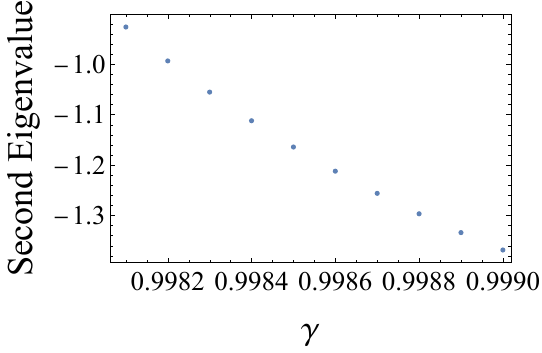}
    \end{subfigure}
    \caption{The two eigenvalues of the pRG flow at the fixed point $P_3$ in the vicinity of
     $\gamma_{c1} = 0.9984$,
     where $P_3$ approaches $(-\infty,-\infty)$ and then re-emerges at $(\infty,\infty)$.
    One eigenvalue diverges at $\gamma_{c1}$ and changes sign, the other evolves continuously through $\gamma_{c1}$.
     As a result, a half-stable fixed point on one side of $\gamma_{c1}$ becomes a stable fixed point on the other side.}
        \label{fig:eigenvalues}

\end{figure*}

\begin{figure}
    \centering
    \includegraphics[width=6.5cm,height=5cm]{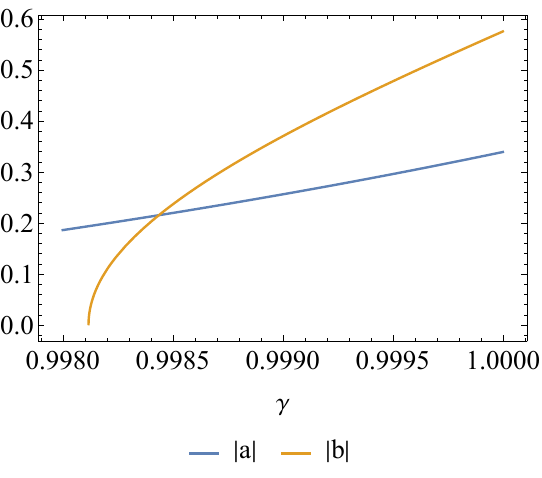}
    \caption{The parameters $a$ (blue) and $b$ (orange), determined by Eq.~\ref{eq:eigenvalues}.  The sign of $a-b$ determines whether the two eigenvalues are of the same or or different sign.
 We see that $a-b$ changes sign  at $\gamma_{c1}$.}
        \label{fig:before}

\end{figure}

\begin{figure}
    \centering
    \includegraphics[width=8cm,height=13.6cm]{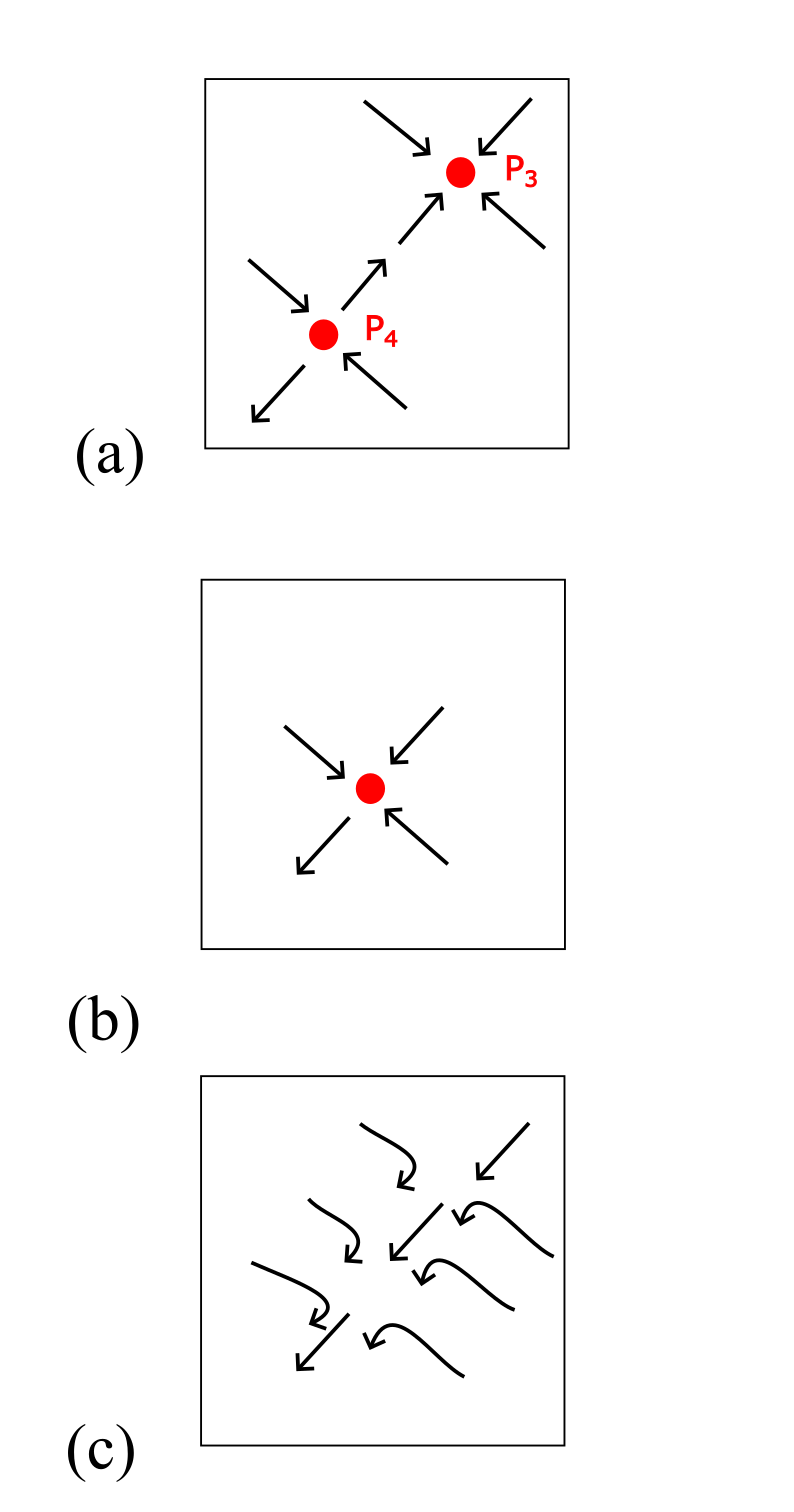}
    \caption{A schematic picture of how the half-stable fixed point $P_4$ and the stable fixed point $P_3$ merge. (a) The two points are close to each other (b) The moment when they merge.  (c) The continuous flow after annihilation.}
    \label{fig:schematic}
\end{figure}

The changes in the phase diagram begin when the half-stable fixed point $P_3$ flows towards $-\infty$.  We find that $P_3$ reaches infinity  at $\gamma_{c1}\approx0.9984$. At infinitesimally smaller $\gamma$, $P_3$  reemerges at  $(+\infty,+\infty)$. We checked the eigenvalues and found (see  Fig.~\ref{fig:eigenvalues}) that one of the eigenvalues passes through infinity and changes sign at $\gamma_{c1}$, i.e., a half-stable fixed point re-emerges as a fully stable fixed point.

We can verify this analytically.  The fixed point $P_3$ moves along the fixed line
 $x_4=\zeta x_2$, where $\zeta\approx1.43$. Substituting this parameterization into Eq.~\ref{eq:stability} and solving for the eigenvalues $\lambda$ at large $|x_2|$,  we obtain the quadratic equation
\begin{equation}
\begin{aligned}
&\lambda^2+2x_2(-2+2d_{ph}(\mathbf{Q})+3d_{ph}(0)\zeta+d_{pp}(\mathbf{Q})\zeta)\lambda\\
&+4x_2^2[(d_{ph}(\mathbf{Q})-1)(d_{ph}(\mathbf{Q})+d_{pp}(\mathbf{Q})\zeta-1)\\&+d_{ph}(0)\zeta(d_{ph}(\mathbf{Q})+
2d_{pp}(\mathbf{Q})\zeta-1)+2d_{ph}(0)^2(\zeta^2-2)]=0,
\label{eq:quadratic}
\end{aligned}
\end{equation}
which yields
\begin{equation}
    \lambda=ax_2\pm\sqrt{b^2x_2^2}=
    \begin{cases}
        (a\pm b)x_2 & \text{if $x_2>0$}\\
        (a\mp b)x_2 & \text{if $x_2<0$},\\
    \end{cases}
    \label{eq:eigenvalues}
\end{equation}
where
\begin{equation}
    \begin{aligned}
        a&=2-2d_{ph}(\mathbf{Q})-3d_{ph}(0)\zeta-d_{pp}(\mathbf{Q})\zeta\\
        b&=[16d_{ph}(0)^2+8d_{ph}(0)(d_{ph}(\mathbf{Q})-1)\zeta\\
        &+(d_{ph}(0)-d_{pp}(\mathbf{Q}))^2\zeta^2]^{1/2}.
    \end{aligned}
    \label{eq:a&b}
\end{equation}
 The sign of $a-b$ determines whether the two eigenvalues are of the same or different sign.
We plot $a$ and $b$ in Fig.~\ref{fig:before}. We see that $a-b$ indeed changes the sign  at $\gamma_{c1}$.

The next event is the annihilation of the half-stable fixed point $P_4$ and the stable fixed point $P_3$ at $\gamma_{c2}\approx 0.997$. Naïvely, one would expect that the annihilation of fixed points resembles the annihilation of two charges with opposite signs in classical electrostatics: the RG flow trajectories follow the same pattern as the field lines between the two charges, which form a dipole. Here the situation is different, however: it looks as if the object with a finite charge merged with an object with zero charge.
To visualize how the local RG flow changes in response to this anomalous merging process, we draw the schematic diagram in Fig. ~\ref{fig:schematic}.  When the two points merge,
the flow across the merging point is in 3-in/1-out, i.e. this point is an inflection point along one direction (the corresponding eigenvalue is zero) and a minimum along the other. At an infinitesimally smaller $\gamma$,  there is no local extremum, resulting in a continuous flow. In algebraic terms, when the two points merge, the solution of Eq.~\ref{RG_equation_x} changes from four distinct real solutions to two degenerate and two non-degenerate solutions. At a smaller $\gamma$, two degenerate roots split and become complex, indicating that a fixed point vanishes.

\subsection{The disappearance of the fixed point $P1$ and the replacement of the PDW by VP}

\begin{figure}
    \centering
    \includegraphics[width=6.5cm,height=5cm]{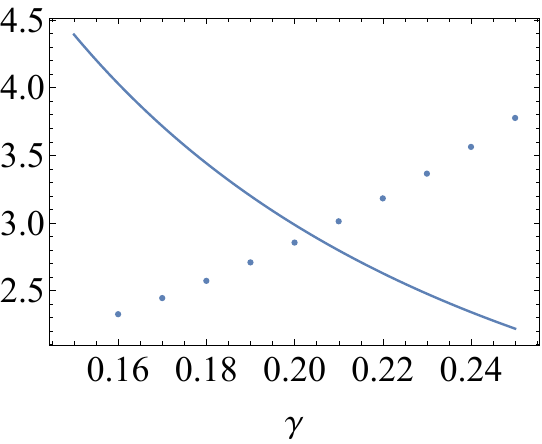}
    \caption{Graphical visualization of the criterion, Eq.~\ref{eq:comparison}, that determines whether VP or PDW is the leading instability in the bottom right corner of the phase diagram. Solid and dashed lines -- left and right sides of Eq.~\ref{eq:comparison}.  The two functions cross at $\gamma \approx0.21$. At smaller $\gamma$,   VP wins over PDW.}
        \label{fig:crossing}
\end{figure}

\begin{figure*}

\begin{subfigure}(a)
    \centering
    \includegraphics[width=8cm,height=5cm]{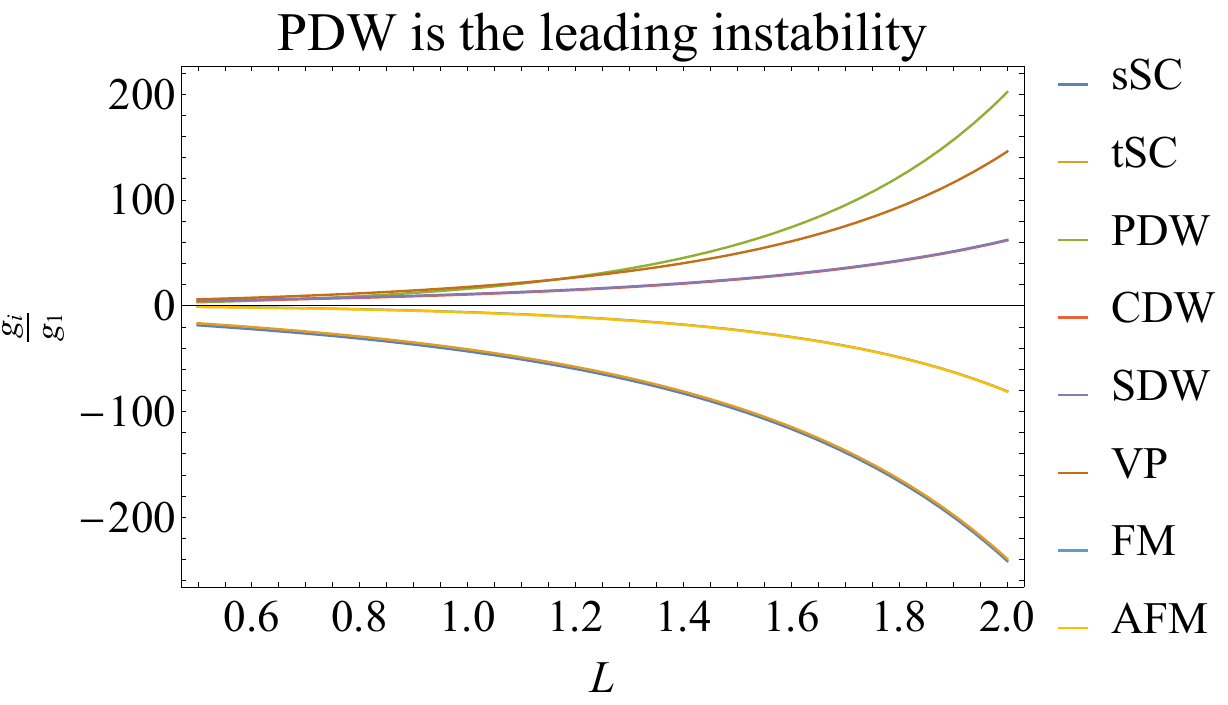}
\end{subfigure}
    \begin{subfigure}(b)
    \centering
    \includegraphics[width=8cm,height=5cm]{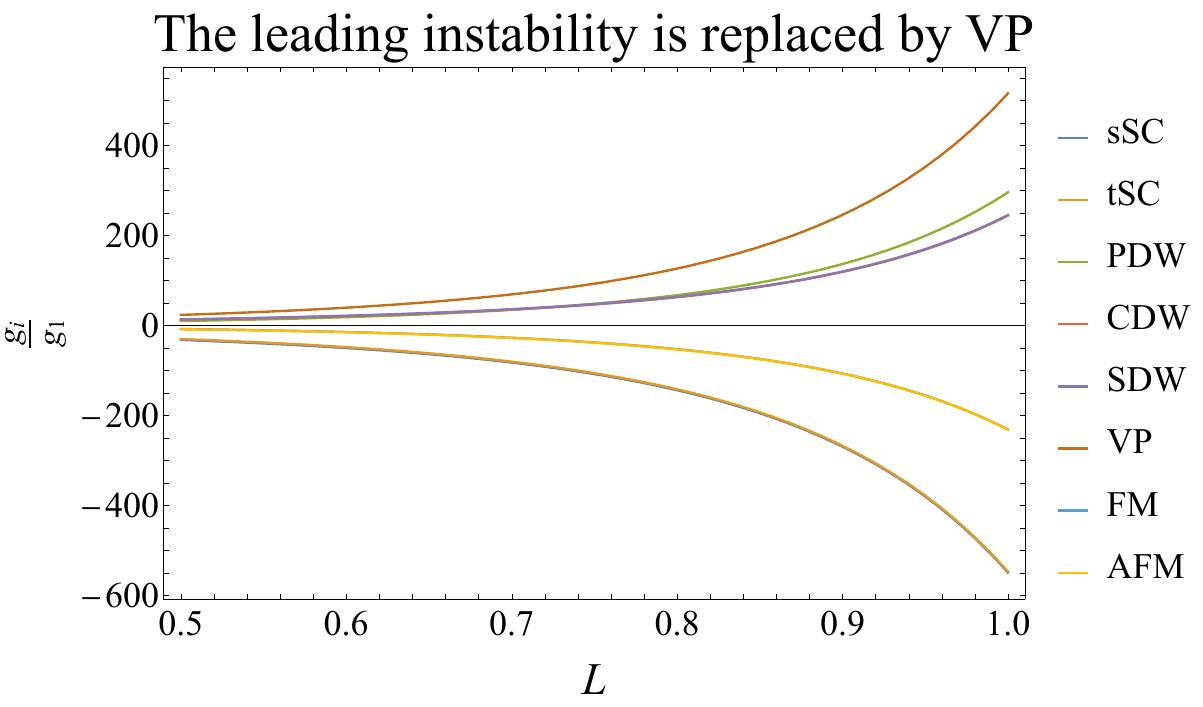}
\end{subfigure}
\caption{The flow of the ratios of couplings with the pRG parameter $L$ for bare values
$x_2^0 =5, x_4^0 =0.1$, which are in the range where the leading ordering tendency at $\gamma =1$ is PDW.
 We argued (see the text) that the ordering tendency changes to VP at $\gamma \approx0.21$.
   We plot the eigenvalues in different channels for somewhat larger and somewhat smaller $\gamma$. For larger $\gamma=0.22$, PDW is  the leading instability;  for smaller $\gamma=0.19$, the leading ordering tendency is VP.}
\label{fig:PDWvsVP}

\end{figure*}

\begin{figure*}
    \begin{subfigure}(a)
        \centering
        \includegraphics[width=8cm,height=5cm]{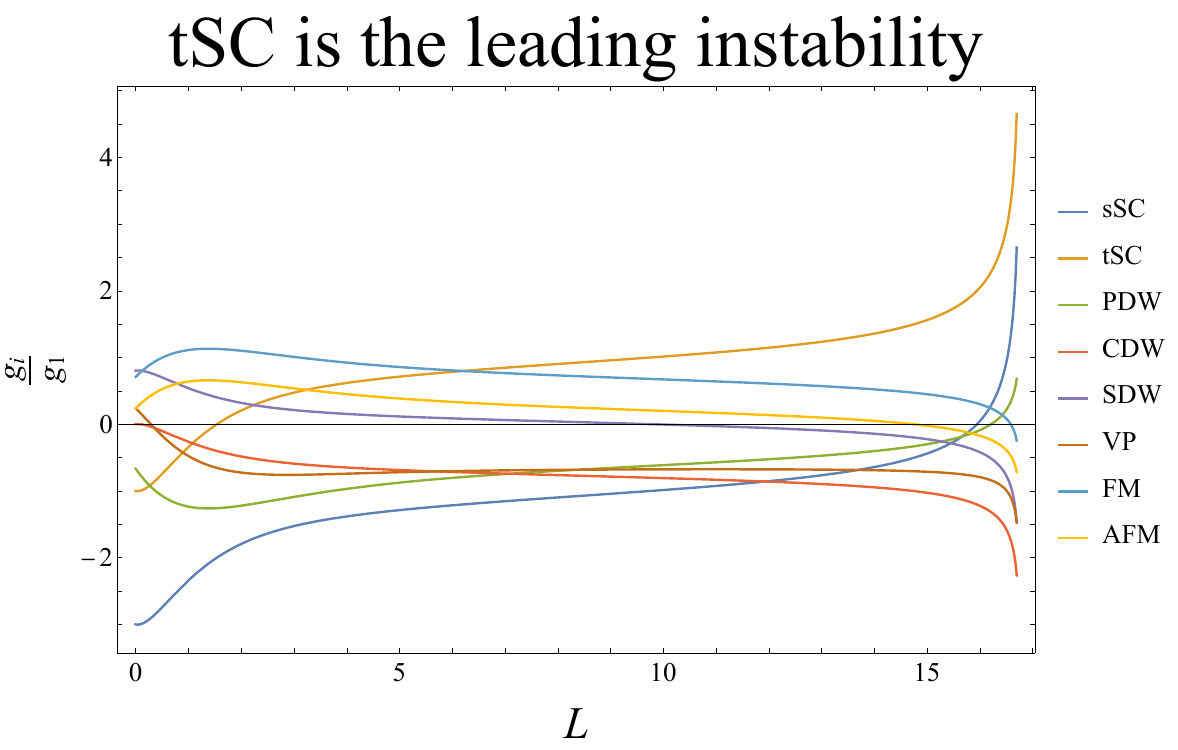}
    \end{subfigure}
    \begin{subfigure}(b)
        \centering
        \includegraphics[width=8.5cm,height=5cm]{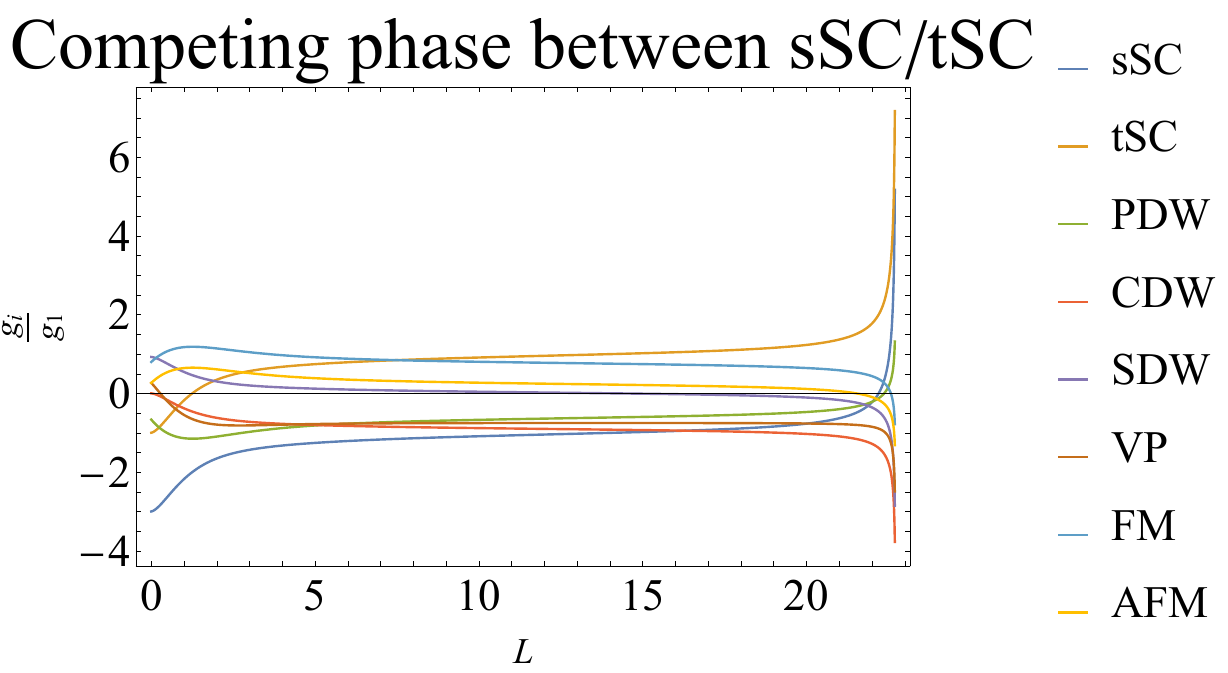}
    \end{subfigure}
    \caption{The flow of the ratios of the coupling as a function of $L$ before ($\gamma=0.18$) and after the disappearance of the stable fixed point $P_1$ ($\gamma=0.12$) using the bare value $x^0_2=x^0_4=2$. One can see that tSC wins over sSC when $P_1$ is present, but the two are degenerate when $P_1$ disappears.}
        \label{fig:after}

\end{figure*}

The behavior of the  fixed point $P_1$ resembles that of $P_3$.  As $\gamma$ decreases towards $\gamma_{c3}\approx 0.14$, it gradually moves towards $(-\infty,-\infty)$. At $\gamma$ infinitesimally smaller than $\gamma_{c3}$,
$P_1$ re-emerges at $(+\infty,+\infty)$ as a half-stable fixed point (one of the eigenvalues changes the sign). This half-stable fixed point remains on the phase diagram for smaller $\gamma$.

Even before $P_1$ disappears and re-appears, the PDW region on the phase diagram is replaced by VP.  One can straightforwardly check that $\lambda_{VP}$ becomes larger than $\lambda_{PDW}$ when
\begin{equation}
    \frac{2d_{ph}(0)}{d_{pp}(\textbf{Q})-d_{ph}(0)}>\frac{|g_4|}{g_2}.
    \label{eq:comparison}
\end{equation}
We plot both sides of this equation in Fig. \ref{fig:crossing}.
We see that they cross at  $\gamma\approx 0.21$.  To verify that at $\gamma$ smaller than this value the ordering tendency changes to VP, we plot in Fig. \ref{fig:PDWvsVP} the eigenvalues in different channels for proper  $x_2$ and $x_4$ at somewhat larger and somewhat smaller $\gamma$. For larger $\gamma=0.22$, PDW is  the leading instability;  for smaller $\gamma=0.19$, the leading ordering tendency is VP.

Next, we show that as $P_1$ disappears and then re-emerges, spin-triplet SC becomes degenerate with spin-singlet SC.
 We demonstrate this in Fig. \ref{fig:after}, where we plot the eigenvalues in different channels for proper $x_2$ and $x_4$. We see that indeed tSC becomes degenerate with sSC once the fixed point $P_1$ disappears from the lower part of the phase diagram.

\subsection{Pair production of the FM and half-stable fixed point}

The final element of the evolution of the pRG flow is the creation of  a pair of fixed points $P_5$ and $P_6$ in the second quadrant of the phase diagram at $\gamma_{c4}\approx0.09$.
The creation mechanism is exactly opposite to the annihilation of $P_3$ and $P_4$ and, like there, one of the fixed points is stable ($P_6$) and one is half-stable ($P_5$).   In our analytic study we find that right at $\gamma =\gamma_{c4}$, there are two degenerate and two non-degenerate solutions of Eq.~\ref{RG_equation_x}. At $\gamma > \gamma_{c4}$, the two degenerate solutions split into two complex solutions (i.e., the pRG flow is continuous), at $\gamma < \gamma_{c4}$ they split into two real solutions. Combining with the other two solutions, we obtain two real eigenvalues for each fixed point.
We analyzed the leading ordering tendency in the basin of attraction of the stable fixed point $P_6$ and found that it is towards global ferromagnetism.  At the same line,  the phase boundary that starts at the unstable fixed point $P_2$ and passes through $P_5$ separates  FM and tSC/sSC orders.
The RG flow along the phase boundary is also somewhat non-trivial. At $\gamma > \gamma_{c5}\approx0.04$,
 the functions $f_i(c_2,c_4)$ in Eq.~\ref{eq:f}, where $c_2$ and $c_4$ are $x_2$ and $x_4$ at $P_5$, are positive, i.e., the couplings diverge under pRG (the leading instability is tSC).  At $\gamma = \gamma_{c5}$, $f_i(c_2,c_4)=0$, i.e., the couplings remain invariant under pRG, similar to super-metal behavior, reported in ~\cite{Isobe2019}.
  At $\gamma < \gamma_{c5}$, $f_i(c_2,c_4) <0$, and the couplings tend to zero.

\bibliographystyle{apsrev4-2}
\bibliography{biblio}

\begin{thebibliography}{57}%
\makeatletter
\providecommand \@ifxundefined [1]{%
 \@ifx{#1\undefined}
}%
\providecommand \@ifnum [1]{%
 \ifnum #1\expandafter \@firstoftwo
 \else \expandafter \@secondoftwo
 \fi
}%
\providecommand \@ifx [1]{%
 \ifx #1\expandafter \@firstoftwo
 \else \expandafter \@secondoftwo
 \fi
}%
\providecommand \natexlab [1]{#1}%
\providecommand \enquote  [1]{``#1''}%
\providecommand \bibnamefont  [1]{#1}%
\providecommand \bibfnamefont [1]{#1}%
\providecommand \citenamefont [1]{#1}%
\providecommand \href@noop [0]{\@secondoftwo}%
\providecommand \href [0]{\begingroup \@sanitize@url \@href}%
\providecommand \@href[1]{\@@startlink{#1}\@@href}%
\providecommand \@@href[1]{\endgroup#1\@@endlink}%
\providecommand \@sanitize@url [0]{\catcode `\\12\catcode `\$12\catcode `\&12\catcode `\#12\catcode `\^12\catcode `\_12\catcode `\%12\relax}%
\providecommand \@@startlink[1]{}%
\providecommand \@@endlink[0]{}%
\providecommand \url  [0]{\begingroup\@sanitize@url \@url }%
\providecommand \@url [1]{\endgroup\@href {#1}{\urlprefix }}%
\providecommand \urlprefix  [0]{URL }%
\providecommand \Eprint [0]{\href }%
\providecommand \doibase [0]{https://doi.org/}%
\providecommand \selectlanguage [0]{\@gobble}%
\providecommand \bibinfo  [0]{\@secondoftwo}%
\providecommand \bibfield  [0]{\@secondoftwo}%
\providecommand \translation [1]{[#1]}%
\providecommand \BibitemOpen [0]{}%
\providecommand \bibitemStop [0]{}%
\providecommand \bibitemNoStop [0]{.\EOS\space}%
\providecommand \EOS [0]{\spacefactor3000\relax}%
\providecommand \BibitemShut  [1]{\csname bibitem#1\endcsname}%
\let\auto@bib@innerbib\@empty
\bibitem [{\citenamefont {Keimer}\ \emph {et~al.}(2015)\citenamefont {Keimer}, \citenamefont {Kivelson}, \citenamefont {Norman}, \citenamefont {Uchida},\ and\ \citenamefont {Zaanen}}]{Keimer2015}%
  \BibitemOpen
  \bibfield  {author} {\bibinfo {author} {\bibfnamefont {B.}~\bibnamefont {Keimer}}, \bibinfo {author} {\bibfnamefont {S.~A.}\ \bibnamefont {Kivelson}}, \bibinfo {author} {\bibfnamefont {M.~R.}\ \bibnamefont {Norman}}, \bibinfo {author} {\bibfnamefont {S.}~\bibnamefont {Uchida}},\ and\ \bibinfo {author} {\bibfnamefont {J.}~\bibnamefont {Zaanen}},\ }\href {https://doi.org/10.1038/nature14165} {\bibfield  {journal} {\bibinfo  {journal} {Nature}\ }\textbf {\bibinfo {volume} {518}},\ \bibinfo {pages} {179} (\bibinfo {year} {2015})}\BibitemShut {NoStop}%
\bibitem [{\citenamefont {Fernandes}\ and\ \citenamefont {Chubukov}(2016)}]{fernandes2016low}%
  \BibitemOpen
  \bibfield  {author} {\bibinfo {author} {\bibfnamefont {R.~M.}\ \bibnamefont {Fernandes}}\ and\ \bibinfo {author} {\bibfnamefont {A.~V.}\ \bibnamefont {Chubukov}},\ }\href@noop {} {\bibfield  {journal} {\bibinfo  {journal} {Reports on Progress in Physics}\ }\textbf {\bibinfo {volume} {80}},\ \bibinfo {pages} {014503} (\bibinfo {year} {2016})}\BibitemShut {NoStop}%
\bibitem [{\citenamefont {Varma}(2020)}]{Varma2020}%
  \BibitemOpen
  \bibfield  {author} {\bibinfo {author} {\bibfnamefont {C.~M.}\ \bibnamefont {Varma}},\ }\href {https://doi.org/10.1103/RevModPhys.92.031001} {\bibfield  {journal} {\bibinfo  {journal} {Rev. Mod. Phys.}\ }\textbf {\bibinfo {volume} {92}},\ \bibinfo {pages} {031001} (\bibinfo {year} {2020})}\BibitemShut {NoStop}%
\bibitem [{\citenamefont {Cao}\ \emph {et~al.}(2018{\natexlab{a}})\citenamefont {Cao}, \citenamefont {Fatemi}, \citenamefont {Fang}, \citenamefont {Watanabe}, \citenamefont {Taniguchi}, \citenamefont {Kaxiras},\ and\ \citenamefont {Jarillo-Herrero}}]{Cao2018SC}%
  \BibitemOpen
  \bibfield  {author} {\bibinfo {author} {\bibfnamefont {Y.}~\bibnamefont {Cao}}, \bibinfo {author} {\bibfnamefont {V.}~\bibnamefont {Fatemi}}, \bibinfo {author} {\bibfnamefont {S.}~\bibnamefont {Fang}}, \bibinfo {author} {\bibfnamefont {K.}~\bibnamefont {Watanabe}}, \bibinfo {author} {\bibfnamefont {T.}~\bibnamefont {Taniguchi}}, \bibinfo {author} {\bibfnamefont {E.}~\bibnamefont {Kaxiras}},\ and\ \bibinfo {author} {\bibfnamefont {P.}~\bibnamefont {Jarillo-Herrero}},\ }\href@noop {} {\bibfield  {journal} {\bibinfo  {journal} {Nature}\ }\textbf {\bibinfo {volume} {556}},\ \bibinfo {pages} {43} (\bibinfo {year} {2018}{\natexlab{a}})}\BibitemShut {NoStop}%
\bibitem [{\citenamefont {Cao}\ \emph {et~al.}(2018{\natexlab{b}})\citenamefont {Cao}, \citenamefont {Fatemi}, \citenamefont {Demir}, \citenamefont {Fang}, \citenamefont {Tomarken}, \citenamefont {Luo}, \citenamefont {Sanchez-Yamagishi}, \citenamefont {Watanabe}, \citenamefont {Taniguchi}, \citenamefont {Kaxiras}, \citenamefont {Ashoori},\ and\ \citenamefont {Jarillo-Herrero}}]{Cao2018insulator}%
  \BibitemOpen
  \bibfield  {author} {\bibinfo {author} {\bibfnamefont {Y.}~\bibnamefont {Cao}}, \bibinfo {author} {\bibfnamefont {V.}~\bibnamefont {Fatemi}}, \bibinfo {author} {\bibfnamefont {A.}~\bibnamefont {Demir}}, \bibinfo {author} {\bibfnamefont {S.}~\bibnamefont {Fang}}, \bibinfo {author} {\bibfnamefont {S.~L.}\ \bibnamefont {Tomarken}}, \bibinfo {author} {\bibfnamefont {J.~Y.}\ \bibnamefont {Luo}}, \bibinfo {author} {\bibfnamefont {J.~D.}\ \bibnamefont {Sanchez-Yamagishi}}, \bibinfo {author} {\bibfnamefont {K.}~\bibnamefont {Watanabe}}, \bibinfo {author} {\bibfnamefont {T.}~\bibnamefont {Taniguchi}}, \bibinfo {author} {\bibfnamefont {E.}~\bibnamefont {Kaxiras}}, \bibinfo {author} {\bibfnamefont {R.~C.}\ \bibnamefont {Ashoori}},\ and\ \bibinfo {author} {\bibfnamefont {P.}~\bibnamefont {Jarillo-Herrero}},\ }\href@noop {} {\bibfield  {journal} {\bibinfo  {journal} {Nature}\ }\textbf {\bibinfo {volume} {556}},\ \bibinfo {pages} {80} (\bibinfo {year} {2018}{\natexlab{b}})}\BibitemShut {NoStop}%
\bibitem [{\citenamefont {Cao}\ \emph {et~al.}(2021)\citenamefont {Cao}, \citenamefont {Rodan-Legrain}, \citenamefont {Park}, \citenamefont {Yuan}, \citenamefont {Watanabe}, \citenamefont {Taniguchi}, \citenamefont {Fernandes}, \citenamefont {Fu},\ and\ \citenamefont {Jarillo-Herrero}}]{Cao2020NematicSC}%
  \BibitemOpen
  \bibfield  {author} {\bibinfo {author} {\bibfnamefont {Y.}~\bibnamefont {Cao}}, \bibinfo {author} {\bibfnamefont {D.}~\bibnamefont {Rodan-Legrain}}, \bibinfo {author} {\bibfnamefont {J.~M.}\ \bibnamefont {Park}}, \bibinfo {author} {\bibfnamefont {N.~F.~Q.}\ \bibnamefont {Yuan}}, \bibinfo {author} {\bibfnamefont {K.}~\bibnamefont {Watanabe}}, \bibinfo {author} {\bibfnamefont {T.}~\bibnamefont {Taniguchi}}, \bibinfo {author} {\bibfnamefont {R.~M.}\ \bibnamefont {Fernandes}}, \bibinfo {author} {\bibfnamefont {L.}~\bibnamefont {Fu}},\ and\ \bibinfo {author} {\bibfnamefont {P.}~\bibnamefont {Jarillo-Herrero}},\ }\href@noop {} {\bibfield  {journal} {\bibinfo  {journal} {Science}\ }\textbf {\bibinfo {volume} {372}},\ \bibinfo {pages} {264} (\bibinfo {year} {2021})}\BibitemShut {NoStop}%
\bibitem [{\citenamefont {Zondiner}\ \emph {et~al.}(2020)\citenamefont {Zondiner}, \citenamefont {Rozen}, \citenamefont {Rodan-Legrain}, \citenamefont {Cao}, \citenamefont {Queiroz}, \citenamefont {Taniguchi}, \citenamefont {Watanabe}, \citenamefont {Oreg}, \citenamefont {von Oppen}, \citenamefont {Stern}, \citenamefont {Berg}, \citenamefont {Jarillo-Herrero},\ and\ \citenamefont {Ilani}}]{Zondiner2020}%
  \BibitemOpen
  \bibfield  {author} {\bibinfo {author} {\bibfnamefont {U.}~\bibnamefont {Zondiner}}, \bibinfo {author} {\bibfnamefont {A.}~\bibnamefont {Rozen}}, \bibinfo {author} {\bibfnamefont {D.}~\bibnamefont {Rodan-Legrain}}, \bibinfo {author} {\bibfnamefont {Y.}~\bibnamefont {Cao}}, \bibinfo {author} {\bibfnamefont {R.}~\bibnamefont {Queiroz}}, \bibinfo {author} {\bibfnamefont {T.}~\bibnamefont {Taniguchi}}, \bibinfo {author} {\bibfnamefont {K.}~\bibnamefont {Watanabe}}, \bibinfo {author} {\bibfnamefont {Y.}~\bibnamefont {Oreg}}, \bibinfo {author} {\bibfnamefont {F.}~\bibnamefont {von Oppen}}, \bibinfo {author} {\bibfnamefont {A.}~\bibnamefont {Stern}}, \bibinfo {author} {\bibfnamefont {E.}~\bibnamefont {Berg}}, \bibinfo {author} {\bibfnamefont {P.}~\bibnamefont {Jarillo-Herrero}},\ and\ \bibinfo {author} {\bibfnamefont {S.}~\bibnamefont {Ilani}},\ }\href@noop {} {\bibfield  {journal} {\bibinfo  {journal} {Nature}\ }\textbf {\bibinfo {volume} {582}},\ \bibinfo {pages} {203} (\bibinfo {year} {2020})}\BibitemShut
  {NoStop}%
\bibitem [{\citenamefont {Wong}\ \emph {et~al.}(2020)\citenamefont {Wong}, \citenamefont {Nuckolls}, \citenamefont {Oh}, \citenamefont {Lian}, \citenamefont {Xie}, \citenamefont {Jeon}, \citenamefont {Watanabe}, \citenamefont {Taniguchi}, \citenamefont {Bernevig},\ and\ \citenamefont {Yazdani}}]{ali_2}%
  \BibitemOpen
  \bibfield  {author} {\bibinfo {author} {\bibfnamefont {D.}~\bibnamefont {Wong}}, \bibinfo {author} {\bibfnamefont {K.~P.}\ \bibnamefont {Nuckolls}}, \bibinfo {author} {\bibfnamefont {M.}~\bibnamefont {Oh}}, \bibinfo {author} {\bibfnamefont {B.}~\bibnamefont {Lian}}, \bibinfo {author} {\bibfnamefont {Y.}~\bibnamefont {Xie}}, \bibinfo {author} {\bibfnamefont {S.}~\bibnamefont {Jeon}}, \bibinfo {author} {\bibfnamefont {K.}~\bibnamefont {Watanabe}}, \bibinfo {author} {\bibfnamefont {T.}~\bibnamefont {Taniguchi}}, \bibinfo {author} {\bibfnamefont {B.~A.}\ \bibnamefont {Bernevig}},\ and\ \bibinfo {author} {\bibfnamefont {A.}~\bibnamefont {Yazdani}},\ }\href@noop {} {\bibfield  {journal} {\bibinfo  {journal} {Nature}\ }\textbf {\bibinfo {volume} {582}},\ \bibinfo {pages} {198} (\bibinfo {year} {2020})}\BibitemShut {NoStop}%
\bibitem [{\citenamefont {Wu}\ \emph {et~al.}(2021)\citenamefont {Wu}, \citenamefont {Zhang}, \citenamefont {Watanabe}, \citenamefont {Taniguchi},\ and\ \citenamefont {Andrei}}]{Wu2021}%
  \BibitemOpen
  \bibfield  {author} {\bibinfo {author} {\bibfnamefont {S.}~\bibnamefont {Wu}}, \bibinfo {author} {\bibfnamefont {Z.}~\bibnamefont {Zhang}}, \bibinfo {author} {\bibfnamefont {K.}~\bibnamefont {Watanabe}}, \bibinfo {author} {\bibfnamefont {T.}~\bibnamefont {Taniguchi}},\ and\ \bibinfo {author} {\bibfnamefont {E.~Y.}\ \bibnamefont {Andrei}},\ }\href@noop {} {\bibfield  {journal} {\bibinfo  {journal} {Nature Materials}\ }\textbf {\bibinfo {volume} {20}},\ \bibinfo {pages} {488} (\bibinfo {year} {2021})}\BibitemShut {NoStop}%
\bibitem [{\citenamefont {Polski}\ \emph {et~al.}(2022)\citenamefont {Polski}, \citenamefont {Zhang}, \citenamefont {Peng}, \citenamefont {Arora}, \citenamefont {Choi}, \citenamefont {Kim}, \citenamefont {Watanabe}, \citenamefont {Taniguchi}, \citenamefont {Refael}, \citenamefont {von Oppen},\ and\ \citenamefont {Nadj-Perge}}]{Nadj_Perge_cascade}%
  \BibitemOpen
  \bibfield  {author} {\bibinfo {author} {\bibfnamefont {R.}~\bibnamefont {Polski}}, \bibinfo {author} {\bibfnamefont {Y.}~\bibnamefont {Zhang}}, \bibinfo {author} {\bibfnamefont {Y.}~\bibnamefont {Peng}}, \bibinfo {author} {\bibfnamefont {H.~S.}\ \bibnamefont {Arora}}, \bibinfo {author} {\bibfnamefont {Y.}~\bibnamefont {Choi}}, \bibinfo {author} {\bibfnamefont {H.}~\bibnamefont {Kim}}, \bibinfo {author} {\bibfnamefont {K.}~\bibnamefont {Watanabe}}, \bibinfo {author} {\bibfnamefont {T.}~\bibnamefont {Taniguchi}}, \bibinfo {author} {\bibfnamefont {G.}~\bibnamefont {Refael}}, \bibinfo {author} {\bibfnamefont {F.}~\bibnamefont {von Oppen}},\ and\ \bibinfo {author} {\bibfnamefont {S.}~\bibnamefont {Nadj-Perge}},\ }\href {https://doi.org/10.48550/ARXIV.2205.05225} {\bibinfo {title} {Hierarchy of symmetry breaking correlated phases in twisted bilayer graphene}} (\bibinfo {year} {2022})\BibitemShut {NoStop}%
\bibitem [{\citenamefont {Zhou}\ \emph {et~al.}(2022)\citenamefont {Zhou}, \citenamefont {Holleis}, \citenamefont {Saito}, \citenamefont {Cohen}, \citenamefont {Huynh}, \citenamefont {Patterson}, \citenamefont {Yang}, \citenamefont {Taniguchi}, \citenamefont {Watanabe},\ and\ \citenamefont {Young}}]{Zhou2022}%
  \BibitemOpen
  \bibfield  {author} {\bibinfo {author} {\bibfnamefont {H.}~\bibnamefont {Zhou}}, \bibinfo {author} {\bibfnamefont {L.}~\bibnamefont {Holleis}}, \bibinfo {author} {\bibfnamefont {Y.}~\bibnamefont {Saito}}, \bibinfo {author} {\bibfnamefont {L.}~\bibnamefont {Cohen}}, \bibinfo {author} {\bibfnamefont {W.}~\bibnamefont {Huynh}}, \bibinfo {author} {\bibfnamefont {C.~L.}\ \bibnamefont {Patterson}}, \bibinfo {author} {\bibfnamefont {F.}~\bibnamefont {Yang}}, \bibinfo {author} {\bibfnamefont {T.}~\bibnamefont {Taniguchi}}, \bibinfo {author} {\bibfnamefont {K.}~\bibnamefont {Watanabe}},\ and\ \bibinfo {author} {\bibfnamefont {A.~F.}\ \bibnamefont {Young}},\ }\href {https://doi.org/10.1126/science.abm8386} {\bibfield  {journal} {\bibinfo  {journal} {Science}\ }\textbf {\bibinfo {volume} {375}},\ \bibinfo {pages} {774} (\bibinfo {year} {2022})},\ \Eprint {https://arxiv.org/abs/https://www.science.org/doi/pdf/10.1126/science.abm8386} {https://www.science.org/doi/pdf/10.1126/science.abm8386} \BibitemShut {NoStop}%
\bibitem [{\citenamefont {Holleis}\ \emph {et~al.}(2023)\citenamefont {Holleis}, \citenamefont {Patterson}, \citenamefont {Zhang}, \citenamefont {Yoo}, \citenamefont {Zhou}, \citenamefont {Taniguchi}, \citenamefont {Watanabe}, \citenamefont {Nadj-Perge},\ and\ \citenamefont {Young}}]{Holleis2023}%
  \BibitemOpen
  \bibfield  {author} {\bibinfo {author} {\bibfnamefont {L.}~\bibnamefont {Holleis}}, \bibinfo {author} {\bibfnamefont {C.~L.}\ \bibnamefont {Patterson}}, \bibinfo {author} {\bibfnamefont {Y.}~\bibnamefont {Zhang}}, \bibinfo {author} {\bibfnamefont {H.~M.}\ \bibnamefont {Yoo}}, \bibinfo {author} {\bibfnamefont {H.}~\bibnamefont {Zhou}}, \bibinfo {author} {\bibfnamefont {T.}~\bibnamefont {Taniguchi}}, \bibinfo {author} {\bibfnamefont {K.}~\bibnamefont {Watanabe}}, \bibinfo {author} {\bibfnamefont {S.}~\bibnamefont {Nadj-Perge}},\ and\ \bibinfo {author} {\bibfnamefont {A.~F.}\ \bibnamefont {Young}},\ }\href@noop {} {\bibinfo {title} {Ising superconductivity and nematicity in bernal bilayer graphene with strong spin orbit coupling}} (\bibinfo {year} {2023}),\ \Eprint {https://arxiv.org/abs/2303.00742} {arXiv:2303.00742 [cond-mat.supr-con]} \BibitemShut {NoStop}%
\bibitem [{\citenamefont {Zhou}\ \emph {et~al.}(2021{\natexlab{a}})\citenamefont {Zhou}, \citenamefont {Xie}, \citenamefont {Taniguchi}, \citenamefont {Watanabe},\ and\ \citenamefont {Young}}]{Zhou2021SC}%
  \BibitemOpen
  \bibfield  {author} {\bibinfo {author} {\bibfnamefont {H.}~\bibnamefont {Zhou}}, \bibinfo {author} {\bibfnamefont {T.}~\bibnamefont {Xie}}, \bibinfo {author} {\bibfnamefont {T.}~\bibnamefont {Taniguchi}}, \bibinfo {author} {\bibfnamefont {K.}~\bibnamefont {Watanabe}},\ and\ \bibinfo {author} {\bibfnamefont {A.~F.}\ \bibnamefont {Young}},\ }\href@noop {} {\bibfield  {journal} {\bibinfo  {journal} {Nature}\ }\textbf {\bibinfo {volume} {598}},\ \bibinfo {pages} {434} (\bibinfo {year} {2021}{\natexlab{a}})}\BibitemShut {NoStop}%
\bibitem [{\citenamefont {Zhou}\ \emph {et~al.}(2021{\natexlab{b}})\citenamefont {Zhou}, \citenamefont {Xie}, \citenamefont {Ghazaryan}, \citenamefont {Holder}, \citenamefont {Ehrets}, \citenamefont {Spanton}, \citenamefont {Taniguchi}, \citenamefont {Watanabe}, \citenamefont {Berg}, \citenamefont {Serbyn},\ and\ \citenamefont {Young}}]{Zhou2021}%
  \BibitemOpen
  \bibfield  {author} {\bibinfo {author} {\bibfnamefont {H.}~\bibnamefont {Zhou}}, \bibinfo {author} {\bibfnamefont {T.}~\bibnamefont {Xie}}, \bibinfo {author} {\bibfnamefont {A.}~\bibnamefont {Ghazaryan}}, \bibinfo {author} {\bibfnamefont {T.}~\bibnamefont {Holder}}, \bibinfo {author} {\bibfnamefont {J.~R.}\ \bibnamefont {Ehrets}}, \bibinfo {author} {\bibfnamefont {E.~M.}\ \bibnamefont {Spanton}}, \bibinfo {author} {\bibfnamefont {T.}~\bibnamefont {Taniguchi}}, \bibinfo {author} {\bibfnamefont {K.}~\bibnamefont {Watanabe}}, \bibinfo {author} {\bibfnamefont {E.}~\bibnamefont {Berg}}, \bibinfo {author} {\bibfnamefont {M.}~\bibnamefont {Serbyn}},\ and\ \bibinfo {author} {\bibfnamefont {A.~F.}\ \bibnamefont {Young}},\ }\href {https://doi.org/10.1038/s41586-021-03938-w} {\bibfield  {journal} {\bibinfo  {journal} {Nature}\ }\textbf {\bibinfo {volume} {598}},\ \bibinfo {pages} {429} (\bibinfo {year} {2021}{\natexlab{b}})}\BibitemShut {NoStop}%
\bibitem [{\citenamefont {Park}\ \emph {et~al.}(2021)\citenamefont {Park}, \citenamefont {Cao}, \citenamefont {Watanabe}, \citenamefont {Taniguchi},\ and\ \citenamefont {Jarillo-Herrero}}]{Park2021TTG}%
  \BibitemOpen
  \bibfield  {author} {\bibinfo {author} {\bibfnamefont {J.~M.}\ \bibnamefont {Park}}, \bibinfo {author} {\bibfnamefont {Y.}~\bibnamefont {Cao}}, \bibinfo {author} {\bibfnamefont {K.}~\bibnamefont {Watanabe}}, \bibinfo {author} {\bibfnamefont {T.}~\bibnamefont {Taniguchi}},\ and\ \bibinfo {author} {\bibfnamefont {P.}~\bibnamefont {Jarillo-Herrero}},\ }\href@noop {} {\bibfield  {journal} {\bibinfo  {journal} {Nature}\ }\textbf {\bibinfo {volume} {590}},\ \bibinfo {pages} {249} (\bibinfo {year} {2021})}\BibitemShut {NoStop}%
\bibitem [{\citenamefont {Jiang}\ \emph {et~al.}(2021)\citenamefont {Jiang}, \citenamefont {Scalapino},\ and\ \citenamefont {White}}]{White2021}%
  \BibitemOpen
  \bibfield  {author} {\bibinfo {author} {\bibfnamefont {S.}~\bibnamefont {Jiang}}, \bibinfo {author} {\bibfnamefont {D.~J.}\ \bibnamefont {Scalapino}},\ and\ \bibinfo {author} {\bibfnamefont {S.~R.}\ \bibnamefont {White}},\ }\href {https://doi.org/10.1073/pnas.2109978118} {\bibfield  {journal} {\bibinfo  {journal} {Proceedings of the National Academy of Sciences}\ }\textbf {\bibinfo {volume} {118}},\ \bibinfo {pages} {e2109978118} (\bibinfo {year} {2021})},\ \Eprint {https://arxiv.org/abs/https://www.pnas.org/doi/pdf/10.1073/pnas.2109978118} {https://www.pnas.org/doi/pdf/10.1073/pnas.2109978118} \BibitemShut {NoStop}%
\bibitem [{\citenamefont {Xie}\ \emph {et~al.}(2021)\citenamefont {Xie}, \citenamefont {Regnault}, \citenamefont {C\ifmmode \u{a}\else \u{a}\fi{}lug\ifmmode~\u{a}\else \u{a}\fi{}ru}, \citenamefont {Bernevig},\ and\ \citenamefont {Lian}}]{Xie2021}%
  \BibitemOpen
  \bibfield  {author} {\bibinfo {author} {\bibfnamefont {F.}~\bibnamefont {Xie}}, \bibinfo {author} {\bibfnamefont {N.}~\bibnamefont {Regnault}}, \bibinfo {author} {\bibfnamefont {D.}~\bibnamefont {C\ifmmode \u{a}\else \u{a}\fi{}lug\ifmmode~\u{a}\else \u{a}\fi{}ru}}, \bibinfo {author} {\bibfnamefont {B.~A.}\ \bibnamefont {Bernevig}},\ and\ \bibinfo {author} {\bibfnamefont {B.}~\bibnamefont {Lian}},\ }\href {https://doi.org/10.1103/PhysRevB.104.115167} {\bibfield  {journal} {\bibinfo  {journal} {Phys. Rev. B}\ }\textbf {\bibinfo {volume} {104}},\ \bibinfo {pages} {115167} (\bibinfo {year} {2021})}\BibitemShut {NoStop}%
\bibitem [{\citenamefont {Chubukov}\ \emph {et~al.}(2008)\citenamefont {Chubukov}, \citenamefont {Efremov},\ and\ \citenamefont {Eremin}}]{CEE}%
  \BibitemOpen
  \bibfield  {author} {\bibinfo {author} {\bibfnamefont {A.~V.}\ \bibnamefont {Chubukov}}, \bibinfo {author} {\bibfnamefont {D.~V.}\ \bibnamefont {Efremov}},\ and\ \bibinfo {author} {\bibfnamefont {I.}~\bibnamefont {Eremin}},\ }\href {https://doi.org/10.1103/PhysRevB.78.134512} {\bibfield  {journal} {\bibinfo  {journal} {Phys. Rev. B}\ }\textbf {\bibinfo {volume} {78}},\ \bibinfo {pages} {134512} (\bibinfo {year} {2008})}\BibitemShut {NoStop}%
\bibitem [{\citenamefont {Chubukov}\ \emph {et~al.}(2016)\citenamefont {Chubukov}, \citenamefont {Khodas},\ and\ \citenamefont {Fernandes}}]{Khodas2016}%
  \BibitemOpen
  \bibfield  {author} {\bibinfo {author} {\bibfnamefont {A.~V.}\ \bibnamefont {Chubukov}}, \bibinfo {author} {\bibfnamefont {M.}~\bibnamefont {Khodas}},\ and\ \bibinfo {author} {\bibfnamefont {R.~M.}\ \bibnamefont {Fernandes}},\ }\href {https://doi.org/10.1103/PhysRevX.6.041045} {\bibfield  {journal} {\bibinfo  {journal} {Phys. Rev. X}\ }\textbf {\bibinfo {volume} {6}},\ \bibinfo {pages} {041045} (\bibinfo {year} {2016})}\BibitemShut {NoStop}%
\bibitem [{\citenamefont {Classen}\ \emph {et~al.}(2017)\citenamefont {Classen}, \citenamefont {Xing}, \citenamefont {Khodas},\ and\ \citenamefont {Chubukov}}]{Classen2017}%
  \BibitemOpen
  \bibfield  {author} {\bibinfo {author} {\bibfnamefont {L.}~\bibnamefont {Classen}}, \bibinfo {author} {\bibfnamefont {R.-Q.}\ \bibnamefont {Xing}}, \bibinfo {author} {\bibfnamefont {M.}~\bibnamefont {Khodas}},\ and\ \bibinfo {author} {\bibfnamefont {A.~V.}\ \bibnamefont {Chubukov}},\ }\href {https://doi.org/10.1103/PhysRevLett.118.037001} {\bibfield  {journal} {\bibinfo  {journal} {Phys. Rev. Lett.}\ }\textbf {\bibinfo {volume} {118}},\ \bibinfo {pages} {037001} (\bibinfo {year} {2017})}\BibitemShut {NoStop}%
\bibitem [{\citenamefont {Metlitski}\ and\ \citenamefont {Sachdev}(2010{\natexlab{a}})}]{Metlitski2010}%
  \BibitemOpen
  \bibfield  {author} {\bibinfo {author} {\bibfnamefont {M.~A.}\ \bibnamefont {Metlitski}}\ and\ \bibinfo {author} {\bibfnamefont {S.}~\bibnamefont {Sachdev}},\ }\href {https://doi.org/10.1103/PhysRevB.82.075128} {\bibfield  {journal} {\bibinfo  {journal} {Phys. Rev. B}\ }\textbf {\bibinfo {volume} {82}},\ \bibinfo {pages} {075128} (\bibinfo {year} {2010}{\natexlab{a}})}\BibitemShut {NoStop}%
\bibitem [{\citenamefont {Metlitski}\ and\ \citenamefont {Sachdev}(2010{\natexlab{b}})}]{Metlitski_2010}%
  \BibitemOpen
  \bibfield  {author} {\bibinfo {author} {\bibfnamefont {M.~A.}\ \bibnamefont {Metlitski}}\ and\ \bibinfo {author} {\bibfnamefont {S.}~\bibnamefont {Sachdev}},\ }\href {https://doi.org/10.1088/1367-2630/12/10/105007} {\bibfield  {journal} {\bibinfo  {journal} {New Journal of Physics}\ }\textbf {\bibinfo {volume} {12}},\ \bibinfo {pages} {105007} (\bibinfo {year} {2010}{\natexlab{b}})}\BibitemShut {NoStop}%
\bibitem [{\citenamefont {Sachdev}\ \emph {et~al.}(2012)\citenamefont {Sachdev}, \citenamefont {Metlitski},\ and\ \citenamefont {Punk}}]{Sachdev_2012}%
  \BibitemOpen
  \bibfield  {author} {\bibinfo {author} {\bibfnamefont {S.}~\bibnamefont {Sachdev}}, \bibinfo {author} {\bibfnamefont {M.~A.}\ \bibnamefont {Metlitski}},\ and\ \bibinfo {author} {\bibfnamefont {M.}~\bibnamefont {Punk}},\ }\href {https://doi.org/10.1088/0953-8984/24/29/294205} {\bibfield  {journal} {\bibinfo  {journal} {Journal of Physics: Condensed Matter}\ }\textbf {\bibinfo {volume} {24}},\ \bibinfo {pages} {294205} (\bibinfo {year} {2012})}\BibitemShut {NoStop}%
\bibitem [{\citenamefont {Wang}\ and\ \citenamefont {Chubukov}(2014)}]{Wang2014}%
  \BibitemOpen
  \bibfield  {author} {\bibinfo {author} {\bibfnamefont {Y.}~\bibnamefont {Wang}}\ and\ \bibinfo {author} {\bibfnamefont {A.}~\bibnamefont {Chubukov}},\ }\href {https://doi.org/10.1103/PhysRevB.90.035149} {\bibfield  {journal} {\bibinfo  {journal} {Phys. Rev. B}\ }\textbf {\bibinfo {volume} {90}},\ \bibinfo {pages} {035149} (\bibinfo {year} {2014})}\BibitemShut {NoStop}%
\bibitem [{\citenamefont {DZYALOSHINSKII}(1987)}]{Dzyaloshinskii_1987}%
  \BibitemOpen
  \bibfield  {author} {\bibinfo {author} {\bibfnamefont {I.}~\bibnamefont {DZYALOSHINSKII}},\ }\href@noop {} {\bibfield  {journal} {\bibinfo  {journal} {ZHURNAL EKSPERIMENTALNOI I TEORETICHESKOI FIZIKI}\ }\textbf {\bibinfo {volume} {93}},\ \bibinfo {pages} {1487} (\bibinfo {year} {1987})}\BibitemShut {NoStop}%
\bibitem [{\citenamefont {Schulz}(1987)}]{Schulz_1987}%
  \BibitemOpen
  \bibfield  {author} {\bibinfo {author} {\bibfnamefont {H.~J.}\ \bibnamefont {Schulz}},\ }\href {https://doi.org/10.1209/0295-5075/4/5/016} {\bibfield  {journal} {\bibinfo  {journal} {Europhysics Letters}\ }\textbf {\bibinfo {volume} {4}},\ \bibinfo {pages} {609} (\bibinfo {year} {1987})}\BibitemShut {NoStop}%
\bibitem [{\citenamefont {Chubukov}(2009)}]{Chubukov2009Review}%
  \BibitemOpen
  \bibfield  {author} {\bibinfo {author} {\bibfnamefont {A.}~\bibnamefont {Chubukov}},\ }\href {https://doi.org/https://doi.org/10.1016/j.physc.2009.03.023} {\bibfield  {journal} {\bibinfo  {journal} {Physica C: Superconductivity}\ }\textbf {\bibinfo {volume} {469}},\ \bibinfo {pages} {640} (\bibinfo {year} {2009})},\ \bibinfo {note} {superconductivity in Iron-Pnictides}\BibitemShut {NoStop}%
\bibitem [{\citenamefont {Shankar}(1994)}]{Shankar1994RMP}%
  \BibitemOpen
  \bibfield  {author} {\bibinfo {author} {\bibfnamefont {R.}~\bibnamefont {Shankar}},\ }\href {https://doi.org/10.1103/RevModPhys.66.129} {\bibfield  {journal} {\bibinfo  {journal} {Rev. Mod. Phys.}\ }\textbf {\bibinfo {volume} {66}},\ \bibinfo {pages} {129} (\bibinfo {year} {1994})}\BibitemShut {NoStop}%
\bibitem [{\citenamefont {Hur}\ and\ \citenamefont {{Maurice Rice}}(2009)}]{Rice2009}%
  \BibitemOpen
  \bibfield  {author} {\bibinfo {author} {\bibfnamefont {K.~L.}\ \bibnamefont {Hur}}\ and\ \bibinfo {author} {\bibfnamefont {T.}~\bibnamefont {{Maurice Rice}}},\ }\href {https://doi.org/https://doi.org/10.1016/j.aop.2009.02.004} {\bibfield  {journal} {\bibinfo  {journal} {Annals of Physics}\ }\textbf {\bibinfo {volume} {324}},\ \bibinfo {pages} {1452} (\bibinfo {year} {2009})},\ \bibinfo {note} {july 2009 Special Issue}\BibitemShut {NoStop}%
\bibitem [{\citenamefont {Metzner}\ \emph {et~al.}(2012)\citenamefont {Metzner}, \citenamefont {Salmhofer}, \citenamefont {Honerkamp}, \citenamefont {Meden},\ and\ \citenamefont {Sch\"onhammer}}]{Metzner2012}%
  \BibitemOpen
  \bibfield  {author} {\bibinfo {author} {\bibfnamefont {W.}~\bibnamefont {Metzner}}, \bibinfo {author} {\bibfnamefont {M.}~\bibnamefont {Salmhofer}}, \bibinfo {author} {\bibfnamefont {C.}~\bibnamefont {Honerkamp}}, \bibinfo {author} {\bibfnamefont {V.}~\bibnamefont {Meden}},\ and\ \bibinfo {author} {\bibfnamefont {K.}~\bibnamefont {Sch\"onhammer}},\ }\href {https://doi.org/10.1103/RevModPhys.84.299} {\bibfield  {journal} {\bibinfo  {journal} {Rev. Mod. Phys.}\ }\textbf {\bibinfo {volume} {84}},\ \bibinfo {pages} {299} (\bibinfo {year} {2012})}\BibitemShut {NoStop}%
\bibitem [{\citenamefont {Qin}\ \emph {et~al.}(2023)\citenamefont {Qin}, \citenamefont {Huang}, \citenamefont {Wolf}, \citenamefont {Wei}, \citenamefont {Blinov},\ and\ \citenamefont {MacDonald}}]{Qin2023}%
  \BibitemOpen
  \bibfield  {author} {\bibinfo {author} {\bibfnamefont {W.}~\bibnamefont {Qin}}, \bibinfo {author} {\bibfnamefont {C.}~\bibnamefont {Huang}}, \bibinfo {author} {\bibfnamefont {T.}~\bibnamefont {Wolf}}, \bibinfo {author} {\bibfnamefont {N.}~\bibnamefont {Wei}}, \bibinfo {author} {\bibfnamefont {I.}~\bibnamefont {Blinov}},\ and\ \bibinfo {author} {\bibfnamefont {A.~H.}\ \bibnamefont {MacDonald}},\ }\href {https://doi.org/10.1103/PhysRevLett.130.146001} {\bibfield  {journal} {\bibinfo  {journal} {Phys. Rev. Lett.}\ }\textbf {\bibinfo {volume} {130}},\ \bibinfo {pages} {146001} (\bibinfo {year} {2023})}\BibitemShut {NoStop}%
\bibitem [{\citenamefont {Furukawa}\ \emph {et~al.}(1998)\citenamefont {Furukawa}, \citenamefont {Rice},\ and\ \citenamefont {Salmhofer}}]{Furukawa1998PRL}%
  \BibitemOpen
  \bibfield  {author} {\bibinfo {author} {\bibfnamefont {N.}~\bibnamefont {Furukawa}}, \bibinfo {author} {\bibfnamefont {T.~M.}\ \bibnamefont {Rice}},\ and\ \bibinfo {author} {\bibfnamefont {M.}~\bibnamefont {Salmhofer}},\ }\href {https://doi.org/10.1103/PhysRevLett.81.3195} {\bibfield  {journal} {\bibinfo  {journal} {Phys. Rev. Lett.}\ }\textbf {\bibinfo {volume} {81}},\ \bibinfo {pages} {3195} (\bibinfo {year} {1998})}\BibitemShut {NoStop}%
\bibitem [{\citenamefont {Nandkishore}\ \emph {et~al.}(2012)\citenamefont {Nandkishore}, \citenamefont {Levitov},\ and\ \citenamefont {Chubukov}}]{Nandkishore2012}%
  \BibitemOpen
  \bibfield  {author} {\bibinfo {author} {\bibfnamefont {R.}~\bibnamefont {Nandkishore}}, \bibinfo {author} {\bibfnamefont {L.~S.}\ \bibnamefont {Levitov}},\ and\ \bibinfo {author} {\bibfnamefont {A.~V.}\ \bibnamefont {Chubukov}},\ }\href {https://doi.org/10.1038/nphys2208} {\bibfield  {journal} {\bibinfo  {journal} {Nature Physics}\ }\textbf {\bibinfo {volume} {8}},\ \bibinfo {pages} {158} (\bibinfo {year} {2012})}\BibitemShut {NoStop}%
\bibitem [{\citenamefont {Yang}\ \emph {et~al.}(2013)\citenamefont {Yang}, \citenamefont {Wang},\ and\ \citenamefont {Lee}}]{FaWang2013}%
  \BibitemOpen
  \bibfield  {author} {\bibinfo {author} {\bibfnamefont {F.}~\bibnamefont {Yang}}, \bibinfo {author} {\bibfnamefont {F.}~\bibnamefont {Wang}},\ and\ \bibinfo {author} {\bibfnamefont {D.-H.}\ \bibnamefont {Lee}},\ }\href {https://doi.org/10.1103/PhysRevB.88.100504} {\bibfield  {journal} {\bibinfo  {journal} {Phys. Rev. B}\ }\textbf {\bibinfo {volume} {88}},\ \bibinfo {pages} {100504} (\bibinfo {year} {2013})}\BibitemShut {NoStop}%
\bibitem [{\citenamefont {Isobe}\ \emph {et~al.}(2018)\citenamefont {Isobe}, \citenamefont {Yuan},\ and\ \citenamefont {Fu}}]{Isobe2018PRX}%
  \BibitemOpen
  \bibfield  {author} {\bibinfo {author} {\bibfnamefont {H.}~\bibnamefont {Isobe}}, \bibinfo {author} {\bibfnamefont {N.~F.~Q.}\ \bibnamefont {Yuan}},\ and\ \bibinfo {author} {\bibfnamefont {L.}~\bibnamefont {Fu}},\ }\href {https://doi.org/10.1103/PhysRevX.8.041041} {\bibfield  {journal} {\bibinfo  {journal} {Phys. Rev. X}\ }\textbf {\bibinfo {volume} {8}},\ \bibinfo {pages} {041041} (\bibinfo {year} {2018})}\BibitemShut {NoStop}%
\bibitem [{\citenamefont {Sherkunov}\ and\ \citenamefont {Betouras}(2018)}]{SherkunovPRB}%
  \BibitemOpen
  \bibfield  {author} {\bibinfo {author} {\bibfnamefont {Y.}~\bibnamefont {Sherkunov}}\ and\ \bibinfo {author} {\bibfnamefont {J.~J.}\ \bibnamefont {Betouras}},\ }\href@noop {} {\bibfield  {journal} {\bibinfo  {journal} {Phys. Rev. B}\ }\textbf {\bibinfo {volume} {98}},\ \bibinfo {pages} {205151} (\bibinfo {year} {2018})}\BibitemShut {NoStop}%
\bibitem [{\citenamefont {You}\ and\ \citenamefont {Vishwanath}(2022)}]{Ashvin}%
  \BibitemOpen
  \bibfield  {author} {\bibinfo {author} {\bibfnamefont {Y.-Z.}\ \bibnamefont {You}}\ and\ \bibinfo {author} {\bibfnamefont {A.}~\bibnamefont {Vishwanath}},\ }\href {https://doi.org/10.1103/PhysRevB.105.134524} {\bibfield  {journal} {\bibinfo  {journal} {Phys. Rev. B}\ }\textbf {\bibinfo {volume} {105}},\ \bibinfo {pages} {134524} (\bibinfo {year} {2022})}\BibitemShut {NoStop}%
\bibitem [{\citenamefont {Nandkishore}\ \emph {et~al.}(2014)\citenamefont {Nandkishore}, \citenamefont {Thomale},\ and\ \citenamefont {Chubukov}}]{Nandkishore2014}%
  \BibitemOpen
  \bibfield  {author} {\bibinfo {author} {\bibfnamefont {R.}~\bibnamefont {Nandkishore}}, \bibinfo {author} {\bibfnamefont {R.}~\bibnamefont {Thomale}},\ and\ \bibinfo {author} {\bibfnamefont {A.~V.}\ \bibnamefont {Chubukov}},\ }\href {https://doi.org/10.1103/PhysRevB.89.144501} {\bibfield  {journal} {\bibinfo  {journal} {Phys. Rev. B}\ }\textbf {\bibinfo {volume} {89}},\ \bibinfo {pages} {144501} (\bibinfo {year} {2014})}\BibitemShut {NoStop}%
\bibitem [{\citenamefont {Van~Hove}(1953)}]{VanHovePhysRev}%
  \BibitemOpen
  \bibfield  {author} {\bibinfo {author} {\bibfnamefont {L.}~\bibnamefont {Van~Hove}},\ }\href {https://doi.org/10.1103/PhysRev.89.1189} {\bibfield  {journal} {\bibinfo  {journal} {Phys. Rev.}\ }\textbf {\bibinfo {volume} {89}},\ \bibinfo {pages} {1189} (\bibinfo {year} {1953})}\BibitemShut {NoStop}%
\bibitem [{\citenamefont {Shtyk}\ \emph {et~al.}(2017)\citenamefont {Shtyk}, \citenamefont {Goldstein},\ and\ \citenamefont {Chamon}}]{Shtyk2017PRB}%
  \BibitemOpen
  \bibfield  {author} {\bibinfo {author} {\bibfnamefont {A.}~\bibnamefont {Shtyk}}, \bibinfo {author} {\bibfnamefont {G.}~\bibnamefont {Goldstein}},\ and\ \bibinfo {author} {\bibfnamefont {C.}~\bibnamefont {Chamon}},\ }\href {https://doi.org/10.1103/PhysRevB.95.035137} {\bibfield  {journal} {\bibinfo  {journal} {Phys. Rev. B}\ }\textbf {\bibinfo {volume} {95}},\ \bibinfo {pages} {035137} (\bibinfo {year} {2017})}\BibitemShut {NoStop}%
\bibitem [{\citenamefont {Yuan}\ \emph {et~al.}(2019)\citenamefont {Yuan}, \citenamefont {Isobe},\ and\ \citenamefont {Fu}}]{Yuan2019}%
  \BibitemOpen
  \bibfield  {author} {\bibinfo {author} {\bibfnamefont {N.~F.~Q.}\ \bibnamefont {Yuan}}, \bibinfo {author} {\bibfnamefont {H.}~\bibnamefont {Isobe}},\ and\ \bibinfo {author} {\bibfnamefont {L.}~\bibnamefont {Fu}},\ }\href {https://doi.org/10.1038/s41467-019-13670-9} {\bibfield  {journal} {\bibinfo  {journal} {Nature Communications}\ }\textbf {\bibinfo {volume} {10}},\ \bibinfo {pages} {5769} (\bibinfo {year} {2019})}\BibitemShut {NoStop}%
\bibitem [{\citenamefont {Classen}\ \emph {et~al.}(2020)\citenamefont {Classen}, \citenamefont {Chubukov}, \citenamefont {Honerkamp},\ and\ \citenamefont {Scherer}}]{Classen2020PRB}%
  \BibitemOpen
  \bibfield  {author} {\bibinfo {author} {\bibfnamefont {L.}~\bibnamefont {Classen}}, \bibinfo {author} {\bibfnamefont {A.~V.}\ \bibnamefont {Chubukov}}, \bibinfo {author} {\bibfnamefont {C.}~\bibnamefont {Honerkamp}},\ and\ \bibinfo {author} {\bibfnamefont {M.~M.}\ \bibnamefont {Scherer}},\ }\href {https://doi.org/10.1103/PhysRevB.102.125141} {\bibfield  {journal} {\bibinfo  {journal} {Phys. Rev. B}\ }\textbf {\bibinfo {volume} {102}},\ \bibinfo {pages} {125141} (\bibinfo {year} {2020})}\BibitemShut {NoStop}%
\bibitem [{\citenamefont {Isobe}\ and\ \citenamefont {Fu}(2019)}]{Isobe2019}%
  \BibitemOpen
  \bibfield  {author} {\bibinfo {author} {\bibfnamefont {H.}~\bibnamefont {Isobe}}\ and\ \bibinfo {author} {\bibfnamefont {L.}~\bibnamefont {Fu}},\ }\href {https://doi.org/10.1103/PhysRevResearch.1.033206} {\bibfield  {journal} {\bibinfo  {journal} {Phys. Rev. Res.}\ }\textbf {\bibinfo {volume} {1}},\ \bibinfo {pages} {033206} (\bibinfo {year} {2019})}\BibitemShut {NoStop}%
\bibitem [{\citenamefont {Link}\ \emph {et~al.}(2019)\citenamefont {Link}, \citenamefont {Forti}, \citenamefont {St\"ohr}, \citenamefont {K\"uster}, \citenamefont {R\"osner}, \citenamefont {Hirschmeier}, \citenamefont {Chen}, \citenamefont {Avila}, \citenamefont {Asensio}, \citenamefont {Zakharov}, \citenamefont {Wehling}, \citenamefont {Lichtenstein}, \citenamefont {Katsnelson},\ and\ \citenamefont {Starke}}]{Link2019}%
  \BibitemOpen
  \bibfield  {author} {\bibinfo {author} {\bibfnamefont {S.}~\bibnamefont {Link}}, \bibinfo {author} {\bibfnamefont {S.}~\bibnamefont {Forti}}, \bibinfo {author} {\bibfnamefont {A.}~\bibnamefont {St\"ohr}}, \bibinfo {author} {\bibfnamefont {K.}~\bibnamefont {K\"uster}}, \bibinfo {author} {\bibfnamefont {M.}~\bibnamefont {R\"osner}}, \bibinfo {author} {\bibfnamefont {D.}~\bibnamefont {Hirschmeier}}, \bibinfo {author} {\bibfnamefont {C.}~\bibnamefont {Chen}}, \bibinfo {author} {\bibfnamefont {J.}~\bibnamefont {Avila}}, \bibinfo {author} {\bibfnamefont {M.~C.}\ \bibnamefont {Asensio}}, \bibinfo {author} {\bibfnamefont {A.~A.}\ \bibnamefont {Zakharov}}, \bibinfo {author} {\bibfnamefont {T.~O.}\ \bibnamefont {Wehling}}, \bibinfo {author} {\bibfnamefont {A.~I.}\ \bibnamefont {Lichtenstein}}, \bibinfo {author} {\bibfnamefont {M.~I.}\ \bibnamefont {Katsnelson}},\ and\ \bibinfo {author} {\bibfnamefont {U.}~\bibnamefont {Starke}},\ }\href {https://doi.org/10.1103/PhysRevB.100.121407} {\bibfield  {journal} {\bibinfo
  {journal} {Phys. Rev. B}\ }\textbf {\bibinfo {volume} {100}},\ \bibinfo {pages} {121407} (\bibinfo {year} {2019})}\BibitemShut {NoStop}%
\bibitem [{\citenamefont {Castro}\ \emph {et~al.}(2023)\citenamefont {Castro}, \citenamefont {Shaffer}, \citenamefont {Wu},\ and\ \citenamefont {Santos}}]{castro2023emergence}%
  \BibitemOpen
  \bibfield  {author} {\bibinfo {author} {\bibfnamefont {P.}~\bibnamefont {Castro}}, \bibinfo {author} {\bibfnamefont {D.}~\bibnamefont {Shaffer}}, \bibinfo {author} {\bibfnamefont {Y.-M.}\ \bibnamefont {Wu}},\ and\ \bibinfo {author} {\bibfnamefont {L.~H.}\ \bibnamefont {Santos}},\ }\href@noop {} {\bibinfo {title} {Emergence of chern supermetal and pair-density wave through higher-order van hove singularities in the haldane-hubbard model}} (\bibinfo {year} {2023}),\ \Eprint {https://arxiv.org/abs/2303.12833} {arXiv:2303.12833 [cond-mat.str-el]} \BibitemShut {NoStop}%
\bibitem [{\citenamefont {Pan}\ \emph {et~al.}(2020)\citenamefont {Pan}, \citenamefont {Wu},\ and\ \citenamefont {Das~Sarma}}]{PanPRR2020}%
  \BibitemOpen
  \bibfield  {author} {\bibinfo {author} {\bibfnamefont {H.}~\bibnamefont {Pan}}, \bibinfo {author} {\bibfnamefont {F.}~\bibnamefont {Wu}},\ and\ \bibinfo {author} {\bibfnamefont {S.}~\bibnamefont {Das~Sarma}},\ }\href {https://doi.org/10.1103/PhysRevResearch.2.033087} {\bibfield  {journal} {\bibinfo  {journal} {Phys. Rev. Res.}\ }\textbf {\bibinfo {volume} {2}},\ \bibinfo {pages} {033087} (\bibinfo {year} {2020})}\BibitemShut {NoStop}%
\bibitem [{\citenamefont {Zang}\ \emph {et~al.}(2021)\citenamefont {Zang}, \citenamefont {Wang}, \citenamefont {Cano},\ and\ \citenamefont {Millis}}]{ZangPRB2021}%
  \BibitemOpen
  \bibfield  {author} {\bibinfo {author} {\bibfnamefont {J.}~\bibnamefont {Zang}}, \bibinfo {author} {\bibfnamefont {J.}~\bibnamefont {Wang}}, \bibinfo {author} {\bibfnamefont {J.}~\bibnamefont {Cano}},\ and\ \bibinfo {author} {\bibfnamefont {A.~J.}\ \bibnamefont {Millis}},\ }\href {https://doi.org/10.1103/PhysRevB.104.075150} {\bibfield  {journal} {\bibinfo  {journal} {Phys. Rev. B}\ }\textbf {\bibinfo {volume} {104}},\ \bibinfo {pages} {075150} (\bibinfo {year} {2021})}\BibitemShut {NoStop}%
\bibitem [{\citenamefont {Szab\'o}\ and\ \citenamefont {Roy}(2022)}]{Szabo2022}%
  \BibitemOpen
  \bibfield  {author} {\bibinfo {author} {\bibfnamefont {A.~L.}\ \bibnamefont {Szab\'o}}\ and\ \bibinfo {author} {\bibfnamefont {B.}~\bibnamefont {Roy}},\ }\href {https://doi.org/10.1103/PhysRevB.105.L201107} {\bibfield  {journal} {\bibinfo  {journal} {Phys. Rev. B}\ }\textbf {\bibinfo {volume} {105}},\ \bibinfo {pages} {L201107} (\bibinfo {year} {2022})}\BibitemShut {NoStop}%
\bibitem [{\citenamefont {Dong}\ \emph {et~al.}(2023{\natexlab{a}})\citenamefont {Dong}, \citenamefont {Chubukov},\ and\ \citenamefont {Levitov}}]{Dong2023a}%
  \BibitemOpen
  \bibfield  {author} {\bibinfo {author} {\bibfnamefont {Z.}~\bibnamefont {Dong}}, \bibinfo {author} {\bibfnamefont {A.~V.}\ \bibnamefont {Chubukov}},\ and\ \bibinfo {author} {\bibfnamefont {L.}~\bibnamefont {Levitov}},\ }\href {https://doi.org/10.1103/PhysRevB.107.174512} {\bibfield  {journal} {\bibinfo  {journal} {Phys. Rev. B}\ }\textbf {\bibinfo {volume} {107}},\ \bibinfo {pages} {174512} (\bibinfo {year} {2023}{\natexlab{a}})}\BibitemShut {NoStop}%
\bibitem [{\citenamefont {Dong}\ \emph {et~al.}(2023{\natexlab{b}})\citenamefont {Dong}, \citenamefont {Levitov},\ and\ \citenamefont {Chubukov}}]{Dong2023}%
  \BibitemOpen
  \bibfield  {author} {\bibinfo {author} {\bibfnamefont {Z.}~\bibnamefont {Dong}}, \bibinfo {author} {\bibfnamefont {L.}~\bibnamefont {Levitov}},\ and\ \bibinfo {author} {\bibfnamefont {A.~V.}\ \bibnamefont {Chubukov}},\ }\href {https://doi.org/10.1103/PhysRevB.108.134503} {\bibfield  {journal} {\bibinfo  {journal} {Phys. Rev. B}\ }\textbf {\bibinfo {volume} {108}},\ \bibinfo {pages} {134503} (\bibinfo {year} {2023}{\natexlab{b}})}\BibitemShut {NoStop}%
\bibitem [{\citenamefont {Chichinadze}\ \emph {et~al.}(2022{\natexlab{a}})\citenamefont {Chichinadze}, \citenamefont {Classen}, \citenamefont {Wang},\ and\ \citenamefont {Chubukov}}]{ChichinadzePRL2022}%
  \BibitemOpen
  \bibfield  {author} {\bibinfo {author} {\bibfnamefont {D.~V.}\ \bibnamefont {Chichinadze}}, \bibinfo {author} {\bibfnamefont {L.}~\bibnamefont {Classen}}, \bibinfo {author} {\bibfnamefont {Y.}~\bibnamefont {Wang}},\ and\ \bibinfo {author} {\bibfnamefont {A.~V.}\ \bibnamefont {Chubukov}},\ }\href {https://doi.org/10.1103/PhysRevLett.128.227601} {\bibfield  {journal} {\bibinfo  {journal} {Phys. Rev. Lett.}\ }\textbf {\bibinfo {volume} {128}},\ \bibinfo {pages} {227601} (\bibinfo {year} {2022}{\natexlab{a}})}\BibitemShut {NoStop}%
\bibitem [{\citenamefont {Chichinadze}\ \emph {et~al.}(2022{\natexlab{b}})\citenamefont {Chichinadze}, \citenamefont {Classen}, \citenamefont {Wang},\ and\ \citenamefont {Chubukov}}]{Chichinadze2022}%
  \BibitemOpen
  \bibfield  {author} {\bibinfo {author} {\bibfnamefont {D.~V.}\ \bibnamefont {Chichinadze}}, \bibinfo {author} {\bibfnamefont {L.}~\bibnamefont {Classen}}, \bibinfo {author} {\bibfnamefont {Y.}~\bibnamefont {Wang}},\ and\ \bibinfo {author} {\bibfnamefont {A.~V.}\ \bibnamefont {Chubukov}},\ }\href {https://doi.org/10.1038/s41535-022-00520-z} {\bibfield  {journal} {\bibinfo  {journal} {npj Quantum Materials}\ }\textbf {\bibinfo {volume} {7}},\ \bibinfo {pages} {114} (\bibinfo {year} {2022}{\natexlab{b}})}\BibitemShut {NoStop}%
\bibitem [{\citenamefont {Murray}\ and\ \citenamefont {Vafek}(2014)}]{Vafek2014}%
  \BibitemOpen
  \bibfield  {author} {\bibinfo {author} {\bibfnamefont {J.~M.}\ \bibnamefont {Murray}}\ and\ \bibinfo {author} {\bibfnamefont {O.}~\bibnamefont {Vafek}},\ }\href {https://doi.org/10.1103/PhysRevB.89.201110} {\bibfield  {journal} {\bibinfo  {journal} {Phys. Rev. B}\ }\textbf {\bibinfo {volume} {89}},\ \bibinfo {pages} {201110} (\bibinfo {year} {2014})}\BibitemShut {NoStop}%
\bibitem [{\citenamefont {Xing}\ \emph {et~al.}(2017)\citenamefont {Xing}, \citenamefont {Classen}, \citenamefont {Khodas},\ and\ \citenamefont {Chubukov}}]{Xing2017}%
  \BibitemOpen
  \bibfield  {author} {\bibinfo {author} {\bibfnamefont {R.-Q.}\ \bibnamefont {Xing}}, \bibinfo {author} {\bibfnamefont {L.}~\bibnamefont {Classen}}, \bibinfo {author} {\bibfnamefont {M.}~\bibnamefont {Khodas}},\ and\ \bibinfo {author} {\bibfnamefont {A.~V.}\ \bibnamefont {Chubukov}},\ }\href {https://doi.org/10.1103/PhysRevB.95.085108} {\bibfield  {journal} {\bibinfo  {journal} {Phys. Rev. B}\ }\textbf {\bibinfo {volume} {95}},\ \bibinfo {pages} {085108} (\bibinfo {year} {2017})}\BibitemShut {NoStop}%
\bibitem [{\citenamefont {Chichinadze}\ \emph {et~al.}(2020)\citenamefont {Chichinadze}, \citenamefont {Classen},\ and\ \citenamefont {Chubukov}}]{Chichinadze2020}%
  \BibitemOpen
  \bibfield  {author} {\bibinfo {author} {\bibfnamefont {D.~V.}\ \bibnamefont {Chichinadze}}, \bibinfo {author} {\bibfnamefont {L.}~\bibnamefont {Classen}},\ and\ \bibinfo {author} {\bibfnamefont {A.~V.}\ \bibnamefont {Chubukov}},\ }\href {https://doi.org/10.1103/PhysRevB.102.125120} {\bibfield  {journal} {\bibinfo  {journal} {Phys. Rev. B}\ }\textbf {\bibinfo {volume} {102}},\ \bibinfo {pages} {125120} (\bibinfo {year} {2020})}\BibitemShut {NoStop}%
\bibitem [{\citenamefont {Klebl}\ \emph {et~al.}(2023)\citenamefont {Klebl}, \citenamefont {Fischer}, \citenamefont {Classen}, \citenamefont {Scherer},\ and\ \citenamefont {Kennes}}]{Klebl2023}%
  \BibitemOpen
  \bibfield  {author} {\bibinfo {author} {\bibfnamefont {L.}~\bibnamefont {Klebl}}, \bibinfo {author} {\bibfnamefont {A.}~\bibnamefont {Fischer}}, \bibinfo {author} {\bibfnamefont {L.}~\bibnamefont {Classen}}, \bibinfo {author} {\bibfnamefont {M.~M.}\ \bibnamefont {Scherer}},\ and\ \bibinfo {author} {\bibfnamefont {D.~M.}\ \bibnamefont {Kennes}},\ }\href {https://doi.org/10.1103/PhysRevResearch.5.L012034} {\bibfield  {journal} {\bibinfo  {journal} {Phys. Rev. Res.}\ }\textbf {\bibinfo {volume} {5}},\ \bibinfo {pages} {L012034} (\bibinfo {year} {2023})}\BibitemShut {NoStop}%
\bibitem [{\citenamefont {Wang}\ \emph {et~al.}(2020)\citenamefont {Wang}, \citenamefont {Shih}, \citenamefont {Ghiotto}, \citenamefont {Xian}, \citenamefont {Rhodes}, \citenamefont {Tan}, \citenamefont {Claassen}, \citenamefont {Kennes}, \citenamefont {Bai}, \citenamefont {Kim}, \citenamefont {Watanabe}, \citenamefont {Taniguchi}, \citenamefont {Zhu}, \citenamefont {Hone}, \citenamefont {Rubio}, \citenamefont {Pasupathy},\ and\ \citenamefont {Dean}}]{Wang2020}%
  \BibitemOpen
  \bibfield  {author} {\bibinfo {author} {\bibfnamefont {L.}~\bibnamefont {Wang}}, \bibinfo {author} {\bibfnamefont {E.-M.}\ \bibnamefont {Shih}}, \bibinfo {author} {\bibfnamefont {A.}~\bibnamefont {Ghiotto}}, \bibinfo {author} {\bibfnamefont {L.}~\bibnamefont {Xian}}, \bibinfo {author} {\bibfnamefont {D.~A.}\ \bibnamefont {Rhodes}}, \bibinfo {author} {\bibfnamefont {C.}~\bibnamefont {Tan}}, \bibinfo {author} {\bibfnamefont {M.}~\bibnamefont {Claassen}}, \bibinfo {author} {\bibfnamefont {D.~M.}\ \bibnamefont {Kennes}}, \bibinfo {author} {\bibfnamefont {Y.}~\bibnamefont {Bai}}, \bibinfo {author} {\bibfnamefont {B.}~\bibnamefont {Kim}}, \bibinfo {author} {\bibfnamefont {K.}~\bibnamefont {Watanabe}}, \bibinfo {author} {\bibfnamefont {T.}~\bibnamefont {Taniguchi}}, \bibinfo {author} {\bibfnamefont {X.}~\bibnamefont {Zhu}}, \bibinfo {author} {\bibfnamefont {J.}~\bibnamefont {Hone}}, \bibinfo {author} {\bibfnamefont {A.}~\bibnamefont {Rubio}}, \bibinfo {author} {\bibfnamefont {A.~N.}\ \bibnamefont {Pasupathy}},\
  and\ \bibinfo {author} {\bibfnamefont {C.~R.}\ \bibnamefont {Dean}},\ }\href {https://doi.org/10.1038/s41563-020-0708-6} {\bibfield  {journal} {\bibinfo  {journal} {Nature Materials}\ }\textbf {\bibinfo {volume} {19}},\ \bibinfo {pages} {861} (\bibinfo {year} {2020})}\BibitemShut {NoStop}%
\end{thebibliography}%
\end{document}